\documentclass[10pt,a4paper,fleqn]{article}
\usepackage{jheppub}

\usepackage[T1]{fontenc}
\usepackage{amsmath}
\usepackage{amssymb}
\usepackage[english]{babel}
\usepackage[dvipsnames]{xcolor}
\definecolor{linkcolor}{rgb}{.17578125,.1875,.5703125}
\usepackage{hyperref}
\hypersetup{
colorlinks=true,
linkcolor=linkcolor,
filecolor=magenta,
urlcolor=linkcolor,
citecolor=BrickRed
}
\usepackage{slashed}
\usepackage{graphicx}
\usepackage{feynmp-auto}
\usepackage{comment}
\usepackage{braket}
\usepackage{bbm}
\usepackage{empheq}
\usepackage{parskip}
\usepackage{tensor}
\usepackage[normalem]{ulem}

\usepackage{tikz}
\usetikzlibrary{decorations.pathmorphing}
\usetikzlibrary{decorations.markings}
\usetikzlibrary{shapes.misc}
\usetikzlibrary{arrows.meta}
\usetikzlibrary{positioning}

\newcommand{\pvec}{{\boldsymbol{p}}}
\newcommand{\xvec}{{\boldsymbol{x}}}
\newcommand{\zvec}{{\boldsymbol{z}}}
\newcommand{\yvec}{{\boldsymbol{y}}}
\newcommand{\numvec}[1]{\boldsymbol{#1}}

\newcommand{\zero}{\boldsymbol{0}}
\newcommand{\Mbold}{\mathbf{M}}
\newcommand{\Gammabold}{\boldsymbol{\Gamma}}

\newcommand{\Scal}{\mathcal{S}}
\newcommand{\Mcal}{\mathcal{M}}

\newcommand{\Hcal}{\mathcal{H}}

\newcommand{\Tcal}{\mathcal{T}}

\newcommand{\gim}{\text{\tiny GIM}}

\newcommand{\dd}{\mathrm{d}}
\newcommand{\e}{\mathrm{e}}
\newcommand{\ii}{\mathrm{i}}
\newcommand{\w}{\mathrm{w}}

\newcommand{\M}{\mathrm{M}}
\newcommand{\F}{\mathrm{F}}

\newcommand{\D}{{D}}
\newcommand{\Dbar}{\widebar{{D}}}
\newcommand{\K}{{K}}
\newcommand{\Kbar}{\widebar{{K}}}

\newcommand{\nn}{\nonumber}

\DeclareFontFamily{U}{mathx}{\hyphenchar\font45}
\DeclareFontShape{U}{mathx}{m}{n}{<-> mathx10}{}
\DeclareSymbolFont{mathx}{U}{mathx}{m}{n}
\DeclareMathAccent{\widebar}{0}{mathx}{"73}

\definecolor{green2}{HTML}{45CB85}

\usepackage{cleveref}
\crefformat{section}{section~#2#1#3}
\crefmultiformat{section}{sections~#2#1#3}
    { and~#2#1#3}{, #2#1#3}{ and~#2#1#3}
\crefrangeformat{section}{sections~#3#1#4\,--\,#5#2#6}
\crefformat{subsection}{section~#2#1#3}
\crefmultiformat{subsection}{sections~#2#1#3}
    { and~#2#1#3}{, #2#1#3}{ and~#2#1#3}
\crefrangeformat{subsection}{sections~#3#1#4\,--\,#5#2#6}
\crefformat{subsubsection}{section~#2#1#3}
\crefmultiformat{subsubsection}{sections~#2#1#3}
    { and~#2#1#3}{, #2#1#3}{ and~#2#1#3}
\crefrangeformat{subsubsection}{sections~#3#1#4\,--\,#5#2#6}
\crefformat{figure}{figure~#2#1#3}
\crefmultiformat{figure}{figures~#2#1#3}
    { and~#2#1#3}{, #2#1#3}{ and~#2#1#3}
\crefrangeformat{figure}{figures~#3#1#4\,--\,#5#2#6}
\crefformat{table}{table~#2#1#3}
\crefmultiformat{table}{tables~#2#1#3}
    { and~#2#1#3}{, #2#1#3}{ and~#2#1#3}
\crefrangeformat{table}{tables~#3#1#4\,--\,#5#2#6}
\crefformat{equation}{eq.~(#2#1#3)}
\crefmultiformat{equation}{eqs.~(#2#1#3)}{ and~(#2#1#3)}{, (#2#1#3)}{ and~(#2#1#3)}
\crefrangeformat{equation}{eqs.~(#3#1#4)--(#5#2#6)}

\newcommand{\cern}{Theoretical Physics Department, CERN, 1211 Geneva 23, Switzerland}
\newcommand{\edin}{School of Physics and Astronomy, University of Edinburgh, Edinburgh EH9 3FD, United Kingdom}

\title{Long distance contributions to neutral $D$-meson mixing from lattice QCD}

\author[a]{Matteo Di Carlo,}
\author[a]{Felix Erben,}
\author[b]{Maxwell T. Hansen}

\affiliation[a]{\cern}
\affiliation[b]{\edin}

\emailAdd{matteo.dicarlo@cern.ch}
\emailAdd{felix.erben@cern.ch}
\emailAdd{maxwell.hansen@ed.ac.uk}

\abstract{
The study of neutral $D$-meson mixing provides a unique probe of long-distance effects in the charm sector, where Standard Model contributions are dominated by nonperturbative effects. In this work, we investigate the feasibility of using spectral reconstruction techniques within lattice QCD to compute the long-distance contributions to $D^0- \widebar{D}^0$ mixing. After outlining the general formalism describing neutral meson mixing in the charm sector, we focus on the determination of the mixing amplitudes and the dimensionless parameters $x = \Delta m_D / \Gamma_D$ and $y = \Delta \Gamma_D /(2 \Gamma_D)$, which respectively encode the mass and width differences between the $D$-meson mass eigenstates. We discuss in detail the required theoretical and computational framework, including the definition and renormalization of the four-quark operators entering the $\Delta C = 1$ weak Hamiltonian, and strategies for evaluating the relevant correlation functions employing variance-reduction techniques. To extract the mixing amplitudes, we explore methods for reconstructing the spectral density from lattice correlators, providing preliminary assessments of the data quality required to reach the scaling regime, where the smearing width is small enough to yield physically meaningful results. Our findings lay the groundwork for future precision determinations of long-distance contributions to $D$-meson mixing from first principles.}

\preprint{CERN-TH-2025-075}

\renewcommand{\beforetochook}{\bigskip\bigskip}

\usepackage{etoolbox}
\appto\appendix{\addtocontents{toc}{\protect\setcounter{tocdepth}{1}}}

\begin{document}

\tikzset{->-/.style={decoration={
markings,
mark=at position .6 with {\arrow{>}}},postaction={decorate}}}
\tikzset{cross/.style={cross out, draw=black, minimum size=2*(#1-\pgflinewidth), inner sep=0pt, outer sep=0pt},
cross/.default={5pt}}

\maketitle

\newpage

\section{Introduction}

The first evidence of neutral $D$-meson mixing was reported by the Belle collaboration~\cite{BELLE:2007dgh} and the BaBar collaboration~\cite{BaBar:2007kib}. These initial findings were subsequently confirmed by CDF~\cite{CDF:2007bdz} and later measured with high precision by LHCb~\cite{LHCb:2012zll}. Since then, multiple independent studies have corroborated these results~\cite{CDF:2013gvz,Belle:2014yoi,LHCb:2016zmn,LHCb:2018zpj,LHCb:2019mxy,LHCb:2021ykz,LHCb:2021dcr,LHCb:2022gnc,LHCb:2024hyb,LHCb:2024,LHCb:2025kch}, with LHCb achieving remarkable precision. More recently, the Belle and Belle II collaborations refined their analysis of the mixing parameters~\cite{Belle-II:2024gfc}, aligning their results with those from LHCb, albeit with slightly lower precision.

In experimental studies, the key observables related to neutral $D$-meson mixing are the parameters $x$ and $y$, and the CP violation parameters $|q/p|$ and $\phi=\mathrm{arg}(q/p)$. The most recent report from the Heavy Flavor Averaging Group (HFLAV)~\cite{HFLAV:2022esi} decisively excludes the no-mixing hypothesis ($x = y = 0$) with a significance exceeding $11.5\sigma$:
\begin{align}
    x &= \frac{\Delta m_\D}{\Gamma_\D} = 0.407(44)\%\,, \qquad
    y = \frac{\Delta\Gamma_\D}{2\Gamma_\D} = 0.643^{+0.024}_{-0.023}\%.
\end{align}
Compatible results are obtained from the recent global fit analysis in ref.~\cite{Betti:2024ldy}. Despite this strong evidence for mixing, no significant indication of indirect CP violation has been observed and measurements remain consistent with $|q/p| = 1$ and $\arg(p/q) = 0$. Interestingly, as in the $\K^0$-$\Kbar^0$ system, the predominantly CP-even state of the neutral $D$ meson is shorter-lived. However, unlike in the kaon system, the CP-even state is heavier than the CP-odd state.

Theoretical understanding of neutral $\D$-mixing remains limited. Short-distance contributions have been explored via lattice QCD calculations~\cite{Carrasco:2015pra,Bazavov:2017weg}, but long-distance effects dominate the mixing process. These effects are typically analysed through three phenomenological approaches: the \emph{inclusive}, \emph{exclusive}, and \emph{dispersive} frameworks.

The inclusive approach, based on the heavy-quark expansion (HQE)~\cite{Khoze:1983yp,Shifman:1984wx,Bigi:1992su}, expresses the mixing parameters as series in $\Lambda_{\text{QCD}} / m_c$. This method has been highly successful in describing $B_{(s)}$-meson mixing, but its application to $D$ mesons faces challenges. In particular, strong Glashow-Iliopoulos-Maiani (GIM)~\cite{Glashow:1970gm} cancellations in the charm sector suppress predictions of $x$ and $y$ by up to four orders of magnitude. Even when next-to-leading order (NLO) corrections are included~\cite{Golowich:2005pt}, significant cancellations between leading and subleading terms persist, especially in the prediction of~$x$. Attempts to resolve this include renormalization schemes with quark-flavour-dependent scales~\cite{Lenz:2020efu} and treatments involving nonlocal QCD condensates~\cite{Melic:2024oqj}, both of which improve agreement with experiments but still underestimate $x$.

The exclusive approach~\cite{Wolfenstein:1985ft, Donoghue:1985hh} calculates mixing parameters by summing over contributions from hadronic intermediate states. This method uses experimental inputs such as branching ratios and strong phases, and it has been suggested that when accounting for $SU(3)$-flavour symmetry breaking, some of the severe cancellations expected in the symmetric limit are lifted~\cite{Falk:2001hx}. However, it requires detailed experimental data and suffers from limited control over unmeasured or poorly known modes.

The dispersive approach~\cite{Falk:2004wg} connects $x$ and $y$ via dispersion relations, using analyticity to relate the mass and width differences to integrals over physical decay amplitudes. It avoids modeling individual intermediate states and instead leverages global features of the decay spectrum. Like the exclusive approach, it is sensitive to $SU(3)$-flavour symmetry breaking, which can enhance both $x$ and $y$ by several orders of magnitude. With these effects included, recent studies find predictions consistent with experiments~\cite{Li:2020xrz,Li:2022jxc}.

Despite progress, all three approaches suffer from sizable theoretical uncertainties and limitations in systematically quantifying errors. For recent overviews of neutral $D$-meson mixing phenomenology, see refs.~\cite{Lenz:2020awd,Petrov:2024ujw}.

In this manuscript, we propose a strategy to compute the $D$-meson mixing amplitudes, $\Delta m_D$ and $\Delta \Gamma_D$, including long-distance contributions, using lattice QCD. While analogous mixing amplitudes have already been computed in the kaon sector~\cite{PhysRevD.109.054501}, the underlying formalism~\cite{Christ:2012se,Bai:2014cva} relies on identifying and subtracting the complete finite-volume energy spectrum below the mass of the neutral meson, followed by its analytic reintegration. For kaon mixing, this typically involves only the two-pion state and, in large volumes or with high precision, perhaps a few additional states. In contrast, the spectrum of multi-hadron states below $m_D \sim 2\,\mathrm{GeV}$ is extremely dense and inaccessible in practice. For this reason, a direct application of the kaon formalism to $D$-meson mixing appears infeasible.

To overcome this challenge, we employ spectral reconstruction techniques that have proven effective in a variety of other lattice contexts~\cite{Bulava:2021fre, ExtendedTwistedMassCollaborationETMC:2022sta,Bailas:2020qmv,Gambino:2022dvu,Barone:2023tbl,Kellermann:2025pzt, Frezzotti:2023nun,Evangelista:2023fmt,ExtendedTwistedMass:2024myu,Frezzotti:2024kqk,Bennett:2024cqv,DeSantis:2025qbb,DeSantis:2025yfm,Bonanno:2023ljc,Panero:2023zdr}. These methods avoid the need for explicit knowledge of the full finite-volume spectrum by employing the reconstruction of a smeared version of the finite-volume spectral density. Once obtained, this smeared density can be evaluated directly at the energy scale of interest -- in our case, the $D$-meson mass. Although this approach introduces new challenges, it offers a viable and credible alternative to the otherwise impractical implementation of traditional methods in the charm sector.

Highlights of this work include the following. First, we decompose the $D$-meson mixing amplitudes into their $U$-spin components and demonstrate how the dominant contributions, which arise at second order in the $U$-spin symmetry breaking, can be efficiently estimated from Euclidean lattice correlators using state-of-the-art variance reduction techniques. Second, we review and compare several methods for extracting the mixing amplitudes from Euclidean correlation functions using spectral reconstruction approaches. We highlight how particular care is required -- especially when employing Chebyshev polynomials -- for the reconstruction of smearing kernels that do not decay exponentially at large energies, such as the principal value kernel relevant in this work. We further discuss how the principal value kernel can be conveniently defined such that the corresponding smeared spectral density exhibits an improved extrapolation to zero smearing width. This construction makes use of the Kramers-Kronig relations, which relate the principal value kernel to smeared Dirac delta functions.
Third, we investigate the specific spectral function relevant for $D$-meson mixing through a numerical study informed by resonance poles near the $D$-meson mass, as observed in experimental data. Finally, we outline strategies for a first-principles lattice QCD calculation of the $D$-meson mixing amplitudes, including a proposal for extrapolating to the physical point by exploiting the fact that the amplitudes vanish in the $U$-spin symmetric limit.

The remainder of this manuscript is structured as follows: in \cref{sec:mixing}, we review the general formalism of $D$-meson mixing within the Standard Model and define the key observables. \Cref{sec:renormalization} discusses the renormalization of $D$-meson mixing matrix elements, with particular emphasis on the breaking of chiral symmetry relevant for certain lattice discretizations. In \cref{sec:lattice}, we outline our proposed strategies for estimating lattice correlation functions, which are then linked to the mixing amplitude using spectral reconstruction methods, as detailed in \cref{sec:reconstruction}.

\newpage
\section{Neutral meson mixing: general formalism}
\label{sec:mixing}

In the Standard Model, neutral mesons with opposite flavour, such as ${\D^0}$ and ${\Dbar^0}$, can mix with each other because the flavour eigenstates differ from the eigenstates of the weak interactions.
To express this mixing, we first define a generic state of the system at time $t$ as
\begin{equation}
    \ket{\Psi(t)} = a(t)\ket{\D^0} + b(t)\ket{\Dbar^0}\,,
\end{equation}
where $\ket{\D^0}$ and $\ket{\Dbar^0}$ are understood as the states in the Heisenberg picture, which are time-independent, or equivalently in the Schr\"odinger picture evaluated at $t=0$.

The time evolution of the coefficients is then governed by a Schr\"odinger equation
\begin{equation}
    i\,\frac{\partial}{\partial t}
    \begin{pmatrix}
        a(t)\\
        b(t)
    \end{pmatrix}
    = \bigg(\Mbold-\frac{\ii}{2}\,\boldsymbol{\Gammabold}\bigg)
    \begin{pmatrix}
        a(t)\\
        b(t)
    \end{pmatrix}\,,
\end{equation}
where $\Mbold$ and $\Gammabold$ are $2\times 2$ Hermitian matrices defined below, whose matrix elements are denoted by $\M_{ij}$ and $\Gamma_{ij}$, respectively.
Invariance under CPT symmetry implies that ${\Gamma_{11}=\Gamma_{22}\equiv \Gamma_{\D}}$ and ${\M_{11}=\M_{22}}$. At leading order in the electroweak interaction ${\M_{11}=\M_{22}} = m_D$ is the mass of the $\D^0$ meson in QCD only. Corrections to this are irrelevant for the leading order mixing amplitude of interest and are not considered in this work.

We next define the right-eigenstates and corresponding eigenvalues of the matrix $(\Mbold-\ii\, \Gammabold/2)$ as
\begin{equation}
    \bigg(\Mbold-\frac{\ii}{2}\,\Gammabold\bigg)
    \begin{pmatrix}
        p\\
        q
    \end{pmatrix}
    = \bigg(m_1 - \frac{\ii}{2}\,\Gamma_1\bigg)
    \begin{pmatrix}
        p\\
        q
    \end{pmatrix}\,,
    \qquad
    \bigg(\Mbold-\frac{\ii}{2}\,\Gammabold\bigg)
    \begin{pmatrix}
        p\\
        -q
    \end{pmatrix}
    = \bigg(m_2 - \frac{\ii}{2}\,\Gamma_2\bigg)
    \begin{pmatrix}
        p\\
        -q
    \end{pmatrix}\,,
\end{equation}
where $p$ and $q$ are complex coefficients and the vectors are normalized such that $|p|^2+|q|^2=1$.
This defines $m_1$, $m_2$, $\Gamma_1$ and $\Gamma_2$.
We stress that these are the real and imaginary parts of the eigenvalues of the Hamiltonian matrix, respectively, and are not the eigenvalues associated to $\Mbold$ and $\Gammabold$ separately. Certain relations do hold, however, such as ${\rm Tr} (\Mbold) = m_1 + m_2$ and
${\rm Tr} (\Gammabold) =   \Gamma_1 + \Gamma_2$.

Then, defining $p_1(t)$ and $q_1(t)$ as the result of the time evolution with initial conditions $p_1(0)=p$ and $q_1(0)=q$ (similarly $p_2(t)$ and $q_2(t)$ with initial conditions $p_2(0)=p$ and $q_2(0)=-q$), we note that the coefficients evolve as
\begin{equation}
    \begin{pmatrix}
        p_1(t)\\
              q_1(t)
         \end{pmatrix}
    = \e^{-\ii m_1 t} \e^{-\Gamma_1 t / 2}
    \begin{pmatrix}
        p\\
        q
    \end{pmatrix}
    \,,
    \qquad
    \begin{pmatrix}
        p_2(t)\\
        q_2(t)  \end{pmatrix}
    = \e^{-\ii m_2 t} \e^{-\Gamma_2 t / 2}
    \begin{pmatrix}
        p\\
        -q
    \end{pmatrix}
    \,,
\end{equation}
such that the vectors do not mix.
This leads to the definition of the mass  eigenstates
\begin{align}
    \ket{\D_1} &= p\ket{\D^0} + q\ket{\Dbar^0}\,, \\
     \ket{\D_2} &= p\ket{\D^0} - q\ket{\Dbar^0}\,,
\end{align}
which then time evolve as
\begin{align}
\ket{\D_1(t)} &= \e^{-\ii m_1 t} \e^{-\Gamma_1 t / 2} \ket{\D_1}\,, \\
\ket{\D_2(t)} &= \e^{-\ii m_2 t} \e^{-\Gamma_2 t / 2} \ket{\D_2}\,.
\end{align}

As directly follows form the eigenvector definition, the ratio $(q/p)^2$ satisfies
\begin{align}
    \left(\frac{q}{p}\right)^2 =
    \frac{\M_{12}^* - (\ii/2)\,\Gamma_{12}^*}
         {\M_{12} - (\ii/2)\,\Gamma_{12}}
    \,.
\end{align}
If we assume that CP~violation is negligible in the mixing then we can apply the approximation that this is exactly conserved by the weak interaction. This corresponds to the statement that $\M_{12}$ and $\Gamma_{12}$ are real, which in turn implies $(q/p)^2 = 1$.
Then the convention defining the $(p, q)$ eigenvector is such that this corresponds to $p = q$.
In this case, the weak eigenstates are also CP~eigenstates, and it is useful to define (adopting the phase convention $\mathrm{CP}\ket{D^0} = -\ket{\bar{D}^0}$, so that $\ket{D_2}$ is the CP-even state)
\begin{align}
    \Delta m_D & = m_2 - m_1 = 2 \M_{12}\,, & \Delta \Gamma_D &= \Gamma_2 - \Gamma_1 = 2 \Gamma_{12}\,,
    \\[5pt]
    x &= \frac{\Delta m_\D}{\Gamma_\D} = 2 \frac{  \M_{12}}{\Gamma_\D}\,, &
    y & =  \frac{\Delta\Gamma_\D}{2 \Gamma_\D} = \frac{ \Gamma_{12}}{\Gamma_\D}\,,
\end{align}
where we also recall the results collected above that $\Gamma_D = (\Gamma_1 + \Gamma_2)/2 = (\Gamma_{11} + \Gamma_{22})/2 = \Gamma_{11} = \Gamma_{22}$.

If, instead, we do not assume CP~conservation, we can define the following parameters
\begin{equation}
    x_{12} = 2\,\frac{|\M_{12}|}{\Gamma_\D}\,, \quad y_{12} = \frac{|\Gamma_{12}|}{\Gamma_\D}\,, \quad \phi_{12} = \arg\bigg(\frac{\M_{12}}{\Gamma_{12}}\bigg)\,.
\end{equation}
The phase $\phi_{12}$ parametrizes CP violation in the mixing, while the CP-conserving parameters $x_{12}$ and $y_{12}$ reduce to $x$ and $y$, respectively, in the absence of CP violation~\cite{Kagan:2020vri,Betti:2024ldy}.
The purpose of this work is to derive a framework for evaluating $\M_{12}$ and $\Gamma_{12}$, referred to as the
dispersive and absorptive mixing amplitudes, respectively,
from first principles in the context of lattice QCD.

To connect this to the Standard Model, the next step is to consider the following element of the $\Scal$-matrix
\begin{equation}
    \langle{\Dbar^0, \boldsymbol{p}'_D|\Scal-\boldsymbol{1}|\D^0, \boldsymbol{p}_D}\rangle = \langle{\Dbar^0, \boldsymbol{p}'_D|\,\mathrm{T}\, \exp\Big\{{-}\ii\int \dd^4 x \,  \Hcal_\w(x)\Big\}\,|\D^0, \boldsymbol{p}_D}\rangle\,,
\end{equation}
with $\Hcal_\w(x)$ denoting the weak Hamiltonian density and $\mathrm{T}$ the time-ordering operator. The states satisfy the usual relativistic normalization
\begin{equation}
    \braket{\D^0 , \boldsymbol p_D |\D^0, \boldsymbol p_D'} = (2\pi)^3 \, 2E_\D  \, \delta^3(\boldsymbol p_D-\boldsymbol p'_D) \label{eq:momentum_norm} \,,
\end{equation}
where $E_\D = \sqrt{m_\D^2 + \boldsymbol p_D^2}$ is the energy of the $\D^0$ meson with momentum $\boldsymbol p_D$.

Rewriting $\Scal=\boldsymbol{1}+\ii \Tcal$ and expanding the above equation to second order in the weak Hamiltonian, one obtains
\begin{equation}
    \langle{\Dbar^0, \boldsymbol{p}'_D|\Tcal|\D^0, \boldsymbol{p}_D}\rangle \equiv (2\pi)^4 \delta^4(P_{\D^0}-P_{\Dbar^0}) \, \Mcal_{\D^0\to\Dbar^0}  \,,
\end{equation}
where $P_{\D^0}=(E_\D,\boldsymbol{p}_D)$ and $P_{\Dbar^0}=(E'_\D,\boldsymbol{p}'_D)$ are the four-momenta of the $\D^0$ and $\Dbar^0$ mesons, respectively and
\begin{equation}
    \Mcal_{\D^0\to\Dbar^0} = - \braket{\Dbar^0, \boldsymbol{p}_D |\Hcal_\w(0)|\D^0, \boldsymbol{p}_D} +\frac{\ii}{2} \, \int\dd^4 x \, \braket{\Dbar^0, \boldsymbol{p}_D |\mathrm{T}\big\{\Hcal_\w(x)\Hcal_\w(0)\big\}|\D^0, \boldsymbol{p}_D}\,.
    \label{eq:amplitude}
\end{equation}
Note that $\Mcal_{\D^0\to\Dbar^0}$ is a Lorentz scalar, so it does not depend on the momentum of the external mesons, provided these are equal.
The quantities $\M_{12}$ and $\Gamma_{12}$ can then be related to the amplitude $\Mcal_{\D^0\to\Dbar^0}$ via
\begin{equation}
    \bigg(\Mbold-\frac{\ii}{2}\,\Gammabold\bigg)_{12} =
    -\frac{1}{2m_{\D}} \, \Mcal_{\D^0\to\Dbar^0}\,.
\end{equation}
The first term in~\cref{eq:amplitude} receives contributions from the short-distance $\Delta C=2$ component of the weak Hamiltonian and contributes only to $\M_{12}$. This contribution, as well as analogous $\Delta C=2$ contributions beyond the Standard Model, can be computed on the lattice, as done in refs.~\cite{Carrasco:2015pra,Bazavov:2017weg}.

Alternative expressions for the long-distance contributions to the dispersive and absorptive parts  of the amplitude can be obtained from the spectral decomposition of the second term in~\cref{eq:amplitude}, which gives
\begin{equation}
    \Mcal_{\D^0\to\Dbar^0} = - \braket{\Dbar^0, \boldsymbol{p}_D |\Hcal_\w(0)|\D^0, \boldsymbol{p}_D} +\lim_{\epsilon\to 0} \int\frac{\dd\omega}{2\pi} \, \frac{\rho(\omega)}{\omega - E_\D - \ii\epsilon}\,,
    \label{eq:amplitude_2}
\end{equation}
where we have defined the spectral density
\begin{equation}
    \rho(\omega) = \braket{\bar{\D}^0, \pvec_\D|\,\Hcal_\w(0) \, (2\pi)^4\delta(\hat{H}-\omega)\delta^3(\hat{\mathbf{P}}-\pvec_\D) \, \Hcal_\w(0)\,|\D^0, \pvec_\D}\,,
    \label{eq:density_inf_vol}
\end{equation}
where $\hat{H}$ and $\hat{\mathbf{P}}$ are the Hamiltonian and momentum operators, respectively. These expressions then decompose into $\M_{12}$ and $\Gamma_{12}$ as
\begin{align}
\M_{12} &= \frac{1}{2m_\D} \, \braket{\Dbar^0, \boldsymbol{p}_D |\Hcal_\w(0)|\D^0, \boldsymbol{p}_D}  - \frac{1}{2m_\D} \, \mathrm{P.V.} \,\int\frac{\dd\omega}{2\pi}\,\frac{\rho(\omega)}{\omega-E_\D}\,,
\label{eq:M12_QFT_def}
\\
\Gamma_{12} &= \frac{1}{2m_\D} \, \rho(E_\D)\,,
\label{eq:Gamma12_QFT_def}
\end{align}
where the symbol `P.V.' denotes the principal value prescription and we have used the functional identity
\begin{equation}
    \lim_{\epsilon\to 0} \frac{1}{x - \ii \epsilon} = \mathrm{P.V.} \,\frac{1}{x} + \ii \pi \delta(x)\,.
\end{equation}

A central aspect of this work is the use of a smeared spectral function, defined via the integral in~\cref{eq:amplitude_2}, but with a non-infinitesimal $\epsilon$ in the denominator. This strategy, building on the approach developed in refs.~\cite{Hansen:2017mnd,Bulava:2019kbi,Briceno:2019opb,Bruno:2020kyl,Patella:2024cto}%
\footnote{Reference~\cite{Patella:2024cto} goes beyond the $i \epsilon$ perspective by making use of Haag-Ruelle scattering theory to provide a non-perturbative relation between scattering amplitudes and infinite-volume Euclidean correlators using axiomatic field theory.}, was proposed in ref.~\cite{Frezzotti:2023nun} for the determination of hadronic amplitudes involving two external currents and either a single hadron or the QCD vacuum as initial and final states.

We define
\begin{equation}
\hat \rho(E_D, \epsilon) = \int\frac{\dd\omega}{2\pi} \, \frac{\rho(\omega)}{\omega - E_\D- \ii\epsilon} \, \qquad \text{(non-infinitesimal $\epsilon$)}\,.
\label{eq:rho_smeared}
\end{equation}
It will be further convenient at various points to break the kernel into its real and imaginary parts:
\begin{equation}
\hat \rho(\bar\omega, \epsilon) = \hat \rho^\mathrm{R}(\bar\omega, \epsilon) + \ii \, \hat \rho^\mathrm{I}(\bar\omega, \epsilon)\,,
\end{equation}
where
\begin{align}
\hat \rho^\mathrm{R}(\bar\omega, \epsilon) = \int\frac{\dd\omega}{2\pi} \, \rho(\omega)\mathcal{K}_\epsilon^\mathrm{R}(\omega,\bar\omega)\,,\qquad
\hat \rho^\mathrm{I}(\bar\omega, \epsilon) = \int\frac{\dd\omega}{2\pi} \, \rho(\omega)\mathcal{K}_\epsilon^\mathrm{I}(\omega,\bar\omega)\,,
\label{eq:kernel-real-imag}
\end{align}
and
we have introduced
\begin{align}
    \mathcal{K}_\epsilon^\mathrm{R}(\omega,\bar\omega) = \frac{\omega - \bar\omega}{(\omega - \bar\omega)^2+\epsilon^2}\,,\qquad
    \mathcal{K}_\epsilon^\mathrm{I}(\omega,\bar\omega) = \frac{\epsilon}{(\omega - \bar\omega)^2+\epsilon^2}\,.
    \label{eq:kernels}
\end{align}
The function $\mathcal{K}_\epsilon^\mathrm{R}(\omega,\bar\omega)$ reduces to $1/(\omega-\bar\omega)$ for $\epsilon\ll|\omega-\bar\omega|$ and vanishes if $\bar\omega\to\omega$, such that in the $\epsilon\to 0$ limit it behaves as the P.V.~functional. By contrast, the function $\mathcal{K}_\epsilon^\mathrm{I}(\omega,\bar\omega)$ behaves as the Dirac delta functional, $\pi\delta(\omega-\bar\omega)$, in the limit $\epsilon\to 0$. Note that both kernels satisfy the following relations $\mathcal{K}_\epsilon^\alpha(\omega,\bar\omega) = \mathcal{K}_\epsilon^\alpha(\omega-\bar\omega,0) = \mathcal{K}_\epsilon^\alpha(0,\bar\omega-\omega)$ ($\alpha\in\{\mathrm{R,I}\}$).

As a consequence, taking the $\epsilon\to 0$ limit, one obtains the following expressions for $\M_{12}$ and $\Gamma_{12}$ in terms of the smeared spectral densities
\begin{align}
    \M_{12} &= \frac{1}{2m_\D} \, \braket{\Dbar^0, \boldsymbol{p}_D |\Hcal_\w(0)|\D^0, \boldsymbol{p}_D}  - \frac{1}{2m_\D} \, \lim_{\epsilon\to 0} \, \hat \rho^\mathrm{R}(E_\D,\epsilon)\,,
    \label{eq:M12_limit}
    \\
    \Gamma_{12} &=   \frac{1}{2m_\D} \lim_{\epsilon\to 0} \, 2  \, \hat \rho^\mathrm{I}(E_\D,\epsilon)\,,
    \label{eq:Gamma12_limit}
\end{align}
where the extra factor of $2$ is required since $\hat \rho^\mathrm{I}(E_\D,\epsilon) \to   \rho(E_\D)/2$ in the limit $\epsilon\to 0$.

It is useful to note that the Kramers-Kronig relations, which generally connect the real and imaginary parts of any complex function that is analytic in the upper half-plane, can be used to relate the smeared spectral densities $\hat{\rho}^\mathrm{R}(\bar\omega,\epsilon)$ and $\hat{\rho}^\mathrm{I}(\bar\omega,\epsilon)$. The relations express one component as the Hilbert transform of the other and for the case at hand give
\begin{equation}
    \hat{\rho}^\mathrm{R}(\bar\omega,\epsilon) = \frac{1}{\pi} \, \mathrm{P.V.} \int \dd\bar\omega' \, \frac{\hat{\rho}^\mathrm{I}(\bar\omega',\epsilon)}{\bar\omega'-\bar\omega}\,,
\end{equation}
with an analogous relation holding between the smearing kernels
\begin{align}
\mathcal{K}_\epsilon^\mathrm{R}(\omega,\bar\omega) = -\frac{1}{\pi} \, \mathrm{P.V.} \int \dd \omega' \, \frac{\mathcal{K}_\epsilon^\mathrm{I}(\omega',\bar\omega)}{\omega' - \omega} \,.
  \label{eq:KK-relation}
\end{align}
Note that the sign difference stems from the fact that $\mathcal{K}_\epsilon(\omega,\bar\omega) =\mathcal{K}_\epsilon^\mathrm{R}(\omega,\bar\omega)  + \ii \mathcal{K}_\epsilon^\mathrm{I}(\omega,\bar\omega) $ is analytic in the lower half plane of $\omega$.

As a final comment we note that any functions $\mathcal{K}_{\epsilon}^\mathrm{R, {\sf x}}(\omega,\bar\omega)$ and $\mathcal{K}_{\epsilon}^\mathrm{I,{\sf x}}(\omega,\bar\omega)$ can be used to define the smeared spectral density provided that the differences $\mathcal{K}_{\epsilon}^\mathrm{R, {\sf x}}(\omega,\bar\omega) - \mathcal{K}_{\epsilon}^\mathrm{R}(\omega,\bar\omega)$ and $\mathcal{K}_{\epsilon}^\mathrm{I, {\sf x}}(\omega,\bar\omega) - \mathcal{K}_{\epsilon}^\mathrm{I}(\omega,\bar\omega)$ become the trivial functional in the limit $\epsilon\to 0$:
\begin{align}
    \lim_{\epsilon\to 0} \int \dd\omega \, \rho(\omega) \big[\mathcal{K}_{\epsilon}^\mathrm{R, {\sf x}}(\omega,\bar\omega) - \mathcal{K}_{\epsilon}^\mathrm{R}(\omega,\bar\omega)\big] &= 0\,,
    \label{eq:kernel_condition_R}
    \\
    \lim_{\epsilon\to 0} \int \dd\omega \, \rho(\omega) \big[\mathcal{K}_{\epsilon}^\mathrm{I, {\sf x}}(\omega,\bar\omega) - \mathcal{K}_{\epsilon}^\mathrm{I}(\omega,\bar\omega)\big] &= 0\,.
    \label{eq:kernel_condition_I}
\end{align}
In the case of $\mathcal{K}_{\epsilon}^\mathrm{I}(\omega,\bar\omega)$, for example, it may be convenient to consider a Gaussian or other Cauchy kernels. This freedom was used successfully in ref.~\cite{Bulava:2021fre}, in the context of the non-linear O(3) sigma model, where the $\epsilon \to 0$ limit was performed via a simultaneous fit to the $\epsilon$-dependence of various smeared spectral densities. For any choice made for $\mathcal{K}_{\epsilon}^{\mathrm{I}, {\sf x}}(\omega,\bar\omega)$, the Kramers-Kronig relation of~\cref{eq:KK-relation} then provides a possible corresponding choice for $\mathcal{K}_{\epsilon}^{\mathrm{R}, {\sf x}}(\omega,\bar\omega)$. We stress that there is no requirement to match the real and imaginary parts in this way, since all choices should yield the same universal result for $\epsilon \to 0$.

\subsection{Weak Hamiltonian and $U$-spin decomposition}
The long distance contributions to neutral $\D$-meson mixing can be studied within an effective theory where the $W$ boson and the $b$ quark have been integrated out~\cite{ PhysRevD.20.2392,Buccella:1994nf,Buchalla:1995vs,Grossman:2006jg,Ryd:2009uf}.
The relevant $\Delta C=1$ four-quark operators are the current current operators
\begin{align}
Q_1^{\bar{q}q'} &= [\bar{q}_a \, \gamma_\mu P_L \, c_a] [\bar{u}_b \, \gamma^\mu  P_L\, q'_b]\,,
\label{eq:Q1}
\\
Q_2^{\bar{q}q'} &= [\bar{q}_a \, \gamma_\mu P_L\, c_b] [\bar{u}_b \, \gamma^\mu P_L\, q'_a] \,,
\label{eq:Q2}
\end{align}
as well as the QCD penguin operators
\begin{align}
Q_3 &= [\bar{u}_a \, \gamma_\mu P_L\, c_a] \sum_q \,[\bar{q}_b \, \gamma^\mu P_L\, q_b]\,,
\label{eq:Q3}\\
Q_4 &= [\bar{u}_a \, \gamma_\mu P_L\, c_b] \sum_q \,[\bar{q}_b \, \gamma^\mu P_L\, q_a]\,,
\label{eq:Q4}\\
Q_5 &= [\bar{u}_a \, \gamma_\mu P_L\, c_a] \sum_q \,[\bar{q}_b \, \gamma^\mu P_R\, q_b]\,,
\label{eq:Q5}\\
Q_6 &= [\bar{u}_a \, \gamma_\mu P_L\, c_b] \sum_q \,[\bar{q}_b \, \gamma^\mu P_R\, q_a]\,.
\label{eq:Q6}
\end{align}
Here $P_{R,L} = (1\pm\gamma_5)/2$ denote the chiral projectors and
a summation over the repeated colour indices $a,b=1,2,3$, as well as the Lorentz index $\mu=1,\dots,4$ is understood. The sum in the operators $Q_{3,\dots,6}$ runs over all active quark flavours $q=\{u,d,s,c\}$.

These operators give rise to the following $\Delta C = 1$ weak Hamiltonian
\begin{equation}
    \Hcal_\w^{\Delta C=1} = \Hcal_\w^\mathrm{SCS} + \Hcal_\w^\mathrm{CF} + \Hcal_\w^\mathrm{DCS}\,,
    \label{eq:Hw}
\end{equation}
where the Cabibbo-favoured (CF), Singly-Cabibbo Suppressed (SCS) and Doubly-Cabibbo Suppressed (DCS) components are defined as
\begin{align}
    \Hcal_\w^\mathrm{SCS} &= \frac{G_\F}{\sqrt{2}}\sum_{q=s,d}  \tensor*{V}{_{uq}} \tensor*{V}{_{cq}^{*}}\big[C_1 Q_1^{\bar{q}q} + C_2 Q_2^{\bar{q}q}\big] - \frac{G_\F}{\sqrt{2}} \tensor*{V}{_{ub}} \tensor*{V}{_{cb}^{*}} \sum_{i=3}^6 \, C_i Q_i + \mathrm{h.c.}\,,
    \label{eq:Hw_SCS}\\
    \Hcal_\w^\mathrm{CF} &= \frac{G_\F}{\sqrt{2}}\, \tensor*{V}{_{ud}} \tensor*{V}{_{cs}^{*}}\sum_{i=1,2}C_i Q_i^{\bar{s}d} + \mathrm{h.c.}\,,
    \label{eq:Hw_CF}\\
    \Hcal_\w^\mathrm{DCS} &= \frac{G_\F}{\sqrt{2}}\, \tensor*{V}{_{us}} \tensor*{V}{_{cd}^{*}} \sum_{i=1,2} C_i Q_i^{\bar{d}s} + \mathrm{h.c.}\,.
    \label{eq:Hw_DCS}
\end{align}
A more general set of operators must be considered when accounting for electromagnetic radiative corrections, including electroweak penguin operators ($Q_{7,\dots,10})$, as well as magnetic penguin operators ($Q_{7\gamma},Q_{8G}$)~\cite{Buchalla:1995vs}. However, in this work, we neglect their contributions.
The Wilson coefficients $C_i$ can be computed in perturbation theory and evaluated at the charm mass scale (see e.g. ref.~\cite{Buchalla:1995vs}). It is also convenient to introduce the two operators
\begin{equation}
    Q_\pm^{\bar{q}q'} = \frac12 \, \big(Q_1^{\bar{q}q'} \pm Q_2^{\bar{q}q'}\big)
    \label{eq:Qplusminus}
\end{equation}
and the corresponding Wilson coefficients $C_\pm=C_1\pm C_2$. As we will discuss in detail in the next section, the operators $Q_{+}^{\bar{q}q'}$ and $Q_{-}^{\bar{q}q'}$ do not mix under renormalization in the absence of chiral symmetry breaking. Consequently, their corresponding matrix elements undergo multiplicative renormalization.

Defining the quantity $\lambda_q=G_\F(\tensor*{V}{_{uq}}\tensor*{V}{_{cq}^*})/\sqrt{2}$ and using the identity $\lambda_d + \lambda_s + \lambda_b = 0$, which follows from CKM unitarity, we can write
\begin{align}
    \Hcal_\w^\mathrm{SCS} =  (\lambda_s - \lambda_d)\sum_{i=1,2} \frac{C_i}{2} (Q_i^{\bar{s}s} - Q_i^{\bar{d}d}) -  \lambda_b \, \bigg\{\sum_{i=1}^2 \frac{C_i}{2} (Q_i^{\bar{s}s} + Q_i^{\bar{d}d}) + \sum_{i=3}^6 \, C_i Q_i \bigg\} +\mathrm{h.c.} \,.
    \label{eq:Hw_SCS_unitarity}
\end{align}
Since $|\lambda_b|/|\lambda_s-\lambda_d|\simeq 10^{-4}$,
this expression shows that the dominant contribution to the mixing amplitude comes from the first term proportional to $(\lambda_s-\lambda_d)$, while the others, which are suppressed by a factor $\lambda_b$, are subdominant and can be neglected.

As shown in~\cref{eq:amplitude}, mixing can also occur via $\Delta C = 2$ transitions mediated by local operators. The $\Delta C = 2$ component of the effective Hamiltonian encapsulates the low-energy effects of high-energy degrees of freedom, such as the $b$ quark and potential contributions from new physics beyond the Standard Model. While their contribution in the Standard Model is very small due to both CKM and GIM suppression~\cite{Hagelin:1981zk,Cheng:1982hq,Buras:1984pq,Datta:1984jx,Falk:2001hx,Burdman:2003rs}, $\Delta C = 2$ transitions can be studied to constrain new physics scenarios beyond the Standard Model.

The effective Hamiltonian for $\Delta C = 2$ transitions is typically expressed in terms of five local four-fermion operators~\cite{Gabbiani:1988rb,Gabbiani:1996hi,Gabrielli:1995bd}. The matrix elements of these operators between $D^0$ and $\bar{D}^0$ states are commonly parameterized in terms of so-called bag parameters $B_i$, which quantify the deviation of the matrix elements from their values in the vacuum saturation approximation. Non-perturbative lattice QCD calculations of these bag parameters have been performed by different collaborations in refs.~\cite{Carrasco:2015pra,Bazavov:2017weg}.
In this work, we will neglect such contributions and focus only on the calculation of long-distance effects.

The mixing amplitudes $\M_{12}$ and $\Gamma_{12}$ are obtained from the real and imaginary parts of~\cref{eq:amplitude}, respectively. Neglecting the contribution of $\Delta C=2$ operators to the weak Hamiltonian, such amplitudes are then obtained from a double insertion of the Hamiltonian $\mathcal{H}_\w^{\Delta C=1}$ defined above in~\cref{eq:Hw}  between a $\D^0$ and a $\Dbar^0$ state.
Given the flavour content of the different components of the Hamiltonian, one can easily show that only two combinations  are relevant for the mixing: the product of two SCS Hamiltonians, $\mathcal{H}_\w^\mathrm{SCS}\times\mathcal{H}_\w^\mathrm{SCS}$, or the product of one CF and one DCS Hamiltonian, $\mathcal{H}_\w^\mathrm{CF}\times\mathcal{H}_\w^\mathrm{DCS}$.
Following the discussion in ref.~\cite{Kagan:2020vri}, we can classify the contributions to the amplitudes $\M_{12}$ and $\Gamma_{12}$ based on their flavour content and, consequently, their properties under $U$-spin symmetry. The latter is a subgroup of the $SU(3)$ flavour symmetry and corresponds to the exchange symmetry between the $d$ and $s$ quarks.
Let us start by writing
\begin{equation}
    \Gamma_{12} = \sum_{i,j=d,s} \, \lambda_i\lambda_j \, \Gamma_{ij} \,, \qquad     \M_{12} = \sum_{i,j=d,s,b} \, \lambda_i\lambda_j \, \M_{ij} \,.
\end{equation}
At the quark level, these mixing amplitudes are associated with box diagrams containing internal $i$ and $j$ quarks, which are on-shell in the case of $\Gamma_{ij}$ and off-shell in the case of $\M_{ij}$ (note, in fact, that also the $b$ quark is included in the sum).
Using again $\lambda_d+\lambda_s+\lambda_b=0$, these can be written as follows
\begin{equation}
    \xi_{12} = \frac{(\lambda_s-\lambda_d)^2}{4} \, \xi_2 +\frac{(\lambda_d-\lambda_s)\lambda_b}{2} \, \xi_1 + \frac{\lambda_b^2}{4} \, \xi_0\,, \qquad \xi = \M,\Gamma\,,
\end{equation}
where
\begin{alignat}{2}
    &\Gamma_2 = \Gamma_{dd}-2\Gamma_{ds}+\Gamma_{ss}\,,
    &&\qquad
    \M_2 = \M_{dd}-2\M_{ds}+\M_{ss}\,,\\
    &\Gamma_1 = \Gamma_{dd}-\Gamma_{ss}\,,
    &&\qquad
    \M_1 = \M_{dd}-\M_{ss}+2(\M_{sb}-\M_{db})\,,\\
    &\Gamma_0 = \Gamma_{dd}+2\Gamma_{ds}+\Gamma_{ss}\,,
    &&\qquad
    \M_0 = \M_{dd}+2\M_{ds}+\M_{ss}-4(\M_{db}+\M_{sb}-\M_{bb})\,.
\end{alignat}

The quantities $\xi_{dd}$ and $\xi_{ss}$ get contributions from a double insertion of the SCS operators $Q_i^{\bar{d}d}$ and $Q_i^{\bar{s}s}$, respectively. On the other hand, $\xi_{sd}$ is given by the two possible ways of inserting one $Q_i^{\bar{d}d}$ and one $Q_i^{\bar{s}s}$, as well as the two possible combinations of the CF operator $Q_i^{\bar{d}s}$ and the DCS operator $Q_i^{\bar{s}d}$. The terms $\M_{sb}$, $\M_{db}$ and $\M_{bb}$ arise instead from penguin operators in the SCS Hamiltonian.
In this form, it is evident that the amplitudes $\xi_{2,1,0}$ correspond to the neutral components ($\Delta U_3 = 0$) of $U$-spin multiplets with $\Delta U = 2,1,0$, respectively.
As discussed in ref.~\cite{Kagan:2020vri}, using the fact that $\lambda_b$ is small, the dominant contributions to $\M_{12}$ and $\Gamma_{12}$ come from the amplitudes $\M_2$ and $\Gamma_2$, respectively, although they are of second order in the $U$-spin symmetry breaking.

The amplitudes $\xi_2$ can be rewritten symbolically in terms of operator insertions as follows:
\begin{align}
    \xi_2 &\sim (Q^{\bar{d}d}Q^{\bar{d}d} + Q^{\bar{s}s}Q^{\bar{s}s} - Q^{\bar{s}s}Q^{\bar{d}d} - Q^{\bar{d}d}Q^{\bar{s}s}) - (Q^{\bar{s}d}Q^{\bar{d}s} + Q^{\bar{d}s}Q^{\bar{s}d}) \nonumber \\
    &= (Q^{\bar{s}s} - Q^{\bar{d}d})(Q^{\bar{s}s} - Q^{\bar{d}d}) - (Q^{\bar{s}d}Q^{\bar{d}s} + Q^{\bar{d}s}Q^{\bar{s}d})\,.
    \label{eq:X2-contrib}
\end{align}

\begin{figure}
    \centering
    \includegraphics[width=\linewidth]{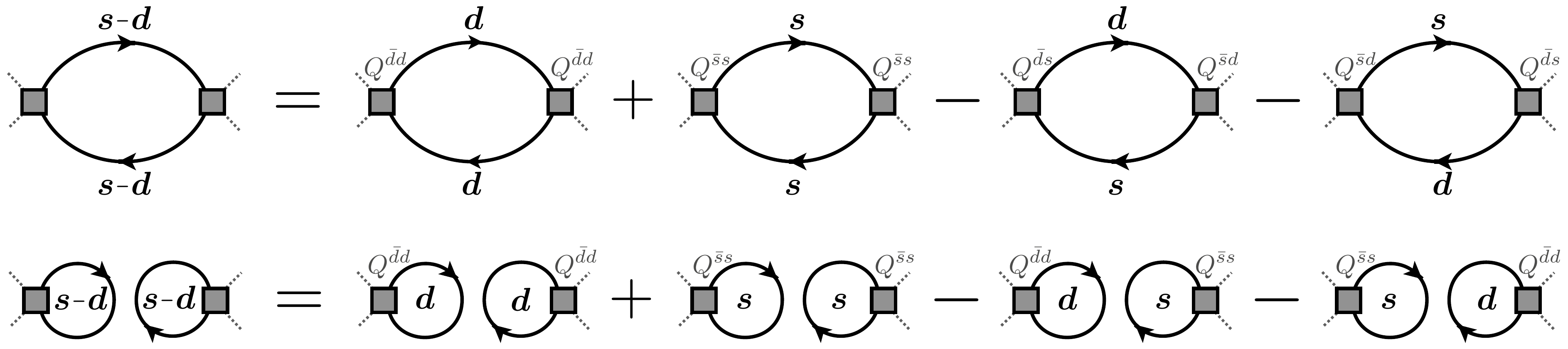}
    \caption{Relevant Wick contractions for the calculation of the dominant $\Delta U=2$ contributions to $\M_{12}$ and $\Gamma_{12}$. As a consequence of the structure of $\xi_2$ in~\cref{eq:X2-contrib}, the internal lines always appear as differences of strange and down propagators.}
        \label{fig:xi2_contractions}
\end{figure}

The structure of $\xi_2$ highlights the effect of the GIM mechanism~\cite{Glashow:1970gm}. In particular, when the differences of operators  $(Q^{\bar{s}s} - Q^{\bar{d}d})(Q^{\bar{s}s} - Q^{\bar{d}d})$
are inserted into the mixing amplitude defined in~\cref{eq:amplitude},
 it leads to a well-behaved, finite integral over spacetime.

In general, the product of two $\Delta C=1$ four-quark operators leads to a quadratic divergence when integrated over the region where the two operators coincide. These divergences are naturally regulated in the Standard Model by the exchange of $W$ bosons, whose effects at low energies are approximated by the effective Hamiltonian $\Hcal_\w$. However, the GIM mechanism removes this quadratic divergence, ensuring that the integral remains convergent.
This implies that no mixing occurs between the long-distance amplitude and the local $\Delta C=2$ local operator in~\cref{eq:amplitude}.
This behaviour is analogous to that of the charm quark contribution of order $\lambda_c^2$ in the long-distance amplitude for $K^0-\widebar{K}^0$ mixing~\cite{Buchalla:1995vs}.

The expression in~\cref{eq:X2-contrib} also implies that the internal quark lines always appear as differences of a strange and a down quark propagator. The possible Wick contractions of the down-type quarks arising from the operator insertions in~\cref{eq:X2-contrib} are shown in~\cref{fig:xi2_contractions}. Moreover, the appearance of differences of quark propagators allows one to take advantage of variance reduction techniques~\cite{Giusti:2019kff,Harris:2023zsl} in the calculation of the four-point correlation functions, as will be discussed below in~\cref{sec:correlators}.

\begin{figure}[t!]
    \centering
    \includegraphics[width=\linewidth]{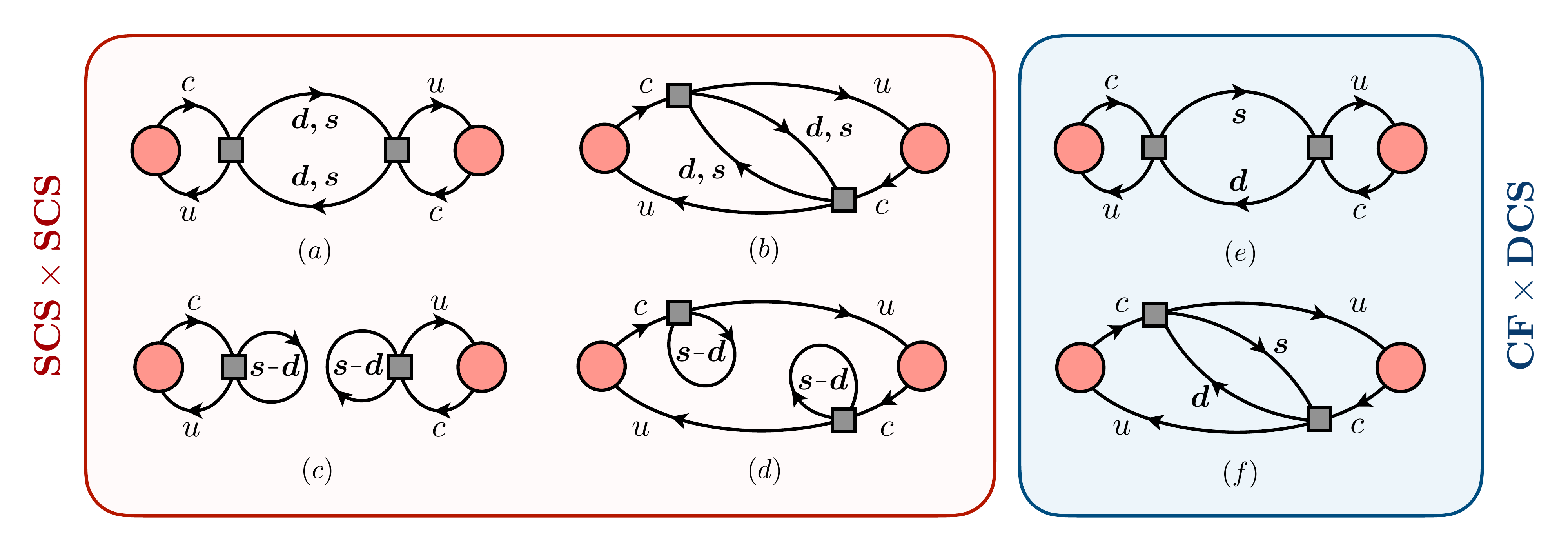}
    \caption{Topologies of the correlation functions relevant for the calculation of $\M_2$ and $\Gamma_2$, arising from a double insertion of two $\mathcal{H}_\w^\mathrm{SCS}$ Hamiltonians (left) or one $\mathcal{H}_\w^\mathrm{CF}$ and one $\mathcal{H}_\w^\mathrm{DCS}$ (right). The gray squares represent the operator insertions, while the red circles are the interpolating operators of the $D^0$ and $\Dbar^0$ mesons.
    Note that only by combining  diagrams (a) and (e) (and similarly (b) and (f)) as in~\cref{fig:xi2_contractions}, the internal lines can be written in terms of $s$-$d$ propagators.
    }
    \label{fig:diags}
\end{figure}

The relevant correlation functions corresponding to the combination of operators in~\cref{eq:X2-contrib} are shown in~\cref{fig:diags}, where the gray squares represent the insertions of the weak Hamiltonians and the red circles the interpolating operators of the $D^0$ and $\bar{D}^0$. Depending on the colour structure of the operator, two different types of contractions can arise. Therefore, each topology in the figure corresponds to four different correlation functions. The strategy for a non-perturbative calculation of such correlation functions in lattice QCD will be discussed below in~\cref{sec:correlators}.

Similar correlation functions, with different flavour content, have been computed by the RBC-UKQCD Collaboration in the context of kaon mixing to estimate the long-distance contributions to the $K_L{-}K_S$ mass splitting~\cite{Christ:2012se,Bai:2014cva} and to $\epsilon_K$~\cite{PhysRevD.109.054501}.
Note that in the case of $\epsilon_K$, contrary to $D$-meson mixing, additional diagrams arise from QCD penguin operators in $\Hcal^\mathrm{SCS}_\w$, due to the dominant top-quark loop contribution and the absence of strong CKM suppression.

We note that the parity-odd $\Delta C = 1$ operators discussed in this work are also relevant for the study of the hadronic decay $D \to K\pi$. A lattice computation of this decay is currently in progress~\cite{Joswig:2022ctr}, and many of the considerations presented here are directly applicable to the $D \to K\pi$ computation and analysis.

\section{Renormalization}
\label{sec:renormalization}

In order to study the renormalization mixing pattern of the operators entering the weak Hamiltonian, let us assume that a mass-independent renormalization scheme is employed and that we use a lattice regularization that potentially breaks chiral symmetry. We will discuss later how the renormalization mixing simplifies when chiral symmetry is preserved. The discussion below follows that in refs.~\cite{PhysRevD.109.054501,Donini:1999sf}.

Considering only $N_f=4$ active flavours, we shall then study the transformation under the group $SU(4)$ of a generic four-quark operator with the following quark content
\begin{equation}
    T{}_{ik}^{jl}=(\bar{q}_i\,q^j\,\bar{q}_k\,q^l)\,,
\end{equation}
where the indices $i,j,k,l$ denote the flavour of the quark field $q=(u,d,s,c)^T$. The tensor $T{}_{ik}^{jl}$ also carries spin and colour indices, which we leave implicit.

Since the quark fields $q$ and $\bar{q}$ belong to the fundamental and conjugate representations of $SU(4)$, respectively, then a four-quark operator of the form of $T{}_{ik}^{jl}$ belongs to the following representation,
\begin{align}
    \numvec{4}\otimes\numvec{\overline{4}}\otimes \numvec{4}\otimes\numvec{\overline{4}} \,
    &= \, (\numvec{1} \oplus \numvec{15})\otimes (\numvec{1} \oplus \numvec{15})\\
    & = \numvec{1}_\mathrm{A} \oplus \numvec{1}_\mathrm{S} \oplus \numvec{15}_\mathrm{A} \oplus \numvec{15}_\mathrm{SA} \oplus \numvec{15}_\mathrm{AS} \oplus \numvec{15}_\mathrm{S} \oplus \numvec{20}_\mathrm{A}\oplus \numvec{45}_\mathrm{SA}\oplus\numvec{\overline{45}}_\mathrm{AS}\oplus\numvec{84}_\mathrm{S}\,. \nn
\end{align}
The 10 irreducible representations correspond to different sets of operators. The subscripts
$ \mathrm{A} $, $ \mathrm{S} $, $ \mathrm{SA} $, and $ \mathrm{AS} $ denote the symmetry properties of the operators in a given representation under the exchange of flavour indices.

\begin{itemize}
    \item The representation $ \numvec{84}_\mathrm{S} $ corresponds to the fully symmetric and traceless operators $ \widehat{T}{}^{\{j,l\}}_{\{i,k\}} $. Here, traceless means that any contraction of an upper and a lower index is zero, and we use the hat to indicate such traceless operators.
    \item The representation $ \numvec{20}_\mathrm{A} $ consists of the fully antisymmetric operators $ \widehat{T}{}^{[j,l]}_{[i,k]} $.
    \item The representations $ \numvec{45}_\mathrm{SA} $ and $ \overline{\numvec{45}}_\mathrm{AS} $ correspond to the operators $ \widehat{T}{}^{\{j,l\}}_{\,[i,k]} $ and $ \widehat{T}{}^{\,[j,l]}_{\{i,k\}} $, respectively.
    \item The trace terms correspond to the penguin operators in the four adjoint representations $ \numvec{15} $:
    \[
    \sum_n \widehat{T}{}^{\{j,n\}}_{\{i,n\}} ~\text{for } \numvec{15}_\mathrm{S}\,, \quad
    \sum_n \widehat{T}{}^{\{j,n\}}_{\,[i,n]}  ~\text{for } \numvec{15}_\mathrm{SA}\,, \quad
    \sum_n \widehat{T}{}^{\,[j,n]}_{\{i,n\}} ~ \text{for } \numvec{15}_\mathrm{AS}\,, \quad
    \sum_n \widehat{T}{}^{[j,n]}_{[i,n]}  ~\text{for } \numvec{15}_\mathrm{A}.
    \]
    In all these cases, the operators are considered traceless in the two free indices $i$ and~$j$.
    \item Finally, the two singlets $ \numvec{1}_\mathrm{S} $ and $ \numvec{1}_\mathrm{A} $ are obtained from $ T{}^{jl}_{ik} $ by contracting all upper and lower indices:
    \[
    \sum_{n,m}T{}^{\{n,m\}}_{\{n,m\}}~ \text{for } \numvec{1}_\mathrm{S}\,, \quad \sum_{n,m}T{}^{[n,m]}_{[n,m]}~ \text{for } \numvec{1}_\mathrm{A}\,.
    \]
    In the following we can ignore the operators in the singlet representations, as they would only contribute to $\Delta C=0$ transitions.
\end{itemize}

Considering now the colour of the quark fields we can distinguish two classes of operators:
\begin{align}
    ({O}_{AB})^{jl}_{ik} = \big[\bar{q}^a_i\, \Gamma_A \, q_a^j\big]\big[\bar{q}^b_k\, \Gamma_B \, q_b^l\big]
    \,,\\[5pt]
    (\widetilde{{O}}_{AB})^{jl}_{ik} = \big[\bar{q}^a_i\, \Gamma_A \, q_b^j\big]\big[\bar{q}^b_k\, \Gamma_B \, q_a^l\big]
    \,,
\end{align}
where $a$,$b$ denote the colour indices,  while  $\Gamma_A$ and $\Gamma_B$ are Dirac matrices whose product forms a Lorentz scalar.

Using the standard notation for Dirac matrices,\footnote{
Here we consider Dirac matrices in Minkowski space, with $\gamma_5=\ii\gamma^0\gamma^1\gamma^2\gamma^3$ and $\sigma^{\mu\nu}=\ii\,[\gamma^\mu,\gamma^\nu]/2$.
}
\begin{align}
    \{S,P,V,A,T,\widetilde{T}\} = \{\mathbbm{1},\gamma_5,\gamma^\mu,\gamma^\mu\gamma_5,\sigma^{\mu\nu},\sigma^{\mu\nu}\gamma_5\},
\end{align}
and assuming implicit summation over Lorentz indices when two Dirac matrices are multiplied (e.g., $VV = \sum_\mu \gamma^\mu\otimes\gamma_\mu$), the two sets of operators are given by:
\begin{align}
    \boldsymbol{O} = \big\{ O_{VV}, O_{AA}, O_{VA}, O_{AV}, O_{SS}, O_{PP}, O_{SP}, O_{PS}, O_{TT}, O_{T\widetilde{T}} \big\}\,,
\end{align}
with an analogous definition for $\boldsymbol{\widetilde{O}}$.

It is convenient to use a Dirac basis for the operators with definite parity quantum numbers. We define the sets of  parity-even and parity-odd operators as follows
\begin{align}
    \boldsymbol{O}{}^+ &= \big\{ O_{VV+AA},\, O_{SS+PP},\, O_{TT},\,  O_{VV-AA},\, O_{SS-PP} \big\}\,,\\
    \boldsymbol{O}{}^- &= \big\{ O_{VA+AV},\, O_{SP+PS},\, O_{T\widetilde{T}},\, O_{VA-AV},\, O_{SP-PS} \big\}\,,
\end{align}
and analogously for the mixed-colour operators $\boldsymbol{\widetilde{O}}{}^\pm$, having used the notation
\begin{equation}
    O_{\Gamma_1\Gamma_2\pm\Gamma_3\Gamma_4} = O_{\Gamma_1\Gamma_2}\pm O_{\Gamma_3\Gamma_4}\,.
\end{equation}
Each parity sector thus consists of 10 operators (5 for each colour structure) resulting in a total of 20 operators.

Using the Fierz rearrangement of the quark fields
\begin{equation}
    q^{j}\bar{q}_k = -\frac14\Big[(\bar{q}_k\, q^{j}) \, \mathbbm{1}+(\bar{q}_k \,\gamma^\mu \,q^{j})\, \gamma_\mu + \frac12 (\bar{q}_k \,\sigma^{\mu\nu} \,q^{j})\sigma_{\mu\nu} - (\bar{q}_k \,\gamma^\mu \gamma_5 \,q^{j})\, \gamma_\mu\gamma_5 + (\bar{q}_k \,\gamma_5 \,q^{j})\,\gamma_5\Big]\,,
\end{equation}
we can rewrite the operators with mixed colour indices $\boldsymbol{\widetilde{O}}$ in terms of the operators $\boldsymbol{O}$  with different flavour indices. Introducing the notation $\boldsymbol{O}_\mathrm{F}$ for Fierz transformed operators, namely
\begin{equation}
    (\boldsymbol{O}_\mathrm{F})^{jl}_{ik} = (\boldsymbol{O})^{lj}_{ik}\,,
\end{equation}
the following relation holds\footnote{
The matrix $\boldsymbol{M}^\pm$ is consistent with the results presented in Appendix B of ref.~\cite{Donini:1999sf}. However, ref.~\cite{Donini:1999sf} employs Euclidean Dirac matrices and defines the operators $O_{TT}$ and $O_{T\widetilde{T}}$ using a sum restricted to only the independent components of $\sigma^{\mu\nu}$. As a consequence, their definitions of $O_{TT}$ and $O_{T\widetilde{T}}$ differ from ours by an overall factor of~$-1/2$.
}
\begin{equation}
    \boldsymbol{\widetilde{O}}{}^\pm = \boldsymbol{M}^\pm \cdot \boldsymbol{O}{}_\mathrm{F}^\pm \,, 
\end{equation}
where $\boldsymbol{M}^{\pm}$ is the unitary matrix
\begin{equation}
    \boldsymbol{M}^{\pm} = (\boldsymbol{M}^{\pm})^{-1} =
    \begin{pmatrix}
        1 & 0 & 0 & 0 & 0 \\
        0 & -1/2 & -1/4 & 0 & 0 \\
        0 & -3 & 1/2 & 0 & 0 \\
        0 & 0 & 0 & 0 & \mp 2 \\
        0 & 0 & 0 & \mp 1/2 & 0
    \end{pmatrix}\,.
\end{equation}
Finally, for each parity sector, we can define the following linear combinations of the operators with their Fierz-transformed counterpart,
\begin{equation}
    \boldsymbol{\mathcal{O}}^\pm = \frac12\,\big(\,\boldsymbol{O}{}^\pm + \boldsymbol{{O}}{}_\mathrm{F}^\pm\,\big)\,, \qquad
    \boldsymbol{\mathcal{Q}}{}^\pm = \frac12\,\big(\,\boldsymbol{{O}}{}^\pm - \boldsymbol{{O}}{}_\mathrm{F}^\pm\,\big)\,.
    \label{eq:sum_diff_ops}
\end{equation}
In order to make contact with the different irreducible representations of $SU(4)$ discussed above, we can  further rewrite the above operators in a slightly different way by symmetrizing and anti-symmetrizing the flavour indices as follows
\begin{align}
    \boldsymbol{O}^\pm &= \frac{1}{4}
    \big(
    \boldsymbol{O}^\pm_\mathrm{S}+\boldsymbol{O}^\pm_\mathrm{A}\big) + \frac{1}{4}
    \big(\boldsymbol{O}^\pm_\mathrm{SA} + \boldsymbol{O}^\pm_\mathrm{AS}
    \big)\,,\qquad \boldsymbol{O}{}_\mathrm{F}^\pm = \frac{1}{4}
    \big(
    \boldsymbol{O}^\pm_\mathrm{S}- \boldsymbol{O}^\pm_\mathrm{A} \big)+\frac14\big( \boldsymbol{O}^\pm_\mathrm{SA} - \boldsymbol{O}^\pm_\mathrm{AS}
    \big)\,,
    \label{eq:symmetrization_ops}
\end{align}
where
\begin{equation}
        (\boldsymbol{O}{}_\mathrm{S})^{jl}_{ik} = (\boldsymbol{O})^{\{j,l\}}_{\{i,k\}}\,, \ \
        (\boldsymbol{O}{}_\mathrm{SA})^{jl}_{ik} = (\boldsymbol{O})^{\{j,l\}}_{\,[i,k]}\,, \ \
        (\boldsymbol{O}{}_\mathrm{AS})^{jl}_{ik} = (\boldsymbol{O})^{\,[j,l]}_{\{i,k\}}\,, \ \
        (\boldsymbol{O}{}_\mathrm{A})^{jl}_{ik} = (\boldsymbol{O})^{[j,l]}_{[i,k]}\,.
\end{equation}
Note that the decomposition of $\boldsymbol{O}^\pm$ in operators with definite flavour symmetry above strictly depends on the Dirac matrices $\Gamma_A$ and $\Gamma_B$ in the bilinears. In fact, it is easy to note that the parity-odd operators $O_{VA-AV}$ and $O_{SP-PS}$, which are antisymmetric under the swap of the Dirac matrices $\Gamma_A$ and $\Gamma_B$ will only have components $\mathrm{SA}$ and $\mathrm{AS}$. Conversely, all other operators will only have $\mathrm{S}$ and $\mathrm{A}$ components.
Rewriting the vectors $\boldsymbol{O}^\pm$ as direct sums,
\begin{align}
    \boldsymbol{O}^+ &= O_{VV+AA} \,\oplus\,
    \big\{O_{SS+PP},O_{TT}\big\}
    \,\oplus\,
    \big\{O_{VV-AA},O_{SS-PP}\big\}
    \equiv {O}_1^+ \,\oplus\, \boldsymbol{O}_2^+ \,\oplus\, \boldsymbol{O}_3^+ \,,\\
    \boldsymbol{O}^- &= O_{VA+AV} \,\oplus\,
    \big\{O_{SP+PS},O_{T\widetilde{T}}\big\}
    \,\oplus\,
    \big\{O_{VA-AV},O_{SP-PS}\big\}
    \equiv {O}_1^- \,\oplus\, \boldsymbol{O}_2^- \,\oplus\, \boldsymbol{O}_3^-\,,
\end{align}
we will have that
\begin{align}
    \{\boldsymbol{O}^+,\, {O}_1^-,\,\boldsymbol{O}_2^-\} & \in \numvec{84}_\mathrm{S}\,\oplus\,\numvec{20}_\mathrm{A}\,\oplus\,\numvec{15}_\mathrm{S}\,\oplus\,\numvec{15}_\mathrm{A}\,\oplus\,\numvec{1}_\mathrm{S}\,\oplus\,\numvec{1}_\mathrm{A}\,,\\
    \boldsymbol{O}_3^- & \in \numvec{45}_\mathrm{SA}\,\oplus\,\overline{\numvec{45}}_\mathrm{AS}\,\oplus\,\numvec{15}_\mathrm{SA}\,\oplus\,\numvec{15}_\mathrm{AS}\,.
\end{align}

From~\cref{eq:sum_diff_ops,eq:symmetrization_ops} directly follows that
\begin{align}
    \boldsymbol{\mathcal{O}}^+ & = \frac{1}{4} \Big[\,({O}_1^+)_\mathrm{S} \, \oplus \, (\boldsymbol{O}_2^+)_\textrm{S}  \, \oplus \, (\boldsymbol{O}_3^+)_\textrm{S} \,\Big]\,, \\
    \boldsymbol{\mathcal{Q}}^+ & = \frac{1}{4} \Big[\,({O}_1^+)_\mathrm{A} \, \oplus \, (\boldsymbol{O}_2^+)_\textrm{A}  \, \oplus \, (\boldsymbol{O}_3^+)_\textrm{A} \,\Big]\,,\\[5pt]
    \boldsymbol{\mathcal{O}}^- & = \frac{1}{4} \Big[\,({O}_1^-)_\mathrm{S} \, \oplus \, (\boldsymbol{O}_2^-)_\textrm{S}  \, \oplus \, (\boldsymbol{O}_3^-)_\textrm{SA} \,\Big]\,, \\
    \boldsymbol{\mathcal{Q}}^- & = \frac{1}{4} \Big[\,({O}_1^-)_\mathrm{A} \, \oplus \, (\boldsymbol{O}_2^-)_\textrm{A}  \, \oplus \, (\boldsymbol{O}_3^-)_\textrm{AS} \,\Big]\,.
\end{align}

From the above expressions, we conclude that the operators $\boldsymbol{\mathcal{O}}^\pm$ and $\boldsymbol{\mathcal{Q}}^\pm$ belong to different representations and thus cannot mix under renormalization. However, within each set, operators sharing the same symmetry properties can mix. This is the case, for example, for the parity-even operators, which all mix under renormalization.

Parity-odd operators, on the other hand, can only mix within the three blocks: $O_1^-$, $\boldsymbol{O}_2^-$, and $\boldsymbol{O}_3^-$. Although $O_1^-$ and $\boldsymbol{O}_2^-$ share the same flavour symmetries, their mixing is prohibited because they are eigenstates of the charge conjugation operator with opposite eigenvalues ($C\,O_1^-\,C^{-1}=-O_1^-$, $C\,\boldsymbol{O}_2^-\,C^{-1}=\boldsymbol{O}_2^-$). This discussion reproduces the results of ref.~\cite{Donini:1999sf}.

We can now study how the mixing simplifies when chiral symmetry is preserved. Instead of studying the transformation of the operators under the flavour group $SU(4)$, we consider the action of the group $SU(4)_L\times SU(4)_R$.
Defining the right- and left-handed quark fields as
\begin{equation}
    q = q_L + q_R\,, \qquad q_{L(R)} = P_{L(R)} \, q \,,\qquad \bar{q}_{L(R)} = \bar{q} \, P_{R(L)}\,,
\end{equation}
it is easy to show that the operators $\boldsymbol{O}^\pm$ take the following form
\begin{align}
    \boldsymbol{O}^\pm &=
    \begin{pmatrix}
         2(\bar{q}_R \, \gamma^\mu  \, q_R)(\bar{q}_R \, \gamma_\mu \, q_R) \pm 2(\bar{q}_L \, \gamma^\mu \, q_L)(\bar{q}_L \, \gamma_\mu \, q_L) \\[6pt]
         2(\bar{q}_L \, q_R)(\bar{q}_L \,  q_R) \pm 2(\bar{q}_R \,  q_L)(\bar{q}_R \,  q_L) \\
         (\bar{q}_L \, \sigma^{\mu\nu}  q_R)(\bar{q}_L \, \sigma_{\mu\nu} \, q_R) \pm (\bar{q}_R \, \sigma^{\mu\nu} \, q_L)(\bar{q}_R \, \sigma_{\mu\nu} \, q_L)\\[6pt]
          2(\bar{q}_L \, \gamma^\mu  q_L)(\bar{q}_R \, \gamma_\mu  q_R) \pm 2(\bar{q}_R \, \gamma^\mu \, q_R)(\bar{q}_L \, \gamma_\mu \, q_L)\\
          2(\bar{q}_R \,  q_L)(\bar{q}_L \,  q_R) \pm 2(\bar{q}_L \,  q_R)(\bar{q}_R \,  q_L)
    \end{pmatrix}
    \,.
\end{align}
From this, it becomes clear that when the left- and right-handed fields transform independently under the groups $SU(4)_L$ and $SU(4)_R$, the sets $O_1^\pm$, $\boldsymbol{O}_2^\pm$, and $\boldsymbol{O}_3^\pm$ belong to different representations of $SU(4)_L \times SU(4)_R$. Consequently, in the presence of chiral symmetry, mixing can only occur within each subset.

If chiral symmetry is preserved and we consider only operators that are singlets under $SU(4)_R$ while containing a charm quark ($c$), an anti-up quark ($\bar{u}$), and a pair of down-type quark and anti-quark, the set of allowed operators reduces to those presented in~\cref{eq:Q1,eq:Q2,eq:Q3,eq:Q4,eq:Q5,eq:Q6}. These operators can be rewritten in the following form to highlight their mixing pattern:
\begin{align}
    & Q_1^{\bar{q}q'} = \frac{1}{16}
    \big[ (O_1^+)_\mathrm{S} - (O_1^+)_\mathrm{A}  -
     (O_1^-)_\mathrm{S} + (O_1^-)_\mathrm{A} \big]{}_{uq}^{cq'}\\
    & Q_2^{\bar{q}q'} = \frac{1}{16}
    \big[ (O_1^+)_\mathrm{S} + (O_1^+)_\mathrm{A}  - (O_1^-)_\mathrm{S} - (O_1^-)_\mathrm{A} \big]{}_{uq}^{cq'}\,,
\end{align}
\begin{align}
    &Q_3 = \frac{1}{16} \sum_{q} Q_2^{\bar{q}q} \,,\\
    &Q_4 = \frac{1}{16} \sum_{q} Q_1^{\bar{q}q} \,,\\
    &Q_5 = \frac{1}{16} \sum_{q}
    \big[ (O_1^+)_\mathrm{S} + (O_1^+)_\mathrm{A} + (O_1^-)_\mathrm{S} + (O_1^-)_\mathrm{A} \big]{}_{uq}^{cq}\,,\\
    &Q_6 = -\frac{1}{8} \sum_{q}
    \big[ (O_5^+)_\mathrm{S} - (O_5^+)_\mathrm{A} + (O_5^-)_\mathrm{AS} - (O_5^-)_\mathrm{SA} \big]{}_{uq}^{cq}\,.
\end{align}
First, one notes that the penguin operators $Q_{3,\dots,6}$ are linear combinations of operators transforming under the four adjoint ($\numvec{15}$) representations of $SU(4)$. Consequently, they mix with each other under renormalization.
On the other hand, the current-current operators $Q_1^{\bar{q}q'}$ and $Q_2^{\bar{q}q'}$ can be decomposed into a traceless part and a trace part. Consequently, they transform as $\numvec{84}_\mathrm{S} \oplus \numvec{20}_\mathrm{A} \oplus \numvec{15}_\mathrm{S} \oplus \numvec{15}_\mathrm{A}$ and, as a result, can mix both with each other and with the penguin operators.
However, the operators~$Q_\pm^{\bar{q}q'}$, defined in~\cref{eq:Qplusminus}, belong to representations with definite flavour symmetries:
\begin{align}
    Q_+^{\bar{q}q'} &\in \numvec{84}_\mathrm{S}\,\oplus\,\numvec{15}_\mathrm{S}\,,\qquad
    Q_-^{\bar{q}q'} \in \numvec{20}_\mathrm{A}\,\oplus\,\numvec{15}_\mathrm{A}\,.
\end{align}
This structure ensures that they cannot mix with each other and mixing can occur only with penguin operators.

A key result relevant to this work is the following. As observed in~\cref{eq:Hw_SCS_unitarity},  the dominant contribution to the singly Cabibbo-suppressed (SCS) part of the weak Hamiltonian arises solely from the difference of the operators $(Q_r^{\bar{s}s} - Q_r^{\bar{d}d})$ for $r=1,2$, while other terms are suppressed by a factor $\lambda_b$ and are thus negligible. Since only the traceless part of the operators survives in this difference,
\begin{equation}
    (Q_r^{\bar{s}s} - Q_r^{\bar{d}d}) \in \numvec{84}_\mathrm{S} \,\oplus\, \numvec{20}_\mathrm{A}\,,
\end{equation}
no mixing with penguin operators is allowed. This significantly simplifies the renormalization of the mixing amplitude.

The non-perturbative renormalization of current--current four-fermion operators on the lattice can be carried out using momentum-subtraction schemes such as RI-MOM~\cite{Martinelli:1994ty} or RI-SMOM~\cite{Sturm:2009kb}. 
Once the operators are renormalized in one of these intermediate schemes, a perturbative matching to the $\overline{\mathrm{MS}}$ scheme can be performed, where the corresponding Wilson coefficients are known~\cite{Buchalla:1995vs}.
However, when using lattice fermion formulations that break chiral symmetry explicitly, additional complications arise due to operator mixing with lower-dimensional operators. These mixings can lead to power divergences and must be carefully subtracted to ensure a well-defined continuum limit.

\section{Relevant lattice correlation functions}
\label{sec:lattice}
\label{sec:correlators}

Up to this point, we have developed and reviewed the formalism for neutral $D$-meson mixing within QCD.  In this section, we outline a lattice QCD framework for computing the mixing amplitudes $\M_{12}$ and $\Gamma_{12}$ from first principles.
Throughout this section we assume a continuum theory in a periodic, cubic, finite volume with physical extent $L$ in each of the three spatial directions. We also ignore the effects of the finite temporal extent $T$ of the spacetime volume, which we assume to be large enough to not affect the results.

We begin by defining the finite-volume correlator,
\begin{equation}
C_L(\tau) = 2 E_D L^3 \int_L \dd^3 \xvec \, \e^{-E_{\D} \tau} \langle\bar{\D}^0, \boldsymbol{p}_D|\,\Hcal_\w(
\tau,\xvec) \Hcal_\w(0)\,|\D^0, \boldsymbol{p}_D\rangle_L\,,
\label{eq:fv-4pt-estimator}
\end{equation}
where the finite-volume states are normalized to unity
\begin{equation}
    \langle \D^0, \boldsymbol{p}'_D | \D^0, \boldsymbol{p}_D \rangle_L =  \delta_{\boldsymbol p'_D \boldsymbol p_D} \,.
\end{equation}
This can be rewritten in the form
\begin{equation}
    C_L(\tau) = \int\frac{\dd\omega}{2\pi} \, \e^{-\omega\tau}  \rho_L(\omega) \,,
    \label{eq:corr_vs_rho}
\end{equation}
where $\rho_L(\omega)$ represents the finite-volume spectral density, analogous to its infinite-volume counterpart in~\cref{eq:density_inf_vol}. It is given by
\begin{equation}
    \rho_L(\omega) = 2 E_D L^3 \langle\bar{\D}^0, \boldsymbol{p}_D|\,\Hcal_\w \, (2\pi)\delta(\hat{H}-\omega) \, L^3\delta_{\hat{\mathbf{P}},\boldsymbol p_D} \, \Hcal_\w\,|\D^0, \boldsymbol{p}_D\rangle_L\,.
    \label{eq:density_finite_vol}
\end{equation}
Since in a finite volume the number of momenta is discrete, the finite-volume spectral density corresponds to a sum of delta functions evaluated in correspondence of the energies of all the possible internal on-shell states. However, it does not give direct access to the amplitudes $\M_{12}$ and $\Gamma_{12}$ of our interest like $\rho(\omega)$ in~\cref{eq:density_inf_vol}, as it either vanishes or diverges when sampled at given energies. Moreover, the extraction of $\rho_L(\omega)$ from the lattice correlator $C_L(\tau)$ requires the numerical solution of an inverse Laplace-transform, which is an ill-posed problem as the data are affected by statistical uncertainty. A possible solution to this issue is provided by the introduction of some smeared spectral densities, which are smooth functions of the energy and are well defined quantities in the infinite volume limit at fixed smearing. As discussed in ref.~\cite{Bulava:2019kbi} for the evaluation of scattering amplitudes, and later in ref.~\cite{Frezzotti:2023nun} for the study of radiative leptonic decays and, more generally, for hadronic observables with two external currents and external single-hadron or vacuum states, the $\ii\epsilon$ prescription used to regulate the amplitude in Minkowski spacetime provides a natural smearing parameter. In fact, defining the smeared hadronic amplitudes
\begin{equation}
\hat{\rho}^\mathrm{R}_L(\bar\omega,\epsilon) = \int\frac{\dd\omega}{2\pi} \, \rho_L(\omega)  \, \mathcal{K}_\epsilon^\mathrm{R}(\omega,\bar\omega)\,,
\qquad
\hat{\rho}^\mathrm{I}_L(\bar\omega,\epsilon) = \int\frac{\dd\omega}{2\pi} \, \rho_L(\omega)  \, \mathcal{K}_\epsilon^\mathrm{I}(\omega,\bar\omega)\,,
\label{eq:smeared_hadr_ampl}
\end{equation}
with the kernels $\mathcal{K}_\epsilon^\mathrm{R,I}(\omega,\bar\omega)$ defined in~\cref{eq:kernels},
one can obtain the desired long-distance absorptive and dispersive contributions to the mixing as
\begin{align}
    \M_{12}^\mathrm{LD}(\bar\omega) &= -\lim_{\epsilon\to 0}\lim_{L\to\infty} \, \frac{1}{2m_\D} \, \hat{\rho}^\mathrm{R}_L(\bar\omega,\epsilon)\,, \qquad
    \Gamma_{12}(\bar\omega) = \lim_{\epsilon\to 0}\lim_{L\to\infty} \, \frac{1}{m_\D} \, \hat{\rho}^\mathrm{I}_L(\bar\omega,\epsilon)\,.
    \label{eq:double_limits}
\end{align}
Evaluating these at the energy $\bar{\omega} = E_D$ gives the desired mixing parameters, where $\M_{12}^\mathrm{LD}$ corresponds to the second term in~\cref{eq:M12_limit} and $\Gamma_{12}$ is defined in~\cref{eq:Gamma12_limit}.

Next, we construct the finite-volume correlator $C_L(\tau)$ starting from two-point and four-point lattice correlation functions.

Let $\phi_\D(x)^\dagger = \bar{c}(x)\, \ii\gamma_5\,  u(x)$ be the creation operator of a $\D^0$ meson, or equivalently the annihilation operator of a $\Dbar^0$ meson.
Taking $x_0>0$, the two-point function for a $D^0$ meson with spatial momentum~$\pvec_D$ is given by
\begin{align}
    C_2(x_0) &= \int_L \dd^3 \xvec \, \e^{-\ii \pvec_D\cdot\xvec} \,   \langle 0 |  \phi_{\D} (x_0,\xvec) \phi_{\D}(0)^\dagger | 0\rangle_L = \sum_n |Z_{n}|^2 \, \e^{-  E_{n}x_0} \,,
\end{align}
where $Z_n = \langle 0 | \phi_D(0)|n,\pvec_D\rangle_L$ and $E_n=E_n(\pvec_D)$ are the overlap factor and the energy of the state $|n,\pvec_D\rangle_L$, respectively.

For large times $x_0\gg 1/\delta E$, where $\delta E$ is the gap between the ground and first excited state, the lattice correlation function approaches the asymptotic form
\begin{align}
    C_2(x_0)  \rightarrow |Z_{D}|^2 \, \e^{- E_{D}x_0}  \,,
\end{align}
where $Z_{D} = \langle 0 |\phi_D(0)|\D^0,\pvec_D\rangle_L $ and $E_{D}$ are the ground-state overlap factor and energy of the $\D^0$ meson, respectively.
We can define analogously the two-point function of the $\Dbar^0$ meson built from the interpolating operator $\phi_{\Dbar}(x) = \phi_\D(x)^\dagger$. Since the two operators are connected by charge conjugation, the same correlation function $C_2(x_0)$ also describes the propagation of a $\Dbar^0$ meson.

We now define the following four-point function
\begin{align}
    C_4(x_0,0,\tau,y_0) &= \int_L \dd^3 \xvec \, \dd^3 \yvec \,\dd^3 \zvec \,
    \e^{\ii\pvec_D\cdot(\xvec-\yvec)} \,\langle 0| \phi_{\Dbar} (y_0,\yvec) \, \Hcal_\w(\tau,\zvec) \Hcal_\w(0) \,\phi_D (x_0,\xvec)^\dagger |0 \rangle_L \\
    &=  \sum_{n,n'}\,Z_{n'} Z^\dagger_{n}\, \int_L \dd^3 \zvec
    \, \langle n',\pvec_D |\Hcal_\w(\tau,\zvec)\Hcal_\w(0) |n,\pvec_D \rangle_L\,   \e^{-E_{n'} y_0} \e^{E_{n} x_0}  \, ,
    \label{eq:C4_def}
\end{align}
where $y_0>\tau>0>x_0$, while $n$ and $n'$ are states with same quantum numbers as the $D^0$ and $\Dbar^0$ mesons, respectively. For the moment, we find it convenient to define the correlation function with one of the Hamiltonians set in the origin. We will discuss later a generalized form of the correlation function, where the Hamiltonians are placed at two generic lattice points.
For large times $y_0\gg\tau$ and $x_0\ll 0$ the function in~\cref{eq:C4_def} approaches the form
\begin{align}
    C_4(x_0,0,\tau,y_0) & \rightarrow  |Z_{\D}|^2 \int_L \dd^3 \zvec \,   \langle \Dbar^0,\pvec_D |\Hcal_\w(\tau,\zvec)\Hcal_\w(0) |\D^0,\pvec_D \rangle_L \, \e^{-E_D(y_0-x_0)} \, ,
\end{align}
having used that $Z_{\Dbar}=Z_{D}$.

To extract the finite-volume correlator of~\cref{eq:fv-4pt-estimator}, we form a ratio with the meson two-point functions,
\begin{align}
    |Z_D|^{2} \frac{C_4(x_0,0,\tau,y_0)}{C_2(y_0-\tau) C_2(-x_0)} & \rightarrow \,   \int_L \dd^3 \zvec  \, \e^{-E_D \tau} \,  \langle \Dbar^0,\pvec_D |\Hcal_\w(\tau,\zvec)\Hcal_\w(0) |\D^0,\pvec_D \rangle_L \, .
\end{align}
This leads to our final expression for the finite-volume correlator,
\begin{align}
    \label{eq:fv-corr}
    C_L(\tau)\equiv \lim_{y_0 \to \infty} \lim_{x_0 \to - \infty} \, 2E_DL^3|Z_D|^2 \,\frac{C_4(x_0,0,\tau,y_0)}{C_2(y_0-\tau) C_2(-x_0) } \,,
\end{align}
where the explicit values of $|Z_D|^2$ and $E_D$ can be extracted from a fit to the two-point function.

While the two-point functions are standard quantities in lattice QCD and their optimal implementation will not be covered in detail, the four-point function requires further discussion.
In the remainder of this section, we outline a possible strategy for the numerical computation of the four-point function $C_4(x_0,0,\tau,y_0)$ and argue that employing different techniques for the various Wick contractions is necessary to achieve an optimal signal-to-noise ratio.

\subsection{Computational estimation of diagrams}

We consider now a discretized Euclidean space-time and define the finite-volume lattice correlation functions in a general form, allowing both insertions of the weak Hamiltonian to be located at arbitrary lattice points.
For simplicity, however, we restrict our attention to the case where the external neutral $D$-mesons are at rest, $\pvec_D=\zero$.

 The four-point function takes the form
\begin{align}
    C_4(x_0,z_0,z_0', y_0) &=
    \frac{1}{L^3}
   \sum_{\zvec,\zvec'}\,
    \langle  \phi_{\Dbar} (y_0) \, \Hcal_\w(z') \Hcal_\w(z) \,\phi_D (x_0)^\dagger  \rangle\,,
    \label{eq:C4_lattice}
\end{align}
where $\phi_{\Dbar} (y_0)$ and $\phi_D (x_0)^\dagger$ denote the meson interpolating operators projected to zero spatial momentum.
Here the angle brackets indicate the usual QCD expectation value, potentially also including the effects of the finite time extent ($T$), though these do not affect the expressions of this section.
To match the notation used in the previous subsection, we can define $\tau = z_0'-z_0$. The additional factor $1/L^3$ compensates the volume factor arising from the integration over the spatial position $\zvec$.

We realize the zero-momentum interpolating field operators in~\cref{eq:C4_lattice} using gauge-fixed wall sources, namely
\begin{equation}
    \phi_{\D}(x_0)^\dagger = \phi_{\Dbar}(x_0) = \bigg[\sum_\xvec \bar{c}(x_0,\xvec)\psi(\xvec)\bigg]\bigg[\sum_{\xvec'} \psi(\xvec')^\dagger \ii \gamma_5 u(x_0,\xvec')\bigg]\,,
    \label{eq:wall_source_interpolator}
\end{equation}
with the source $\psi(\xvec)$ being defined such that
\begin{equation}
    \psi(\xvec)_{\alpha,a} \psi(\xvec')_{\beta,b}^\dagger = \delta_{\xvec,\xvec'} \, \delta_{\alpha,\beta}\, \delta_{a,b}\,,
\end{equation}
where $\alpha,\beta$ denote Dirac indices, while $a,b$ are colour indices.

\begin{figure}[h!]
    \centering
    \includegraphics[width=0.5\linewidth]{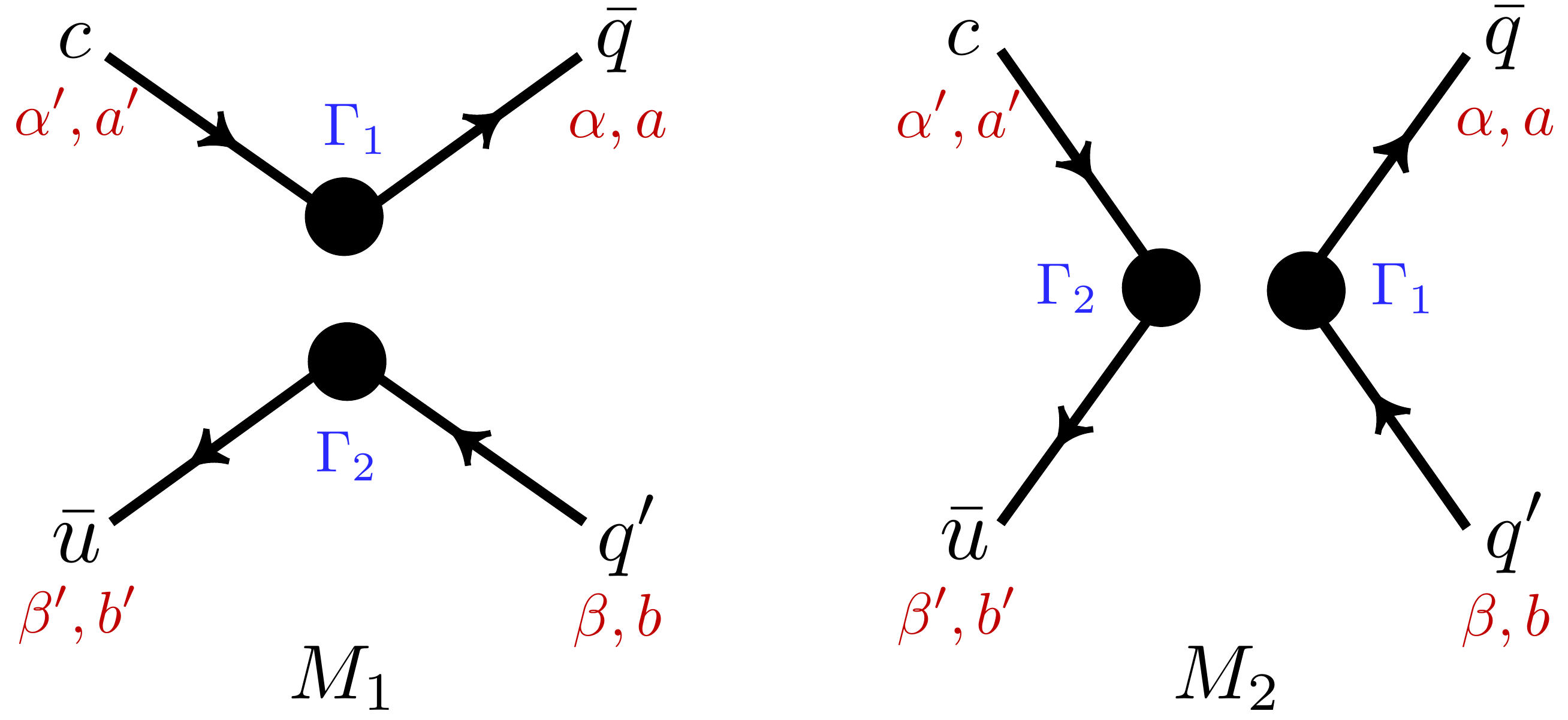}
    \caption{Dirac and colour structures $M_1$ and $M_2$ of the operators $Q_1^{\bar{q}q'}$ and $Q_2^{\bar{q}q'}$.}
    \label{fig:M1_M2}
\end{figure}

As discussed in~\cref{sec:mixing,sec:renormalization}, the relevant operators in the weak Hamiltonians for the calculation of the mixing amplitude are the current-current operators $Q_1^{\bar{q}q'}$ and $Q_2^{\bar{q}q'}$, defined in~\cref{eq:Q1,eq:Q2}. Using Fierz rearrangement, we can express $Q_2^{\bar{q}q'}$ as the product of two colour-singlet bilinear operators, allowing us to introduce the following useful  notation for the two possible Dirac-colour structures:
\begin{align}
    Q_r^{\bar{q}q'}\!(z) = \bar{q}(z)_{\alpha,a} \, c(z)_{\alpha',a'} \, \bar{u}(z)_{\beta',b'} \, q'(z)_{\beta,b} \ [M_r]^{aa'b'b}_{\alpha\alpha'\beta'\beta}\,,
    \label{eq:Q_spincolor_indices}
\end{align}
with
\begin{equation}
    [M_1]^{aa'b'b}_{\alpha\alpha'\beta'\beta} = (\Gamma_1)_{\alpha\alpha'}(\Gamma_2)_{\beta'\beta} \;
    \delta_{a,a'}\delta_{b,b'}
    \,,\qquad
    [M_2]^{aa'b'b}_{\alpha\alpha'\beta'\beta} = (\Gamma_1)_{\alpha\beta}(\Gamma_2)_{\beta'\alpha'}\;
    \delta_{a,b}\delta_{a',b'}\,.
\end{equation}
In the case of interest, the Dirac matrices are given by $\Gamma_1 = \Gamma_2 = \gamma^\mu(1 - \gamma_5)$, with an implicit sum over repeated Lorentz indices.
Visual representations of the structures $M_1$ and $M_2$ are shown in~\cref{fig:M1_M2}.

Neglecting contributions of $\mathcal{O}(\lambda_b)$, the double insertions of the weak Hamiltonians yield
\begin{align}
    \Hcal_\w^\mathrm{SCS}(z')\Hcal_\w^\mathrm{SCS}(z) &=  \frac{(\lambda_s - \lambda_d)^2}{4}\sum_{r,r'}  C_rC_{r'}\, Q_r^\gim(z')\, Q_{r'}^\gim(z)\,,\\
    \Hcal_\w^\mathrm{CF}(z')\Hcal_\w^\mathrm{DCS}(z) &= -\frac{(\lambda_s - \lambda_d)^2}{4}\sum_{r,r'} C_r C_{r'} \, Q_r^{\bar{s}d}(z')\,Q_{r'}^{\bar{d}s}(z)
    \,,\\
    \Hcal_\w^\mathrm{DCS}(z')\Hcal_\w^\mathrm{CF}(z) &=
    -\frac{(\lambda_s - \lambda_d)^2}{4}\sum_{r,r'}  C_r C_{r'} \, Q_r^{\bar{d}s}(z')\,Q_{r'}^{\bar{s}d}(z)
    \,,
\end{align}
where $r,r'=1,2$ and
where we have introduced the shorthand notation for the difference of operators in the SCS Hamiltonian,
\begin{equation}
    Q_r^\gim(x) = Q_r^{\bar{s}s}(x) - Q_r^{\bar{d}d}(x)\,.
\end{equation}
The correlator $C_4(x_0,z_0,z_0', y_0)$ can then be written as
\begin{equation}
    C_4(x_0,z_0,z_0', y_0) = \frac{(\lambda_s - \lambda_d)^2}{4}\sum_{r,r'}  C_rC_{r'} \, D_{r,r'}(x_0,z_0,z_0', y_0)\,,
\end{equation}
with
\begin{equation}
    D_{r,r'}(x_0,z_0,z_0', y_0) = D^\mathrm{SCS\times SCS}_{r,r'}(x_0,z_0,z_0', y_0) - D^\mathrm{CF\times DCS}_{r,r'}(x_0,z_0,z_0', y_0)
    \label{eq:D_correlator}
\end{equation}
and
\begin{align}
    D^\mathrm{SCS\times SCS}_{r,r'}(x_0,z_0,z_0', y_0) &= \frac{1}{L^3}
   \sum_{\zvec,\zvec'}\,
    \langle  \phi_{\Dbar} (y_0) \, Q_r^\gim(z') Q_{r'}^\gim(z) \,\phi_D (x_0)^\dagger  \rangle\,,
    \label{eq:D_scs_scs}\\
    D^\mathrm{CF\times DCS}_{r,r'}(x_0,z_0,z_0', y_0) &= \frac{1}{L^3}
   \sum_{\zvec,\zvec'}\,
    \langle  \phi_{\Dbar} (y_0) \, Q_r^{\bar{s}d}(z') Q_{r'}^{\bar{d}s}(z) \,\phi_D (x_0)^\dagger  \rangle + (s\leftrightarrow d)\,.
    \label{eq:D_cf_dcs}
\end{align}

\begin{figure}
    \centering
    \includegraphics[width=0.9\linewidth]{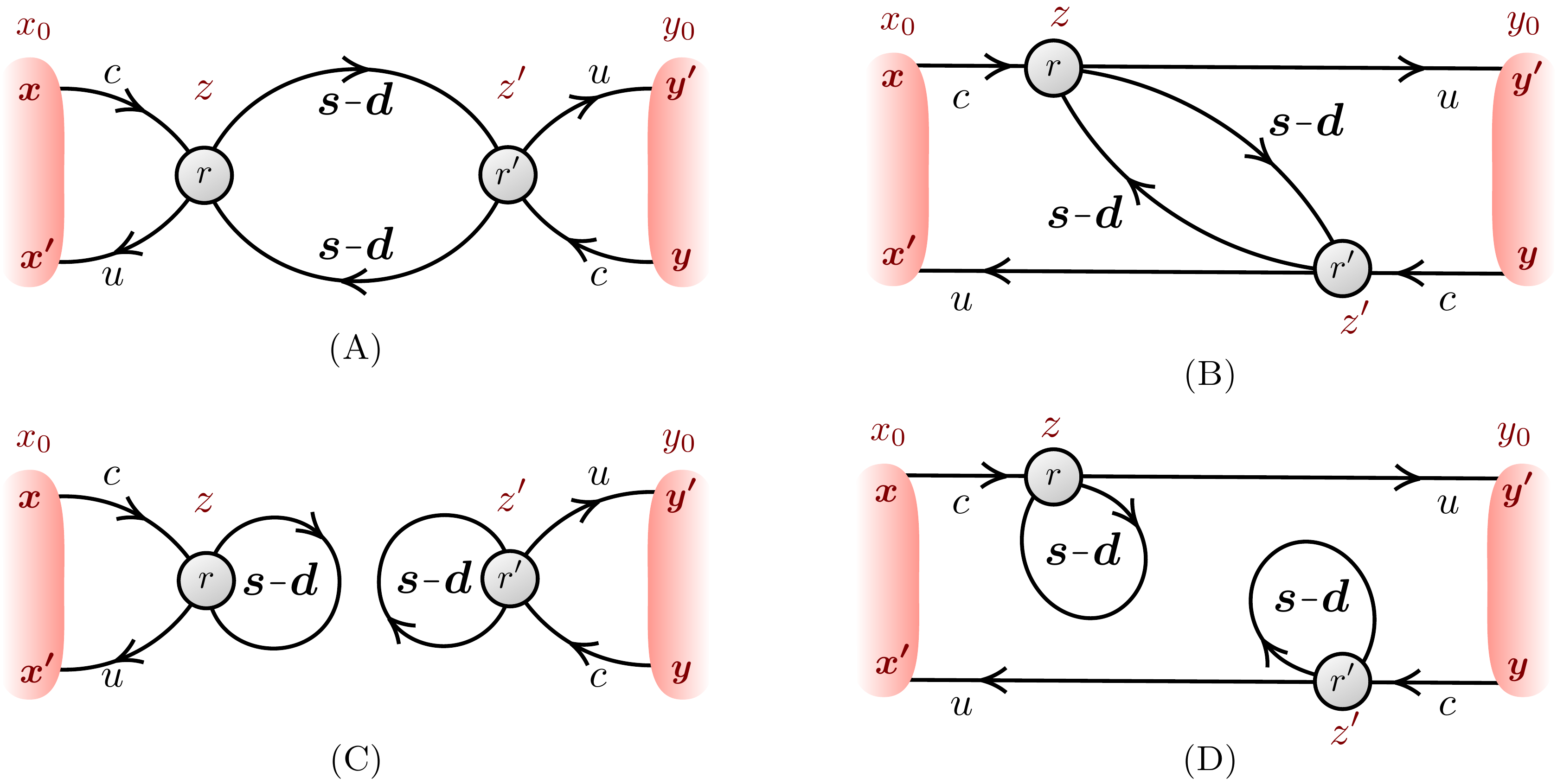}
    \caption{All four topologies contributing to the correlation function $D_{r,r'}(x_0,z_0,z_0', y_0)$. Topology (A) corresponds to the combination of diagrams (a) and (e) in \cref{fig:wick_contractions}, and topology (B) to the combination of (b) and (f). Red endcaps indicate wall sources at timeslices $x_0$ and $y_0$, while grey blobs represent insertions of the weak Hamiltonian with Dirac structure $r$ at positions $z$ and $z'$. All propagators that originate and terminate at weak operator insertions correspond to $s-d$ lines and can be estimated using the split-even technique. This allows for summing over all spatial positions $\zvec$, $\zvec'$ and accessing all weak-Hamiltonian time combinations $z_0$, $z_0'$ at the cost of contractions.
}
    \label{fig:wick_contractions}
\end{figure}

As discussed in~\cref{sec:mixing}, the combination of SCS$\times$SCS and CF$\times$DCS correlators in~\cref{eq:D_correlator} ensures that the internal lines of the correlator $D_{r,r'}(x_0,z_0,z_0', y_0)$ always appear as differences of strange and down quark propagators. For brevity, we refer to this difference as an ``$s-d$ propagator'' in the following discussion.
Rather than classifying the contributions to the correlator based on the operators inserted, it is convenient for the remainder of the discussion to divide the correlator into contributions from the four topologies arising from the Wick contractions of the internal quark fields. These are shown in~\cref{fig:wick_contractions} and are referred to in the rest of the section as topologies (A), (B), (C) and (D). We split the correlator accordingly as
\begin{equation}
        D_{r,r'}(x_0,z_0,z_0', y_0) = \sum_{k\in\{\text{A,B,C,D}\}} \, D_{r,r'}^{(k)}(x_0,z_0,z_0', y_0)\,.
\end{equation}
Clearly, the topologies (C) and (D) arise only when two Hamiltonians $\mathcal{H}_\w^\mathrm{SCS}$ are inserted, while (A) and (B) receive contributions from all operator insertions.

\subsubsection{Split-even approach for $s-d$ diagrams}

A particular challenge arises with the topologies (C) and (D) in~\cref{fig:wick_contractions}, where internal quark lines both originate and terminate at the same spacetime location. These diagrams exhibit an exponentially deteriorating signal-to-noise ratio as the temporal separation between the two weak Hamiltonians increases.
As demonstrated in ref.~\cite{Giusti:2019kff} for Wilson fermions and extended to Domain-Wall fermions in ref.~\cite{Harris:2023zsl}, by replacing
\begin{align}
    S_s(x,x)-S_d(x,x) = (m_d-m_s) \sum_w S_d(x,w) S_s(w,x)
\end{align}
a more efficient estimator for these diagrams can be obtained, which has a variance that is suppressed by a factor $(m_s-m_d)^4$, leading to drastically improved signals in these noisy contributions.

The application of this approach to the calculation of the rare kaon decay  $K \to \pi \ell^+ \ell^-$ has been proposed in ref.~\cite{Hodgson:2025iit}, where differences between correlators with an internal loop of charm and up quark propagators arise as a consequence of GIM mechanism.

Although diagrams (A) and (B) do not benefit from the same variance reduction as (C) and (D), we propose here a method to compute all four topologies using the split-even approach. In~\cref{sec:correlators_point_sources}, we also present an alternative, more conventional strategy for computing the fully connected diagrams, which provides better statistical precision.

We start by noting that an internal $s-d$ propagator can be rewritten using the split-even approach mentioned above in terms of the following estimator
\begin{align}
    S_{s-d}(z,z')_{\alpha\beta} &= S_{s}(z,z')_{\alpha\beta} -S_{d}(z,z')_{\alpha\beta} \\
    &= (m_d-m_s) \, \frac{1}{N_\mathrm{s}}\sum_{i=1}^{N_\mathrm{s}}\bigg[\sum_w S_d(z,w)\eta_i(w)\bigg]_{\alpha}\bigg[\sum_{w'}\eta_i(w')^\dagger S_s(w',z')\bigg]_{\beta}\,,
\end{align}
where $\eta_i^{\alpha,a}(x)$ are random noise sources such that
\begin{equation}
    \langle \eta^{\alpha,a}(x) \eta^{\beta,b}(x')^\dagger\rangle_\mathrm{src} = \delta_{x_0,x_0'}\delta_{\xvec,\xvec'} \delta_{\alpha,\beta}\delta_{a,b}\,, \qquad \langle \eta^{\alpha,a}(x) \rangle_\mathrm{src} = 0\,,
\end{equation}
having used the notation $\langle\dots\rangle_\mathrm{src} = \lim_{N_\mathrm{s}\to \infty} \tfrac{1}{N_\mathrm{s}}\sum_{i=1}^{N_\mathrm{s}}(\dots)$.

Let us define now the wall-source and noise-source propagators as
\begin{align}
    S_f^\psi(z,y_0) &= \sum_{\yvec} S_f(z,y)\,\psi(\yvec)\,, \qquad \widetilde{S}_f^\psi(z,y_0) = \sum_{\yvec} S_f(z,y)\,\gamma_5\psi(\yvec)\,,
    \label{eq:prop_psi}\\
    S_f^{\eta_i}(z) &= \sum_{w} S_f(z,w)\,\eta_i(w)\,, \qquad \ \
    \widetilde{S}_f^{\eta_i}(z) = \sum_{w} S_f(z,w)\,\gamma_5\eta_i(w)\,,
    \label{eq:prop_eta}
\end{align}
which act as vectors in Dirac and colour space.

In this way, the split-even propagator can be written as
\begin{equation}
\begin{tikzpicture}[baseline={-2pt}]
\draw [->-, line width=.7mm] (0,0) -- (2,0) node [above] at (1,0.2) {$s-d$};
\end{tikzpicture}
\ = \
\begin{tikzpicture}[baseline={-2pt}]
\draw [->-, line width=.7mm ] (0,0) -- (1,0) node [above] at (0.5,0.2) {$s$};
\node at (1,0) [circle,draw=black,fill=green2,line width=.7mm, inner sep=0pt, minimum size=7pt] {};
\draw [->-, line width=.7mm ] (1.3,0) -- (2.3,0) node [above] at (1.8,0.2) {$d$};
\end{tikzpicture}
\ = \ (m_d-m_s) \, \big[S^{\eta_i}_d(z)\big]\big[\widetilde{S}^{\eta_i}_s(z')^\dagger\gamma_5\big]\,,
\end{equation}
and all the diagrams can then be constructed starting from the following building block
\begin{align}
    \mathcal{B}^{ij}_r(x_0,z_0,y_0) &=
    \begin{tikzpicture}[baseline={-2pt},blob/.style={rectangle,minimum size=6mm,rounded corners=3mm,very thick,draw=black,top color=white,bottom color=black!20}]
    \draw [|->, line width=.7mm] (0,0) -- (0.7,0) node [above] at (0.5,0.2) {$c$};
    \draw [-, line width=.7mm] (0.5,0) -- (1,0);
    \draw [-, line width=.7mm] (1.5,0) -- (2,0);
    \draw [>-|, line width=.7mm] (1.8,0) -- (2.5,0) node [above] at (2,0.2) {$u$};
    \draw [->-, line width=.7mm ] (1.25,0) -- (0.75,-1)
    node [above] at (0.65,-0.8) {$s$}
    node [below] at (0.75,-1.1) {$i$};
    \node at (0.75,-1) [circle,draw=black,fill=green2,line width=.7mm, inner sep=0pt, minimum size=7pt] {};
    \draw [->-, line width=.7mm ] (1.75,-1) -- (1.25,0)
    node [above] at (1.85,-0.8) {$d$}
    node [below] at (1.75,-1.1) {$j$};
    \node (E) [blob] at (1.25,0) {$r$};
    \node [above] at (1.25,0.3) {$z_0$};
    \node [below] at (0.1,-0.2) {$x_0$};
    \node [below] at (2.5,-0.2) {$y_0$};
    \end{tikzpicture}
    \label{eq:Bcal}
    \\
    &
    = [M_r]^{aa'b'b}_{\alpha\alpha'\beta'\beta} \sum_\zvec \,
    [S_u^\psi(z,y_0)^\dagger \gamma_5]{}_{\beta',b'} \, [S_d^{\eta_j}(z)]{}_{\beta,b} [\widetilde{S}_s^{\eta_i}(z)^\dagger\gamma_5]{}_{\alpha,a} [S_c^\psi(z,x_0)]{}_{\alpha',a'}\,,\nn
\end{align}
where $r=1,2$ denotes the insertion of the current-current operators $Q_r^{\bar{q}q'}$.

All diagrams shown in~\cref{fig:wick_contractions} can then be obtained as follows
\begin{align}
    D_{r,r'}^\mathrm{(A)}(x_0,z_0,z_0',y_0) &= -(m_s-m_d)^2  \, \frac{1}{L^3 N_\mathrm{s}^2}\sum_{i,j=1}^{N_\mathrm{s}}\Big[ \mathcal{B}^{ij}_r(x_0,z_0,x_0) \,\mathcal{B}^{ji}_{r'}(y_0,z_0',y_0)\Big]\,,\\
    D_{r,r'}^\mathrm{(B)}(x_0,z_0,z_0',y_0) &= -(m_s-m_d)^2 \, \frac{1}{L^3 N_\mathrm{s}^2}\sum_{i,j=1}^{N_\mathrm{s}}\Big[ \mathcal{B}^{ij}_{r}(x_0,z_0,y_0) \,\mathcal{B}^{ji}_{r'}(y_0,z_0',x_0)\Big]\,,\\
    D_{r,r'}^\mathrm{(C)}(x_0,z_0,z_0',y_0) &= -(m_s-m_d)^2  \, \frac{1}{L^3 N_\mathrm{s}^2}\sum_{i,j=1}^{N_\mathrm{s}}\Big[ \mathcal{B}^{ii}_{r}(x_0,z_0,x_0) \,\mathcal{B}^{jj}_{r'}(y_0,z_0',y_0)\Big]\,,\\
    D_{r,r'}^\mathrm{(D)}(x_0,z_0,z_0',y_0) &= -(m_s-m_d)^2 \, \frac{1}{L^3 N_\mathrm{s}^2}\sum_{i,j=1}^{N_\mathrm{s}} \Big[ \mathcal{B}^{ii}_{r}(x_0,z_0,y_0) \,\mathcal{B}^{jj}_{r'}(y_0,z_0',x_0)\Big]\,.
\end{align}

Due to diagram (C) being quark-line disconnected, we expect its gauge variance to be constant with time and the noise-to-signal ratio to grow exponentially. In a computation of the $K_L-K_S$ mass difference, these diagrams were assumed to be heavily suppressed by the OZI rule~\cite{Okubo:1963fa} (due to the removal of the vacuum-state contribution by a shift of the weak Hamiltonian), but found to be much larger than anticipated~\cite{Bai:2014cva}. Special attention will therefore need to be brought to the effective computation of this diagram. While the above mentioned split-even method has the advantage of the lattice volume average at both weak Hamiltonians, the method has not been tested for these diagrams yet.

We observe that the use of stochastic sources is essential for diagrams (C) and (D), which involve self-connected quark loops. In these cases, the split-even approach effectively reduces the variance of the correlator by a factor $(m_s-m_d)^4$.
On the other hand, although it is in principle possible to use stochastic sources for diagrams (A) and (B), as shown above, this is not recommended. This would in fact introduce contributions to the variance of the correlators that are comparable to those of two-loop diagrams (C) and (D) before any variance reduction, ultimately leading to a significant degradation in the statistical quality of the correlators.

We therefore propose to use the split-even approach exclusively for the evaluation of diagrams $\mathrm{(C)}$ and $\mathrm{(D)}$ containing self-connected quark loops, while adopting a different strategy, based on point sources and outlined in the following subsection, for the fully-connected diagrams (A) and (B).

\subsubsection{Point-source approach for fully-connected diagrams}
\label{sec:correlators_point_sources}

We now focus on the correlation functions corresponding to topologies (A) and (B) in~\cref{fig:wick_contractions}, which have fully connected quark lines. These correlators arise from either the insertion of two $\mathcal{H}_\w^\mathrm{SCS}$ Hamiltonians or one $\mathcal{H}_\w^\mathrm{CF}$ and one $\mathcal{H}_\w^\mathrm{DCS}$ Hamiltonian. For clarity, we detail the strategy for the latter case; the former follows analogously by simply changing the flavour of the internal quarks.
The methodology for computing these diagrams follows the approach outlined in ref.~\cite{Bai:2014cva}, where the spatial position of one of the weak Hamiltonians is fixed to a specific lattice point, foregoing one volume average.
We fix one of the Hamiltonians at the origin. This choice is arbitrary and does not influence the final result, as any lattice point could be used equivalently. Indeed, we will later advocate for sampling these source points randomly from a uniform distribution over the lattice volume to achieve optimal noise reduction in practical lattice QCD calculations.
With this setup, we define
\begin{align}
    D^{ds}_{r,r'}(x_0,z_0,z_0', y_0) &=
   \sum_{\zvec'}\,
    \langle  \phi_{\Dbar} (y_0) \, Q_r^{\bar{s}d}(z_0',\zvec') Q_{r'}^{\bar{d}s}(z_0,\zero) \,\phi_D (x_0)^\dagger  \rangle\,.
\end{align}
It is convenient to define the operator $Q_{r'}^{\bar{d}s}(z_0,\zero)$ in terms of point sources $\xi(w)$, which are defined such that
\begin{equation}
    \xi(x)_{\alpha,a} \xi(x')^\dagger_{\beta,b} = \delta_{x_0,x_0'}\,\delta_{\xvec,\xvec'}\,\delta_{\alpha,\beta}\,\delta_{a,b}\,.
\end{equation}
With this definition, the operator $Q_{r'}^{\bar{d}s}(z_0,\zero)$ can be expressed as follows,
\begin{align}
    Q_{r'}^{\bar{d}s}(z_0,\zero) = F_{r'}(z_0) \, \Big[\sum_v \bar{u}(v) \xi(v) \Big]
    \Big[\sum_w \xi(w)^\dagger c(w) \Big]
    \Big[\sum_{v'} \bar{d}(v') \xi(v') \Big]
    \Big[\sum_{w'} \xi(w')^\dagger s(w') \Big]\,,
\end{align}
where spin indices are implicitly contracted inside the square brackets and we have defined
\begin{equation}
    F_{r'}(z_0) = [M_{r'}]^{aa'b'b}_{\alpha\alpha'\beta'\beta} \
    \xi^\dagger(z_0,\boldsymbol{0})_{\alpha,a} \,
    \xi(z_0,\boldsymbol{0})_{\alpha',a'} \,
    \xi^\dagger(z_0,\boldsymbol{0})_{\beta',b'} \,
    \xi(z_0,\boldsymbol{0})_{\beta,b}\,.
\end{equation}
In the computation of the correlation function, the operator $Q_{r'}^{\bar{d}s}(z_0,\zero)$ will be multiplied from the left by $Q_r^{\bar{s}d}(z')$, which contracts the $\bar{d}$ and $s$ fields.

In addition to the wall-source and noise-source propagators introduced in~\cref{eq:prop_psi,eq:prop_eta}, let us define the following point-source propagators
\begin{align}
    S_f^\xi(z) &= \sum_{y} S_f(z,y)\,\xi(y)\,, \qquad \quad \ \; \widetilde{S}_f^\xi(z) = \sum_{y} S_f(z,y)\,\gamma_5\xi(y)\,,
    \label{eq:prop_xi}
\end{align}
which also act as vectors in Dirac and colour space.

In this way, the correlators of interest can be obtained from the contraction of the following two quantities:
\begin{align}
   A_r(x_0,z_0,y_0) &=
    \begin{tikzpicture}[baseline={-2pt},square/.style={rectangle,minimum size=5mm,rounded corners=0mm,very thick,draw=black,top color=white,bottom color=blue!20}]
    \draw [|->, line width=.7mm] (0,0) -- (0.7,0) node [above] at (0.5,0.2) {$c$};
    \draw [-, line width=.7mm] (0.5,0) -- (1,0);
    \draw [-, line width=.7mm] (1.5,0) -- (2,0);
    \draw [>-|, line width=.7mm] (1.8,0) -- (2.5,0) node [above] at (2,0.2) {$u$};
    \node (E) [square] at (1.25,0) {$r$};
    \node [above] at (1.25,0.3) {$z_0$};
    \node [below] at (0.1,-0.2) {$x_0$};
    \node [below] at (2.5,-0.2) {$y_0$};
    \end{tikzpicture}
    \\[5pt]
   &=
   F_{r}(z_0) \Big[\sum_v S^\psi_u(v,y_0)^\dagger \gamma_5 \xi(v)\Big] \Big[\sum_w \xi(w)^\dagger S_c^\psi(w,x_0)\Big] \,,
   \nonumber
   \end{align}
   \begin{align}
   B_{r'}^{ds}(x_0,z_0',y_0)
   &=
   \begin{tikzpicture}[baseline={-2pt},blob/.style={rectangle,minimum size=6mm,rounded corners=3mm,very thick,draw=black,top color=white,bottom color=black!20}]
    \draw [|->, line width=.7mm] (0,0) -- (0.7,0) node [above] at (0.5,0.2) {$c$};
    \draw [-, line width=.7mm] (0.5,0) -- (1,0);
    \draw [-, line width=.7mm] (1.5,0) -- (2,0);
    \draw [>-|, line width=.7mm] (1.8,0) -- (2.5,0) node [above] at (2,0.2) {$u$};
    \draw [->-, line width=.7mm ] (1.25,0) to [bend right=45] (1.25,-1.25) node [left] at (0.9,-0.8) {$s$};
    \draw [->-, line width=.7mm ] (1.25,-1.25) to [bend right=45] (1.25,0) node [right] at (1.5,-0.8) {$d$};
    \filldraw [black] (1.25,-1.25) circle [radius=3pt];
    \node (E) [blob] at (1.25,0) {$r'$};
    \node [above] at (1.25,0.3) {$z_0'$};
    \node [below] at (0.1,-0.2) {$x_0$};
    \node [below] at (2.5,-0.2) {$y_0$};
    \end{tikzpicture}
    \label{eq:B_ds}
   \\
   &= [M_{r'}]^{aa'b'b}_{\alpha \alpha' \beta' \beta} \sum_{\zvec'} [S_{u}^\psi(z',y_0)^\dagger \gamma_5]_{\beta',b'}
 [S_{d}^\xi(z')]_{\beta,b}  [\widetilde{S}_{s}^\xi(z')^\dagger \gamma_5 ]_{\alpha,a} [S_{c}^\psi(z',x_0)]_{\alpha',a'}\,.
 \nonumber
\end{align}
Note that the function $B_r^{ds}$ is analogous to $\mathcal{B}_r^{ij}$ in~\cref{eq:Bcal}, with the noise sources being substituted by point sources.

Defining the functions $B_{r'}^{sd}$, $B_{r'}^{ss}$ and $B_{r'}^{dd}$, in analogy with $B_{r'}^{ds}$ in~\cref{eq:B_ds}, their contractions with the function $A_r$ give diagrams (A) and (B) in~\cref{fig:wick_contractions} with internal $s-d$ quark propagators
\begin{align}
    D^{(\mathrm{A})}_{r,r'}(x_0,z_0,z_0', y_0) &=  -A_{r}(x_0,z_0,x_0)  [ B_{r'}^{ss+dd}(y_0,z'_0,y_0)
    - B_{r'}^{sd+ds}(y_0,z'_0,y_0)
    ]\,, \\
    D^{(\mathrm{B})}_{r,r'}(x_0,z_0,z_0', y_0) &= -A_{r}(x_0,z_0,y_0) [B_{r'}^{ss+dd}(y_0,z'_0,x_0) - B_{r'}^{sd+ds}(y_0,z'_0,x_0)] \,.
\end{align}

For any selection of propagator source positions $x_0$, $y_0$ and $z_0$, the four-point correlation function can be evaluated for any $z'_0$ with additional contractions, which are vastly less costly compared to extra inversions.
In the approach of ref.~\cite{Bai:2014cva}, the lattice correlation function is computed by summing over both $z_0$ and~$z'_0$, which requires evaluating the diagram with a single point source on each timeslice~$z_0$. Owing to the high computational cost of this method, their analysis is limited to a fixed meson separation~$\Delta T = y_0 - x_0$.
In contrast, in our approach we do not require summation over $z_0$ and $z'_0$, enabling us to explore the dependence on $\tau = z'_0 - z_0$ and multiple meson separations~$\Delta T$.

To leverage the full benefit of this point-source approach, one may consider field sparsening for the evaluation of diagrams (A) and (B). This technique, proposed in ref.~\cite{Li:2020hbj}, involves randomly sampling the point-source position uniformly across the spatial volume on each gauge configuration and timeslice. It significantly reduces computational cost while only mildly increasing gauge noise compared to computing correlation functions from all available point sources, due to the strong spatial correlations in lattice gauge ensembles. Although the method has so far been tested only for two- and three-point functions, where the additional noise introduced by the random field selection is shown to be subdominant to gauge noise, we expect it to be similarly effective for the four-point functions required in $D$-meson mixing.

A promising alternative to the point-source technique could be the use of multigrid low-mode averaging (MG-LMA)~\cite{Gruber:2024cos}, a technique shown to be especially effective for estimating quark-line connected diagrams. Its benefits are most significant at large Euclidean time separations, which is precisely the regime of interest for our four-point function, where the two meson sources/sinks and the two current insertions must be well separated to suppress excited-state contamination.  It is worth investigating whether MG-LMA can aid in the evaluation of diagrams (A) and (B).

\section{Spectral reconstruction of the mixing amplitudes}
\label{sec:reconstruction}

In this section, we detail how the neutral $D$-meson mixing amplitudes can be determined via spectral reconstruction. We first review various algorithms available to achieve the basic task, focusing completely on linear reconstruction methods. Next, we examine sources of systematic uncertainty in these analyses and propose strategies to mitigate them, with a particular focus on the zero-smearing-width extrapolations. We then discuss freedom in whether the reconstruction is performed for the full correlator or for correlators projected onto definite parity. Special attention is given to the challenges specific to the spectral density relevant for $D$-meson mixing, guided by the experimentally known resonance spectrum. We illustrate these challenges with a model spectral density. Finally, we present considerations for a first-principles lattice QCD calculation, including a proposal to exploit the vanishing mixing amplitudes at the $U$-spin symmetric point to extract the physical amplitude from simulations at unphysical quark masses.

\subsection{Numerical methods for reconstructing the spectral density}

The goal is to efficiently reconstruct the smeared finite-volume spectral densities $\hat{\rho}^\mathrm{R}_L(E_\D,\epsilon)$ and $\hat{\rho}^\mathrm{I}_L(E_\D,\epsilon)$ starting from the finite-volume correlator $C_L(\tau)$, using linear reconstruction techniques. In all cases, the essential idea is to approximate the smearing kernels $\mathcal{K}_\epsilon^\mathrm{R}(\omega,\bar\omega)$ and $\mathcal{K}_\epsilon^\mathrm{I}(\omega,\bar\omega)$ defined in~\cref{eq:kernels} with a linear combination of decaying exponentials as follows
\begin{equation}
\widetilde{\mathcal{K}}_\epsilon^\mathrm{\alpha}(\boldsymbol g^{\alpha} \vert \omega,\bar\omega) = \sum_{ n= n_\mathrm{min}}^{ n_\mathrm{max}} g_n^{\alpha}(\bar\omega,\epsilon) \, \e^{-\omega a n}
=
\sum_{ n= n_\mathrm{min}}^{ n_\mathrm{max}} g_n^{\alpha}(\bar\omega,\epsilon) \, y^n \bigg\vert_{y = \e^{-\omega a}} \,,
\label{eq:kernel_approx}
\end{equation}
where $\alpha \in \{\mathrm{R}, \mathrm{I}\}$ labels -- here and in the following -- the real and imaginary parts of the smearing kernel, respectively, and the tilde indicates that  $\widetilde{\mathcal{K}}_\epsilon^\alpha$ is an imperfect approximation of the target smearing kernel $\mathcal{K}_\epsilon^\alpha(\omega,\bar\omega)$.
Here we have defined $\tau=an$, where $a$ is the lattice spacing, and the sum runs over a set of non-negative integers $n_\mathrm{min} \leq n \leq n_\mathrm{max}$.
To avoid contact terms one should generally require $n_\mathrm{min} \geq 1$.
In principle the range can be freely chosen, up to the caveat that it will affect the quality of $\widetilde{\mathcal{K}}_\epsilon^\mathrm{\alpha}(\boldsymbol g^{\alpha} \vert \omega,\bar\omega)$ as an approximation of $\mathcal{K}_\epsilon^\alpha(\omega,\bar\omega)$.
With the second equality in~\cref{eq:kernel_approx} we have emphasized the fact that we are considering a generic polynomial in the variable~$y = \e^{-\omega a}$.

It then directly follows that
\begin{equation}
{\widetilde \rho}_{L}^{\,\alpha}(\boldsymbol{g}^\alpha \vert \bar\omega,\epsilon) = \sum_{n=n_\mathrm{min}}^{n_\mathrm{max}} g_n^\alpha(\bar\omega,\epsilon) \, C_L(a n)\,,
\end{equation}
where we have defined
\begin{equation}
{\widetilde \rho}_{L}^{\,\alpha}(\boldsymbol{g}^\alpha \vert \bar\omega,\epsilon) = \int \frac{\dd\omega}{2\pi} \, \widetilde{\mathcal{K}}_\epsilon^{\alpha}(\boldsymbol{g}^\alpha \vert \omega,\bar\omega) \, {\rho}_L(\omega)\,.
\end{equation}
If one can reliably assign a systematic uncertainty associated with
\begin{equation}
\widetilde{\mathcal{K}}_\epsilon^{\alpha}(\boldsymbol{g}^\alpha \vert \omega,\bar\omega) - {\mathcal{K}}_\epsilon^{\alpha}(\omega,\bar\omega) \neq 0 \,,
\end{equation}
then ${\widetilde \rho^{\,\alpha}}_{L}(\boldsymbol{g}^\alpha \vert \bar\omega,\epsilon)$ can be viewed as an estimator for the smeared, finite-volume spectral density $\hat{\rho}_{L}^{\,\alpha}(\bar\omega,\epsilon)$. By applying the ordered double limit described by~\cref{eq:double_limits}, one can then recover the mixing amplitude and determine $\M_{12}$ and $\Gamma_{12}$.

In the following subsections we describe various specific approaches for determining $\boldsymbol{g}$. As these discussions apply quite generally, we drop the $\alpha$ label throughout, with the understanding that any target smearing kernel can be used (provided that the resulting integral with the spectral density is convergent).
In addition to the linear methods considered in this work, various non-linear approaches to spectral reconstruction have also been proposed and used extensively, see for example~refs.~\cite{Rothkopf:2022ctl,Bergamaschi:2023xzx}.

Before turning to the specific approaches we note that, if we consider the intermediate goal of constructing $\hat \rho_L(\bar \omega, \epsilon)$ as well as possible, then an ideal criterion for the coefficients can be identified as the minimization of the following functional:
\begin{align}
W^{\sf ideal}[\boldsymbol g] & = \Big (\rho_L(\bar\omega,\epsilon)-\tilde\rho_L(\boldsymbol g \vert \bar\omega,\epsilon)  \Big )^2 + \text{Var}[\tilde\rho_L(\boldsymbol g \vert \bar\omega,\epsilon)  ] \,.
\end{align}
The first term is the squared systematic uncertainty associated with the difference between the target smearing kernel and the approximation, while the second term is the variance of the estimator of $\tilde\rho_L(\boldsymbol g \vert \bar\omega,\epsilon)$. Thus $W^{\sf ideal}[\boldsymbol g]$ is a measure of the total uncertainty associated with the estimator with systematic and statistical contributions added in quadrature.

The two terms above can be written explicitly as
\begin{align}
\begin{split}
    \Big (\rho_L(\bar\omega,\epsilon)-\tilde\rho_L(\boldsymbol g \vert \bar\omega,\epsilon)  \Big )^2 & = \int_{E_0}^\infty \dd\omega_1 \dd\omega_2 \, \rho_L(\omega_1) \rho_L(\omega_2) \,
    \\ & \qquad \times \big(\mathcal{K}_\epsilon(\omega_1, \bar{\omega}) - \widetilde{\mathcal{K}}_\epsilon(\boldsymbol{g} \vert \omega_1, \bar{\omega})\big) \big(\mathcal{K}_\epsilon(\omega_2, \bar{\omega}) - \widetilde{\mathcal{K}}_\epsilon(\boldsymbol{g} \vert \omega_2, \bar{\omega})\big) \,,
\end{split}
\\
\text{Var}[\tilde\rho_L(\boldsymbol g \vert \bar\omega,\epsilon)  ] & = \sum_{n_1 = n_\mathrm{min}}^{n_\mathrm{max}} \sum_{n_2 = n_\mathrm{min}}^{n_\mathrm{max}} g_{n_1} g_{n_2} \, \mathbb C_{n_1 n_2} \,,
\end{align}
where ${\mathbb C}_{n_1n_2}$ is the covariance of $C_L(\tau)$ between the $n_1$ and $n_2$ timeslices. The parameter $E_0$ may be freely chosen, as long as it lies outside the support of the spectral density and hence $\rho_L(\omega)=0$ for $\omega \leq E_0$.

In practice this expression is not particularly useful because $\rho_L(\omega)$ is unknown and thus $W^{\sf ideal}[\boldsymbol g]$ cannot be computed. However, as a number of the methods below make reference to a specific minimising functional, we think it instructive to provide this as a basis for comparison.

\subsubsection{Backus-Gilbert method and variants}

The seminal work of Backus, Gilbert, and Bullard~\cite{Backus:1968svk,Backus:1970afdsyiu}\footnote{Though ref.~\cite{Backus:1968svk} is commonly cited, also by some of us in previous works, the method is actually much more clearly recognisable in the subsequent work of Backus, Gilbert, and Bullard~\cite{Backus:1970afdsyiu}. In addition the description in Numerical recipes \cite{Press:2007asdfasdf} provides a very clear and compact summary that has clearly motivated the discussion found in lattice QCD related literature.} provides a framework for reconstructing spectral densities from a finite set of data points with statistical uncertainties. The key idea of the original work is to determine the coefficients $\boldsymbol g$ by minimising a functional, with the explicit goal of making
$\widetilde {\mathcal K}_{\epsilon}(\boldsymbol g \vert \omega, \bar \omega)$ as close as possible to a the Dirac delta function: $\delta(\omega - \bar \omega)$.
A second essential observation is that the equation to minimize the functional often relies on inverting an ill-conditioned matrix, and this can be improved by using the covariance matrix of the noisy data to stabilize the inversion. The application of the approach in lattice QCD was discussed as early as 2007 by Meyer in ref.~\cite{Meyer:2007dy}.

An important insight from Hansen, Lupo and Tantalo (HLT)~\cite{Hansen:2019idp}, is that it is more interesting to target a particular smearing kernel (e.g.~a Gaussian with width $\epsilon$ of order $m_\pi$) rather than a delta function.
Following the approach presented in that work, one method to obtain the coefficients is to minimize a linear combination of two functionals
\begin{equation}
W[\boldsymbol{g}] = (1 - \lambda)   A[\boldsymbol{g}]
+ \lambda  \, B[\boldsymbol{g}] \,,
\end{equation}
where $A[\boldsymbol{g}]$ denotes the target functional
\begin{equation}
A[\boldsymbol{g}] = \frac{1}{N_A} \int_{E_0}^\infty \mathrm{d}\omega \Big ( \widetilde{\mathcal{K}}_\epsilon(\boldsymbol{g} \vert \omega,\bar{\omega}) - {\mathcal{K}}_\epsilon(\omega,\bar{\omega}) \Big )^2 \,,
\end{equation}
while $B[\boldsymbol{g}]$ is the error functional, defined in terms of the covariance of $C_L(\tau)$\,,
\begin{equation}
B[\boldsymbol{g}] = \frac{1}{N_B} \sum_{n_1 = n_\mathrm{min}}^{n_\mathrm{max}}\sum_{n_2 = n_\mathrm{min}}^{n_\mathrm{max}} g_{n_1} g_{n_2} \,  {\mathbb C}_{n_1n_2} \,,
\end{equation}
Here we have also included the normalization factors $N_A$ and $N_B$, which can be optionally included to ensure that the two functionals are of similar size. For example in ref.~\cite{Hansen:2019idp} the authors set $N_A$ as the area of $\mathcal{K}_\epsilon(\omega,\bar{\omega})$ on the positive real line and $N_B = C_L(0)$.

Finally, $\lambda$ is a regularization parameter that can be tuned to control the trade-off between the two functionals.
Note that the first term ($A[\boldsymbol{g}]$) is a measure of the distance between the achieved smearing kernel and the target, while the second term is proportional to the statistical uncertainty of the resulting $\widetilde {\rho}_{L}(\boldsymbol{g} \vert \bar\omega,\epsilon)$. Since these are precisely the quantities we aim to minimize, the choice of functional is a natural one.

Another feature is that the functional is quadratic in $\boldsymbol g$ meaning that the minimizing solution is a simple linear result. One finds
\begin{align}
{\boldsymbol g}(\bar{\omega},\epsilon \, | \lambda)_{\sf HLT} &= \mathbf{W}^{-1}(\lambda)\,
\boldsymbol f(\bar{\omega},\epsilon \, | \lambda) \,,
\end{align}
where $\mathbf{W}(\lambda)$ is a matrix with elements
\begin{align}
\mathrm{W}_{n_1 n_2}(\lambda) =   \frac{1{-}\lambda}{N_A} \,\int_{E_0}^{\infty} \dd\omega\,  \e^{- \omega a (n_1 + n_2)}  +  \frac{\lambda}{N_B} \mathbb C_{n_1 n_2}
 = \frac{1{-}\lambda}{N_A} \,\frac{\e^{{-}E_0 a (n_1 + n_2)}}{a(n_1 + n_2)}  + \frac{\lambda}{N_B} \mathbb C_{n_1 n_2} \,,
\end{align}
and $\boldsymbol f(\bar{\omega},\epsilon \, | \lambda)$ is a vector with elements
\begin{align}
f_n(\bar{\omega}, \epsilon \, \vert \lambda) = (1 - \lambda) \frac{1}{N_A} \int_{E_0}^{\infty} \dd\omega\, \e^{- \omega a n}\, \mathcal K_\epsilon(\omega,\bar{\omega}) \,.
\end{align}
Note that the indices $n$, $n_1$, and $n_2$ run over the range $n_\mathrm{min} \leq n, n_1, n_2 \leq n_\mathrm{max}$. So, for example, the first row and column of the matrix $\mathbf{W}(\lambda)$ correspond to $n_1 = n_2 = n_\mathrm{min}$.

Many alternatives are possible. For example a Lagrange multiplier can be used to enforce the normalization condition (or some other condition) exactly. In this case the solution for $\boldsymbol g$ will not generically be proportional to the smearing kernel $\mathcal K_\epsilon(\bar{\omega}, \omega)$, but rather to a linear combination of the smearing kernel and the constraint, as in the solution directly presented in ref.~\cite{Hansen:2019idp}. In addition one has a great deal of freedom with the definition of $A[\boldsymbol g]$. For example the replacement $\dd \omega \to \dd\omega \, h(\omega)$ for some positive function $h(\omega)$ can be used to weight the integral and enforce agreement with the target kernel in a particular region of the spectrum.

Ideally, the reconstruction should be insensitive to choice of the regulator $\lambda$. In~ref.~\cite{Hansen:2019idp} a prescription is proposed for the choice of an optimal parameter $\lambda^*$, which is determined by requiring that the predicted spectral density remains stable within the statistical uncertainty under variations of the value of $\lambda$.

A Bayesian variant of the Backus-Gilbert method has recently been proposed in ref.~\cite{DelDebbio:2024lwm}, reformulating the spectral reconstruction problem in terms of a probability distribution over a functional space of spectral densities. In this framework, the standard Backus-Gilbert resolution parameter $\lambda$ is reinterpreted as a hyperparameter controlling the likelihood in a Bayesian inference setup. For fixed~$\lambda$,\footnote{The authors denote their parameter by $\lambda$, while referring to the standard Backus-Gilbert parameter as $\lambda'$.} both methods yield equivalent central values, but the Bayesian formulation results in more conservative error estimates. Consequently, the optimal choice of $\lambda$ may differ in practice between the two approaches. The authors view the Bayesian-Backus-Gilbert method as complementary to the standard Backus-Gilbert technique, particularly since both share the same computational bottleneck -- the inversion of the Laplace transform.
This compatibility could make a dual or comparative use of the two approaches a promising strategy for better controlling systematic uncertainties in the reconstruction.

\subsubsection{Chebyshev polynomials}

An alternative reconstruction strategy involves the use of Chebyshev polynomials. First proposed by Barata and Fredenhagen in ref.~\cite{Barata:1990rn}, this approach has recently been applied to the study of inclusive semileptonic decays~\cite{Bailas:2020qmv,Gambino:2022dvu,Barone:2023tbl,Kellermann:2025pzt}.

In this method, the smearing kernels $\mathcal{K}_\epsilon(\omega,\bar{\omega})$ are approximated as linear combinations of shifted Chebyshev polynomials. To reconstruct the spectral density over the range $\omega \in [E_0, \Lambda ]$, it is convenient to define a family of orthogonal polynomials in the variable $y = \e^{-a\omega}$
that are orthogonal over this range and thus over $y \in [\e^{- a \Lambda}, \e^{- a E_0}]$. Since the standard Chebyshev polynomials are orthogonal over $[-1, 1]$, the required set is given by
\begin{align}
\widetilde{T}_n(y)  = T_n \! \left ( \frac{2y \, \e^{aE_0}-1-\e^{-a(\Lambda-E_0)}}{1-\e^{-a(\Lambda-E_0)}} \right ) \,,
\end{align}
where $T_n(x)$ denotes the Chebyshev polynomial of the first kind of degree $n$.

Since spectral functions generally have support up to infinity, one might expect this to be the physically relevant choice. However, as we discuss more below, convergence issues arise, especially for the P.V.~kernel. For this reason the cutoff is useful in certain cases, and we will keep it in the following.

The shifted polynomial $\widetilde{T}_n(y)$ then satisfy the orthogonality condition
\begin{equation}
    \int_{\e^{-a \Lambda}}^{\,\e^{-a\omega_0}} \dd y\, \widetilde{T}_n(y)\widetilde{T}_m(y)\,\widetilde{\Omega}(y) = \delta_{m,n} \, h_n,
    \qquad h_n = \frac{\pi}{2}(1 + \delta_{n,0}),
\end{equation}
with weight function
\begin{align}
\widetilde{\Omega}(y) = \big[(y-\e^{-aE_0})(\e^{-a\Lambda}-y)\big]^{-1/2} \,.
\end{align}

This yields an approximation of the form of~\cref{eq:kernel_approx}, which we write as
\begin{align}
\widetilde{\mathcal{K}}_\epsilon(\boldsymbol{g}_{\sf C}|\omega,\bar\omega)
&= \sum_{n = n_\mathrm{min}}^{n_\mathrm{max}} g_n(\bar\omega, \epsilon)_{\sf C} \, \e^{-a\omega n} = y^{n_\mathrm{min}} \sum_{n = n_\mathrm{min}}^{n_\mathrm{max}} g_n(\bar\omega, \epsilon)_{\sf C} \, y^{n-n_\mathrm{min}}
 \bigg \vert_{y = \e^{-a\omega}} \,,
\end{align}
where the subscript ${\sf C}$ indicates that the coefficients $g_n(\bar\omega, \epsilon)_{\sf C}$ are determined via the Chebyshev polynomials.

To obtain the values for the coefficients, we next substitute
\begin{align}
    y^{n-n_\mathrm{min}} = \sum_{j=0}^{n-n_\mathrm{min}} \tilde{a}_j^{(n-n_\mathrm{min})} \, \widetilde{T}_j(y) \,,
\end{align}
where the coefficients $\tilde{a}_j^{(\alpha)}$ are determined using the orthogonality condition:
\begin{equation}
    \tilde{a}_k^{(\alpha)} = \frac{1}{h_k} \int_{\e^{-a \Lambda}}^{\,\e^{-a\omega_0}} \dd y\, y^{\alpha} \, \widetilde{T}_k(y) \, \widetilde{\Omega}(y).
\end{equation}
This gives
\begin{align}
\widetilde {\mathcal{K}}_\epsilon(\boldsymbol{g}_{\sf C}|\omega,\bar\omega)
&= y^{n_\mathrm{min}} \sum_{n = n_\mathrm{min}}^{n_\mathrm{max}} g_n(\bar\omega, \epsilon)_{\sf C} \sum_{j = 0}^{n - n_\mathrm{min}} \tilde{a}_j^{(n - n_\mathrm{min})} \, \widetilde{T}_j(y)\bigg \vert_{y = \e^{-a\omega}}  \,.
\end{align}
By interchanging the order of summation, the kernel can be rewritten in the form
\begin{align}
y^{- n_\mathrm{min}}\, \widetilde{\mathcal{K}}_\epsilon(\boldsymbol{g}_{\sf C}|\omega,\bar\omega)
&=   \sum_{j=0}^{n_\mathrm{max}-n_\mathrm{min}} \widetilde{T}_j(\e^{-a\omega}) \, \tilde{c}_j(\boldsymbol{g}_{\sf C}|n_\mathrm{min},n_\mathrm{max})\,,
\label{eq:exp_times_K_cheby}
\end{align}
where the coefficient of the truncated Chebyshev expansion is related to $g_n(\bar\omega, \epsilon)_{\sf C}$ via
\begin{equation}
    \tilde{c}_j(\boldsymbol{g}_{\sf C}|n_\mathrm{min},n_\mathrm{max}) = \sum_{n=j+n_\mathrm{min}}^{n_\mathrm{max}} \tilde{a}_j^{(n-n_\mathrm{min})} g_n(\bar\omega,\epsilon)_{\sf C}\,.
\end{equation}
The coefficients ${\tilde{c}}_j$ can be obtained imposing the orthogonality of Chebyshev polynomials as
\begin{equation}
    \tilde{c}_j(\boldsymbol{g}_{\sf C}|n_\mathrm{min},n_\mathrm{max}) = \frac{1}{h_j} \, \int_{\e^{- a \Lambda}
    }^{\,\e^{- a E_0}} \, \dd y \, \Big[y^{- n_\mathrm{min}
    }\, {\mathcal{K}}_\epsilon(- \tfrac{1}{a} \ln y,\bar\omega) \Big]\, \widetilde{T}_j(y) \widetilde\Omega(y)\,,
    \label{eq:cj_coeff_def}
\end{equation}
which in turn defines the coefficients $\boldsymbol{g}_{\sf C}$.

If we could send $n_{\rm{max}} \to \infty$, then the Chebyshev expansion would minimize the following $L^\infty$-norm functional:
\begin{equation}
    W[\boldsymbol{g}] = \underset{\omega\in[E_0, \Lambda]}{\mathrm{max}}\Big|\e^{a\omega n_\mathrm{min}}\, \Big(\mathcal{K}_\epsilon(\omega,\bar\omega) -\widetilde{\mathcal{K}}_\epsilon(\boldsymbol{g}|\omega,\bar\omega) \Big)\Big|\,.
\end{equation}
However, when truncating the series to $n_\mathrm{max}-n_\mathrm{min}+1$ terms, as in~\cref{eq:exp_times_K_cheby}, the condition in~\cref{eq:cj_coeff_def} instead corresponds to the minimization of the following $L_2$-norm functional~\cite{Gambino:2022dvu}:
\begin{equation}
    W[\boldsymbol{g}] = \int_{\e^{- a \Lambda}
    }^{\,\e^{- a E_0}} \, \dd y \,\widetilde{\Omega}(y)\, \Big[ y^{-n_\mathrm{min}} \, \Big(\mathcal{K}_\epsilon( {-}\tfrac{1}{a} \ln y,\bar\omega) -
    \widetilde{\mathcal{K}}_\epsilon(\boldsymbol{g}| {-}\tfrac{1}{a} \ln y,\bar\omega) \Big)\Big]^2\,.
    \label{eq:chebyshev_minimization}
\end{equation}
One can understand this by thinking of $\mathcal{K}_\epsilon( {-}\tfrac{1}{a} \ln y,\bar\omega)$ as a vector in a vector space with basis elements given by the Chebyshev polynomials. Projecting onto a subset of Chebyshev polynomials amounts to projecting the vector into the subspace spanned by those elements. Because the vector is perfectly represented in the subspace, what is minimized is the norm of the difference between the vector and its projection onto the subspace. This is equivalent to minimizing the $L_2$-norm of the difference between the two vectors, which is what is shown in~\cref{eq:chebyshev_minimization}.

The expressions above make the role of the ultraviolet regulator $\Lambda$ explicit. In the limit $\Lambda \to \infty$ and for $n_\mathrm{min} > 0$, the factor $y^{-n_{\rm min}}$ causes the integral to diverge at small $y$ (corresponding to a large $\omega$ divergence from $\e^{a\omega n_\mathrm{min}}$) , unless this divergence is sufficiently suppressed by the kernel function $\mathcal{K}_\epsilon(\omega,\bar\omega)$. For both smearing kernels considered in this work (see \cref{eq:kernels}), the integral in~\cref{eq:cj_coeff_def} would indeed diverge, as these kernels decay only as inverse powers of $\omega$ at large $\omega$, which is insufficient to suppress the exponential growth from the factor~$\e^{a\omega n_\mathrm{min}}$.
While it is, in principle, possible to define smearing kernels $\mathcal{K}_\epsilon^\mathrm{I}(\omega,\bar\omega)$ that both reproduce the Dirac delta in the $\epsilon \to 0$ limit and decay exponentially at large $\omega$ (e.g. as a Gaussian centered at $\bar\omega$ with width $\epsilon$), this construction is not straightforward for the principal value kernel $\mathcal{K}_\epsilon^\mathrm{R}(\omega,\bar\omega)$. One possibility would be to define a modified kernel that behaves as $1/(\omega - \bar\omega)$ up to a given cut-off $\Lambda$, beyond which it decays exponentially. We note that no issue arises in the study of inclusive semileptonic decays~\cite{Bailas:2020qmv,Gambino:2022dvu,Barone:2023tbl,Kellermann:2025pzt}. In this case, the smeared kernel employed is a smeared Heaviside theta function $\theta_\epsilon(\bar{\omega}-\omega)$ implemented using a sigmoid function. Since this decays as $\e^{-(\omega-\bar\omega)/\epsilon}$ at large $\omega$, the convergence of the integral is guaranteed, provided that $a n_\mathrm{min} < 1/\epsilon$.

Finally, in order to keep statistical errors under control in the reconstruction, a trade-off parameter $\lambda$ can be introduced also in this case, modifying the functional to $W[\boldsymbol{g}]$ to incorporate, for instance, the covariance matrix of the data.

\subsubsection{Mellin method}

A recently proposed method~\cite{Bruno:2024fqc} offers an analytical construction of the coefficients $\boldsymbol{g}$, explicitly optimized to reduce discretization effects.
The starting point is the following Fredholm integral equation
\begin{equation}
    \int \dd t \, \e^{-\omega \tau} \, C(\tau) = \int \dd\omega' \, \mathcal{H}(\omega,\omega')\, \rho(\omega')\,,
\end{equation}
where $\mathcal{H}(\omega,\omega')=(\omega+\omega')^{-1}$ is the Carleman operator.
Inverting this operator constitutes an ill-posed problem and thus requires regularization.
Introducing the orthonormal Mellin basis
\begin{equation}
    u_s(\tau)=\frac{\e^{\ii s \log(\tau)}}{\sqrt{2\pi \tau}}\,, \qquad s\in\mathbb{R}\,,
\end{equation}
the Carleman operator can be expressed as
\begin{equation}
    \mathcal{H}(\omega,\omega') = \int \dd s \, u_s^*(\omega)\,  |\lambda_s|^2\,  u_s(\omega')\,, \qquad \lambda_s = \Gamma\big(1/2+\ii s\big)\,.
\end{equation}
We now define a regulated inverse operator,
\begin{equation}
    \mathcal{H}_\alpha^{-1}(\omega,\omega') = \int \dd s \, u_s^*(\omega)\,  |\lambda_s|^2\,G_s(\alpha)\,  u_s(\omega')\,,
\end{equation}
where the function $G_s(\alpha)$ is chosen such that the composition
\begin{equation}
    \delta_\alpha(\omega,\omega') = \int\dd\bar\omega \, \mathcal{H}(\omega,\bar\omega)\,\mathcal{H}_\alpha^{-1}(\bar\omega,\omega')  = \int \dd s \, u_s^*(\omega)\,  |\lambda_s|^4\,G_s(\alpha)\,  u_s(\omega')
\end{equation}
reproduces a Dirac delta function in the limit $\alpha \to 0$. This condition is satisfied if
$
    \lim_{\alpha\to 0} G_s(\alpha) = |\lambda_s|^{-4}\,.
$
In ref.~\cite{Bruno:2024fqc}, the Tikhonov regularization was adopted, corresponding to the choice
\begin{equation}
    G_s(\alpha) = \frac{1}{|\lambda_s|^2(|\lambda_s|^2+\alpha)}\,.
\end{equation}

A given smearing kernel can then be obtained from the convolution of a kernel function with the smeared delta function, namely
\begin{equation}
    \mathcal{K}_\epsilon(\omega,\bar\omega|\alpha) = \int \dd\omega' \, \delta_\alpha(\omega,\omega')\,\mathcal{K}_\epsilon(\omega',\bar\omega)  = \int \dd \tau \, g_\tau(\bar\omega,\epsilon|\alpha)_{\sf M} \, \e^{-\omega\tau}\,,
\end{equation}
with the coefficients $g_\tau$ being defined as
\begin{equation}
    g_\tau(\bar\omega,\epsilon|\alpha)_{\sf M} = \int \dd\omega' \, \mathcal{K}_\epsilon(\omega',\bar\omega) \, \int \dd s \, u_s^*(\omega')\,  \lambda_s|\lambda_s|^2\,G_s(\alpha)\,  u_s^*(t)\,.
\end{equation}
Such coefficients minimize the following functional
\begin{equation}
    W[\boldsymbol{g}] = \int \dd\omega \, \Big[\mathcal{K}_\epsilon(\omega,\bar\omega) - \int \dd \tau \, g_\tau(\bar\omega,\epsilon|\alpha)_{\sf M} \, \e^{-\omega\tau} \Big]^2 + \alpha \, \int \dd \tau \, g_\tau(\bar\omega,\epsilon|\alpha)_{\sf M}^2\,,
\end{equation}
with is similar to the Backus-Gilbert case discussed above, but with the covariance matrix substituted by the identity matrix.
The proposal of ref.~\cite{Bruno:2024fqc} is to consider a different version of the kernel, $\mathcal{K}_{\epsilon}(\omega,\bar\omega|\alpha,a)$, which is defined at fixed lattice spacing $a$, and to obtain a set of coefficients $\boldsymbol{g}$  constructed from the correlator $C(an)$ sampled at discrete Euclidean times ($n\in \mathbb{N}$). This amounts to solving the discretized Fredholm equation
\begin{equation}
    a\sum_{n=1}^\infty C(an) \,\e^{-a\omega n} = \int \dd\omega'\,  \overline{\mathcal{H}}_a(\omega,\omega') \, \bar{\rho}_a(\omega')\,.
\end{equation}
In this setup, the coefficients obtained in ref.~\cite{Bruno:2024fqc} using the Tikhonov regularization are given by
\begin{align}
g_{n}(\bar\omega,\epsilon \vert \alpha, a)_{\sf M} = \int_0^{\infty} \dd\omega\, {\mathcal{K}}_\epsilon(\omega,\bar\omega)\,
\int_{0}^\infty \dd s \, v_s(\omega,a)\, \frac{|\lambda_s|}{|\lambda_s|^2 + \alpha}\, \overline{v}_s(n-1,a)\,,
\end{align}
where $v_s(\omega,a)$  and $\overline{v}_s(n-1,a)$ are the eigenfunctions of the discretized Carleman operator at fixed $a$, given explicitly in ref.~\cite{Bruno:2024fqc}. These coefficients differ from the ones above by cut-off effects of~$\mathcal{O}(a^2)$. However, the authors of ref.~\cite{Bruno:2024fqc} argue that discretization effects introduced by the inverse Laplace transform on the spectral density itself are instead minimized when using the set of coefficients $g_{n}(\bar\omega,\epsilon \vert \alpha, a)_{\sf M}$ defined at finite lattice spacing, rather than the continuum counterpart $g_{n}(\bar\omega,\epsilon \vert \alpha)_{\sf M}$.

The regulator $\alpha$ plays a role analogous to the trade-off parameter $\lambda$ in Backus-Gilbert methods, helping to control statistical uncertainties in the reconstructed spectral density. As in that case, it can be tuned to achieve an optimal balance between statistical and systematic errors.

\subsection{Systematic effects in the spectral reconstruction}
When reconstructing spectral densities from numerically determined Euclidean lattice correlators with statistical uncertainties, several systematic effects must be carefully assessed and quantified. These include discretization effects, finite volume effects, uncertainties associated with the extrapolation required to remove the smearing and extract the true underlying spectral density, and biases due to discrepancies between the target and reconstructed smearing kernels.
While a thorough discussion of discretization and finite-volume effects goes beyond the scope of this work, it is worth commenting on the last two points.

When approximating the smearing kernel, it is crucial to verify that different kernel choices yield consistent results. This can be tested by exploring various parameterizations, such as those proposed in ref.~\cite{Bulava:2021fre}, and by varying the smearing width $\epsilon$. At least one choice of $\epsilon$ should be smaller than the width of the narrowest resonance in the spectrum, and smaller than the energy difference between the scale of interest -- here, the $D$-meson mass $m_D$ -- and the nearest resonance.

 No matter which method is chosen to approximate the smearing kernel, there will always be some degree of mismatch between the true kernel and its approximation. While this discrepancy may appear visually small in most regions, any rapid variations in the kernel will inevitably lead to larger mismatches, regardless of the chosen approximation method. Furthermore, since the underlying finite-volume spectral function consists of a sum of delta functions, any sharp changes in the smearing kernel can amplify errors, particularly if a finite-volume energy level lies near a region where the kernel approximation is imperfect.

To reduce the risk that such mismatches distort the value of the spectral density, one can compare multiple approximations $\widetilde {\mathcal{K}}_\epsilon(\boldsymbol{g} \vert \omega,\bar\omega)$ of a given target smearing function $\mathcal{K}_\epsilon( \omega,\bar\omega)$ and assess the stability. In the case of Backus-Gilbert and related modifications, this is achieved in part by varying the $\lambda$ parameter in the minimizing functional. Other approaches might also be worth exploring. For example, given one choice $\widetilde {\mathcal{K}}_\epsilon(\boldsymbol{g}_1 \vert \omega,\bar\omega)$ one could design a second chocie $\widetilde {\mathcal{K}}_\epsilon(\boldsymbol{g}_2 \vert \omega,\bar\omega)$ such that the differences $\widetilde {\mathcal{K}}_\epsilon(\boldsymbol{g}_1 \vert \omega,\bar\omega) -  {\mathcal{K}}_\epsilon( \omega,\bar\omega)$ and $\widetilde {\mathcal{K}}_\epsilon(\boldsymbol{g}_2 \vert \omega,\bar\omega) -  {\mathcal{K}}_\epsilon( \omega,\bar\omega)$ are orthogonal with respect to an inner product defined by an integral (potentially with some weight function). This could be achieved in practice by using a Lagrange multiplier to enforce the orthogonality condition. Since $\rho_L(\omega)$ itself can be decomposed as a sum of orthogonal functions, the two differences would overlap distinct components. If one finds that the two reconstructions $\widetilde \rho_L(\boldsymbol{g}_1 \vert \bar\omega,\epsilon)$ and $\widetilde \rho_L(\boldsymbol{g}_2 \vert \bar\omega,\epsilon)$ yield consistent results within statistical uncertainties, this would provide additional evidence that the smeared finite-volume spectral density is being accurately reconstructed.

All reconstruction methods considered in this work aim to estimate the smeared finite-volume spectral density, denoted by $\hat{\rho}_L(  \bar{\omega},\epsilon)$, and then to estimate the ordered double limit
\begin{equation}
\rho(\bar{\omega}) = \lim_{\epsilon \to 0} \lim_{L \to \infty} \hat{\rho}_L( \bar{\omega} , \epsilon) \,.
\end{equation}
In practice, this limit must be approximated from a finite set of data at fixed values of $L$ and $\epsilon$, requiring a careful assessment of the range that must be covered to ensure that the underlying spectral density is reliably recovered. The aim of this section is to better understand the systematic uncertainties associated with the second extrapolation
\begin{equation}
\rho(\bar{\omega}) = \lim_{\epsilon \to 0}  \, \hat{\rho}( \bar{\omega}, \epsilon) \,,
\end{equation}
where $\hat{\rho}(\bar{\omega}, \epsilon )$ is the smeared spectral density in infinite volume, defined in~\cref{eq:rho_smeared} above:
\begin{equation}
\hat{\rho}(\bar{\omega}, \epsilon) = \lim_{L \to \infty}  \hat{\rho}_L(\bar{\omega}, \epsilon) =
\int\frac{\dd\omega}{2\pi} \, \frac{\rho(\omega)}{\omega -  \bar \omega - \ii\epsilon} \,.
\end{equation}
Two key observations are useful here.

The first is that $\hat \rho(\bar{\omega}, \epsilon)$ admits a series expansion in $\epsilon$ at fixed $\bar{\omega}$. Dividing into the real and imaginary parts this can be written as
\begin{align}
\hat \rho^{\mathrm{I},{\sf x}}(\bar{\omega}, \epsilon) & = \frac{\rho(\bar{\omega})}{2} + \sum_{k=1}^\infty c_k^{\mathrm{I},{\sf x}}(\bar \omega) \epsilon^k,
\label{wq:sd-imag-exp}
\\
\hat \rho^{\mathrm{R},{\sf x}}(\bar{\omega}, \epsilon) & =\mathrm{P.V.} \,\int\frac{\dd\omega}{2\pi}\,\frac{\rho(\omega)}{\omega-\bar\omega} + \sum_{k=1}^\infty c_k^{\mathrm{R},{\sf x}}(\bar \omega) \epsilon^k,
\end{align}
where we used spectral functions smeared with the general kernels
\begin{align}
\hat \rho^{\mathrm{I},{\sf x}}(\bar{\omega}, \epsilon) = \int\frac{\dd\omega}{2\pi} \, \rho(\omega)\mathcal{K}_\epsilon^{\mathrm{I},{\sf x}}(\omega,\bar{\omega})\,,
\qquad
\hat \rho^{\mathrm{R},{\sf x}}(\bar{\omega}, \epsilon) = \int\frac{\dd\omega}{2\pi} \, \rho(\omega)\mathcal{K}_\epsilon^{\mathrm{R},{\sf x}}(\omega,\bar{\omega})\,,
\end{align}
recalling the observation of~\cref{eq:kernel_condition_I,eq:kernel_condition_R} that any choice is valid provided the difference to $\mathcal K_\epsilon^{\mathrm{I}}$ and $\mathcal K_\epsilon^{\mathrm{R}}$ vanishes as $\epsilon \to 0$. As is explained in refs.~\cite{Bulava:2021fre,Frezzotti:2023nun} the coefficients $c_k^{\mathrm{I},{\sf x}}(\bar{\omega})$ and $c_k^{\mathrm{R},{\sf x}}(\bar{\omega})$ can be expressed as a product of a spectral-function-dependent part and a part dictated only by the smearing kernel.

For the imaginary part, we can define the smearing kernel as any function that reduces to $\pi\delta(\omega-\bar\omega)$ in the limit $\epsilon \to 0$. Therefore, by writing
\begin{equation}
    \mathcal{K}_\epsilon^{\mathrm{I},{\sf x}}(\omega,\bar{\omega}) = \pi \, \delta_\epsilon^{\sf x}(\omega-\bar\omega)\,,
\end{equation}
we can define the kernel via the following smeared delta functions, which generalize those in ref.~\cite{Bulava:2021fre},
\begin{equation}
    \delta_\epsilon^{\sf g}(x) = \frac{1}{\sqrt{2\pi}\epsilon} \, \exp\bigg[{-}\frac{x^2}{2\epsilon^2}\bigg]\,, \qquad
    \delta_\epsilon^{{\sf c}s}(x) = \frac{\Gamma[s+1]}{\Gamma[s+1/2]\sqrt{\pi}} \, \frac{\epsilon^{2s+1}}{(\epsilon^2+x^2)^{s+1}}\,.
    \label{eq:other_kernels_I}
\end{equation}
As is stressed in ref.~\cite{Bulava:2021fre}, using a Gaussian kernel sets $c_k^{\mathrm{I},{\sf g}}(\bar{\omega}) = 0$ for all odd $k$ and thus improves the extrapolation. Similarly, one can show that for $s\geq1$, the Cauchy kernels satisfies $c_1^{\mathrm{I},{\sf c}s}(\bar{\omega}) = 0$, leading to corrections that start at $\mathcal{O}(\epsilon^2)$.

Having defined the imaginary part of the smearing kernel, one can construct the associated real part using the Kramers-Kronig relation, as given in~\cref{eq:KK-relation}, leading to the expressions%
\footnote{Note that the Gaussian kernel does not go to zero as $|x|\to \infty$ along the imaginary axis and therefore does not satisfy all conditions of the Kramers-Kronig relation. Nonetheless, the relation still gives a kernel that behaves as a P.V. functional as $\epsilon\to 0$.}
\begin{align}
    \mathcal{K}_\epsilon^{\mathrm{R},{\sf g}}(\omega,\bar{\omega}) = \frac{\sqrt{2}}{\epsilon}\, D\bigg(\frac{\omega-\bar{\omega}}{\sqrt{2}\, \epsilon}\bigg)
\,, \quad \mathcal{K}_\epsilon^{\mathrm{R},{\sf c}s}(\omega,\bar{\omega}) = \frac{1}{(\omega-\bar\omega)}\,{}_2F_1\Big(\tfrac{1}{2},\, 1;\, \tfrac{1}{2} - s; -\frac{\epsilon^2}{(\omega-\bar\omega)^{2}}\Big)
 \,,
 \label{eq:other_kernels_R}
\end{align}
where $D(x)$ denotes the Dawson function and $_2F_1(a,b;c;z)$ is the ordinary hypergeometric function.
Through this relation, the coefficients $c_k^\mathrm{R,{\sf x}}(\bar\omega)$ in the expansion of the smeared spectral density can be expressed as Hilbert transforms of the corresponding coefficients $c_k^\mathrm{I,{\sf x}}(\bar\omega)$.
Notably, if the kernel $\mathcal{K}_\epsilon^{\mathrm{I},{\sf x}}(\omega,\bar{\omega}) $ is chosen such that the expansion of the smeared spectral density $\hat{\rho}^{\mathrm{I},{\sf x}}(\bar\omega,\epsilon)$ begins at $\mathcal{O}(\epsilon^2)$, with no linear term, then the same behavior will be inherited by the real part.

The second important point is that one requires sufficiently small $\epsilon$ to enter the scaling regime where this expansion is useful. This follows from the fact that the underlying spectral density can contain variations, such as narrow peaks, which arise from the resonance content of the channel, as well from branch points associated with multi-particle thresholds and other effects known to influence the line shape, described below. Loosely speaking, if complicated features exist with characteristic width $\delta$ at a distance $\Delta$ from a target $\bar{\omega}$, then the scaling regime only occurs when $\epsilon \ll \delta$ or $\epsilon \ll \Delta$.

\subsection{Infinite-volume spectrum}

To assess the systematic uncertainties of spectral reconstruction, it is useful to understand as much as possible about the spectrum, both in a finite and an infinite volume. In short we are interested in states which overlap
$\bra{\Dbar^0} \, \Hcal_\w(0)$ and $\Hcal_\w(0) \, \ket{\D^0}$.
Each of these states is given by the action of a $J=0$ operator ($\Hcal_\w$) acting on a $J=0$ pseudoscalar state (either $\D^0$ or $\Dbar^0$). As a result, each carries total angular momentum $J=0$, which constrains the spectrum contributing to the reconstruction.

The weak Hamiltonian can be separated into a parity-positive and a parity-negative component:
\begin{align}
\Hcal_\w = \Hcal_\w^+ + \Hcal_\w^- \, .
\end{align}
Since the $|\D^0\rangle$ has negative parity, the resulting states have quantum numbers
\begin{align}
\Hcal_\w^+ \,|\D^0\rangle \, : \, J^P = 0^- \, ,\\
\Hcal_\w^- \,|\D^0\rangle \, : \, J^P = 0^+ \, .
\end{align}
The finite-volume correlator in~\cref{eq:fv-4pt-estimator} has a nonzero expectation value only if both weak Hamiltonians share the same parity quantum number.
This results in two distinct sectors:
\begin{align}
C_L^-(\tau) = 2m_DL^3 \,\int_L \dd^3 \xvec \, \e^{-m_\D \tau} \langle\bar{\D}^0|\,\Hcal_\w^+(\tau,\xvec) \Hcal_\w^+(0)\,|\D^0\rangle_L\, , \\
C_L^+(\tau) = 2m_DL^3 \, \int_L \dd^3 \xvec \, \e^{-m_\D \tau} \langle\bar{\D}^0|\,\Hcal_\w^-(\tau,\xvec) \Hcal_\w^-(0)\,|\D^0\rangle_L\, ,
\end{align}
where we denote the two correlators by $C_L^-$ and $C_L^+$, respectively. In each case this refers to the parity of the states that arise when we insert a complete set of states between the two currents:
\begin{align}
C^P_L(\tau) = 2m_DL^6 \, \sum_n \e^{- E_n^P(L) \tau} \langle\bar{\D}^0|\,\Hcal_\w(0) |n^P \rangle_L \langle n^P | \Hcal_\w(0)\,|\D^0\rangle_L \,, \qquad (P=\pm) \,.
\end{align}
The definite-parity spectral function can then similarly be written as
\begin{align}
{\rho}^P_L(\omega) = 2m_P L^6 \sum_n \langle\bar{\D}^0|\,\Hcal_\w(0) |n^P \rangle_L \langle n^P | \Hcal_\w(0)\,|\D^0\rangle_L \delta(\omega - E_n^P(L)) \,, \qquad (P=\pm) \,,
\end{align}
where $|n^P\rangle_L$ denotes the $n$th finite-volume state with definite parity $P$ and zero total spatial momentum.

All operators contributing to the weak Hamiltonian $\Hcal_\w$ include a pair of $c, \bar{u}$ fields, that are contracted with the corresponding quark fields of the $D^0$ mesons.
In the operators $Q_1$ and $Q_2$, which are the most relevant for our study, the remaining two fields consist of $s$ and $d$ quarks and antiquarks. This implies that the intermediate states will have zero charm quantum number, $C=0$.

In summary, the states appearing between the two weak Hamiltonians belong to the $J=0$ sector and are either unflavoured or else carry downness and strangeness quantum numbers.
The infinite-volume spectral function $\rho(\omega)$ is characterized by resonances with these quantum numbers, which are listed in Table~\ref{table:pdg-resonances}.

\begin{table}[b]
\centering
\begin{minipage}[t]{0.48\textwidth}
\centering
\small
\begin{tabular}{|c|c|c|c|}
\hline
\multicolumn{4}{|c|}{$J^P = 0^+$} \\
\hline
\textbf{Name} & $\Delta$ [MeV] & $M$ [MeV] & $\Gamma$ [MeV] \\
\hline
$f_0(500)$ & -1390 & 400--550 & 400--700 \\
$f_0(980)$ & -870 & 980--1010 & 40--70 \\
$a_0(980)$ & -870 & 970--1020 & 60--140 \\
$f_0(1370)$ & -520 & 1250--1440 & 120--600 \\
$a_0(1450)$ & -470 & 1290--1500 & 60--280 \\
$f_0(1500)$ & -385 & 1430--1530 & 80--180 \\
$a_0(1710)^\star$ & \textbf{-152} & 1713(19) & \textbf{107(15)} \\
$f_0(1710)$ & \textbf{-115} & 1680--1820 & \textbf{100--360} \\
$\pi(1800)^\star$ & \textbf{-55} & 1810(11) & \textbf{215(8)} \\
$a_0(1950)^\star$ & ${\bf  66}$ & 1931(26) & \textbf{270(40)} \\
$f_0(2020)$ & $110$ & 1870--2080 & 240--480 \\
$f_0(2100)^\star$ & $230$ & 2095(19) & 287(32) \\
$f_0(2200)^\star$ & $322$ & 2187(14) & 210(40) \\
$f_0(2330)^\ddagger$ & $500$ & 2312--2419 & 65--274 \\
$f_0(2470)^\star$ & $605$ & 2470(7) & 75(14) \\
\hline
$K_0^*(700)$ & -1185 & 630--730 & 520--680 \\
$K_0^*(1430)$ & -434 & 1431(6) & 220(38) \\
$K_0^*(1950)^\star$ &  ${\bf 92}$ & 1957(14) & {\bf 170(50)} \\
\hline
$\chi_{c0}(1P)^\star$ & $1550$ & 3414.71(30) & 10.7(6) \\
\hline
\end{tabular}
\end{minipage}
\hfill
\begin{minipage}[t]{0.48\textwidth}
\centering
\small
\begin{tabular}{|c|c|c|c|}
\hline
\multicolumn{4}{|c|}{$J^P = 0^-$} \\
\hline
\textbf{Name} & $\Delta$ [MeV] & $M$ [MeV] & $\Gamma$ [MeV] \\
\hline
$\eta'(958)^\star$ & -907 & 957.78(6) & 0.188(6) \\
$\eta(1295)^\star$ & -571 & 1294(4) & 55(5) \\
$\pi(1300)^\star$ & -565 & 1300(100) & 200--600 \\
$\eta(1405)^\star$ & -456 & 1408.7(2.0) & 50.3(2.5) \\
$\eta(1475)^\star$ & -389 & 1476(4) & 96(9) \\
$\eta(1760)^\star$ & \textbf{-114} & 1751(15) & \textbf{240(30)} \\
$X(1835)^\star$ & \textbf{-38} & 1827(13) & \textbf{242(15)} \\
$\eta(2225)^\star$ & $356$ & 2221(13) & 185(40) \\
\hline
$K(1460)^\ddagger$ & -383 & $1482(16)$ & $336(11)$ \\
$K(1830)^\ddagger$ & ${\bf 9}$ & $1874(123)$ & ${\bf 168(294)}$ \\
\hline
$\eta_c(1S)^\star$ & $1119$ & 2984.1(4) & 30.5(5) \\
$\eta_c(2S)^\star$ & $1773$ & 3637.7(9) & 11.8(1.6) \\
\hline
\multicolumn{4}{c}{\vspace{72.5pt}}
\end{tabular}
\end{minipage}
\caption{Summary of scalar (left) and pseudoscalar (right) resonances based on PDG data~\cite{ParticleDataGroup:2024cfk}. Asymmetric errors are symmetrized by taking the larger value. While there is no direct mapping between the spectrum and $\rho(\omega)$, resonances shape the line profile and influence smearing. We highlight those with $\Delta = M - m_D$ and $\Gamma$ both below $2M_\pi \approx 280\,\mathrm{MeV}$, where $M$ is the central mass and $m_D = 1865\,\mathrm{MeV}$ (with $\Delta$ rounded to the nearest MeV).
Values are taken from T-matrix pole positions, $M - \ii\Gamma/2$, when available. If not, entries are marked $^\star$ and use the PDG’s ``(Breit-Wigner) mass / width'' averages. Special cases without PDG averages are marked~$^\ddagger$, with details provided in the main text.
\label{table:pdg-resonances}}
\end{table}

To ensure a reliable $\epsilon \to 0$ limit, the smearing parameter $\epsilon$ should be smaller than the resonance's width $\Gamma$ or the gap of the resonance's mass from the $D^0$, $\Delta = M - m_D$. Since the finite-volume effects on $\hat{\rho}^\alpha_L(\bar\omega,\epsilon)$ are governed by the parameter $\epsilon L$, one is particularly interested in the case where both $M$ and $\Gamma$ are smaller than the pion mass, $m_\pi$. This may require $\epsilon < m_\pi$, in order to be in the regime where a small $\epsilon$ expansion is useful. But this, in turn, could lead to enhanced finite-volume effects.

The following resonances merit dedicated discussion:
\begin{itemize}
\item The $\eta'(958)$ has the smallest width of all resonances listed, $\Gamma = 0.188(6)$ MeV. This narrowness arises from the fact that the $\eta'$ is stable in isosymmetric QCD. Since foreseeable lattice QCD calculations will be performed in the isosymmetric theory, the $\eta'$ will manifest as a zero-width particle corresponding to a single finite-volume state. This allows it to be directly subtracted from the $C^-_L(\tau)$ correlator before spectral reconstruction.
\item The $f_0(980)$ has a highly uncertain width, with PDG estimates ranging from $10$~MeV to $100$~MeV. The most recent determination by the BESIII collaboration~\cite{BESIII:2015you} reports a width of~$15.3(4.7)$ MeV.
\item The charmonium resonances $\chi_{c0}(1P)$,  $\eta_c(1S)$  and $\eta_c(2S)$ have relatively narrow widths of $10$~MeV, $30$~MeV and $12$~MeV, respectively. However, with masses at $3.41$~GeV, $2.98$~GeV, and $3.64$~GeV, these states lie well above $m_D$, and are therefore expected to have minimal impact on the spectral reconstruction.
\item The $K(1830)$ lacks a PDG average due to limited historical data, but a recent measurement from LHCb~\cite{LHCb:2016axx} reports a mass of $M = 1874 \pm 43^{+59}_{-115}$ MeV, just 9 MeV above $m_D$, with large uncertainties. Its width, $\Gamma = 168 \pm 90^{+280}_{-104}$ MeV, is broad and poorly constrained. The PDG notes it as seen in partial-wave analysis but not yet confirmed. Similarly, the $K(1460)$ has no PDG average, though LHCb~\cite{LHCb:2017swu} provides a value far from $m_D$ and with a large width. In the table, statistical and systematic uncertainties for both $K(1460)$ and $K(1830)$ are combined in quadrature, using the larger value for asymmetric systematic errors.
\item The $X(1835)$ has a well-determined mass  $M  = 1827(13)$~MeV, with a central value about $38$~MeV below $m_{D}$ and a broad width of $\Gamma  = 242(15)$~MeV.
\item For the $f_0(2330)$, the PDG does not quote an average value. This is due to inconsistencies among available measurements; the values listed in the table encompass the full range of central values from the individual experiments cited by the PDG.
\end{itemize}
All other allowed resonances either have a width of at least $\Gamma \gtrsim 50$ MeV or are at least $50$~MeV away from $m_D$. This scale aligns with the smallest practical smearing parameter $\epsilon$, allowing to reconstruct the spectral function meaningfully around $m_D$.

 We note that while the parity sectors of the weak Hamiltonians can be treated independently\footnote{In a finite-volume this statement holds only in the case $\boldsymbol{p}_D = \boldsymbol{0}$, which forms the basis of all discussion in this section. Extending our spectral reconstruction formalism to evaluate \cref{eq:fv-4pt-estimator} at non-zero momentum would result in mixing between the parity sectors of the weak Hamiltonians. Consequently, one would need to reconstruct a combined spectral density encompassing resonances from both parity sectors.}, it is not possible to isolate specific sub-diagrams of the finite-volume correlator to restrict the intermediate states to a particular isospin content. As a result, the reconstruction of the spectral density must account for the combined spectra of QCD states composed solely of light quarks as well as those that include strange-quark contributions.

\subsection{Model for the spectral density}
\label{sec:model}

The aim of this section is to give a reasonable model for $\rho(\bar{\omega})$, to understand how the scaling regime is approached and how one might apply systematic uncertainties to the data.
Our model is based on an all-orders diagrammatic representation of the $D^0 \to \widebar D^0$ transition amplitude, similar to those used to derive $S$-matrix unitarity relations as well as relations between amplitudes and finite-volume data~\cite{Luscher:1986pf,Luscher:1990ux}.

Still the framework is far from a rigorous description for two important reasons. First, we do not include the effect of multi-hadron channels with more than two particles in the final state. Second, we have to choose the form of various meromorphic functions (the two-to-two K-matrix and its counterpart for one-to-two transitions) that are not uniquely determined by the singularities. We take values motivated by a numerical lattice QCD calculation~\cite{Briceno:2017qmb} with others arbitrarily chosen.

In this framework the spectral function has the following form:
\begin{equation}
\rho(\omega) = \text{Im} \bigg [ \boldsymbol {\mathcal{T}}^T(\omega^2) \cdot \boldsymbol{\rho}(\omega^2) \cdot \left [  1 - \boldsymbol{\mathcal{K}}(\omega^2) \cdot \boldsymbol{\rho}(\omega^2) \right ]^{-1} \cdot \boldsymbol{\mathcal{T}}(\omega^2) \bigg ] \,.
\end{equation}
Here $\boldsymbol{\mathcal{K}}(\omega^2)$ is the K-matrix for a set of coupled two-to-two channels, which is a real meromorphic function of $\omega^2$ and is defined through its relation to the amplitudes:
\begin{equation}
\boldsymbol{\mathcal{M}}(\omega^2) =  \left [  1 - \boldsymbol{\mathcal{K}}(\omega^2) \cdot \boldsymbol{\rho}(\omega^2) \right ]^{-1} \cdot \boldsymbol{\mathcal{K}}(\omega^2) \,.
\end{equation}
The matrix $\boldsymbol{\mathcal{M}}(\omega^2)$, in turn, has entries given by the usual definition of the two-to-two scattering amplitude:
\begin{equation}
(2 \pi)^4 \delta^4(p_i - p_j) \times \mathcal{M}_{ij}(\omega^2) = \langle i, \text{out} \vert j, \text{in} \rangle_{\sf conn} \,,
\end{equation}
where $\ket{j, \text{in}}$ and $\ket{i, \text{out}}$ are the incoming and outgoing states, respectively, for the $j$th and $i$th two-particle channels. These are each projected to a definite angular momentum which leads to another index, left implicit in the notation. The subscript ${\sf conn}$ indicates that only connected diagrams are included in the amplitude.

The matrix $\boldsymbol{\rho}(\omega^2)$ is diagonal and contains the phase space factors for the two-particle channels. Its entries are given by
\begin{equation}
\boldsymbol{\rho}_{ij}(\omega^2) = \ii \delta_{ij} \xi_i \frac{\sqrt{\lambda (\omega^2, m_{i1}^2, m_{i2}^2)}}{8 \pi \omega} \,,
\end{equation}
where $\lambda(x,y,z) = x^2 + y^2 + z^2 - 2xy - 2xz - 2yz$ is the K\"all\'en function and $m_{ij}$ are the masses of the two particles in the $i$th channel. The $\xi_i$ are symmetry factors: $\xi_i = 1$ for distinguishable particles, $\xi_i = 1/2$ for identical particles.

Finally $\boldsymbol{\mathcal{T}}(\omega^2)$ is a vector whose entries are the K-matrix analogues for the $D^0 \to j$ decay amplitudes.
It is defined through its relation to $\boldsymbol{\mathcal{A}}(\omega^2)$, which is the vector of $D^0 \to j$ amplitudes:
\begin{equation}
\boldsymbol{\mathcal{A}}(\omega^2) =  \left [  1 - \boldsymbol{\mathcal{K}}(\omega^2) \cdot \boldsymbol{\rho}(\omega^2) \right ]^{-1} \cdot \boldsymbol{\mathcal{T}}(\omega^2) \,.
\end{equation}
The entries of $\boldsymbol{\mathcal{A}}(\omega^2)$ are given, in analogy to $\boldsymbol{\mathcal M}(\omega^2)$, by
\begin{equation}
(2 \pi)^4 \delta^4(p_D - p_i) \times \mathcal{A}_{i}(\omega^2) = \langle i, \text{out} \vert D^0 \rangle_{\sf conn} \,.
\end{equation}

To implement the framework above in practice, we first need to choose which two-particle channels to consider.
Here we simplify the problem by restricting ourselves to the case of two-particle channels with vacuum quantum numbers. We note that the part of the neutral $D$-meson mixing amplitude built form two singly-Cabibbo-suppressed weak Hamiltonians does, in fact, admit these quantum numbers, together with isospin-one states. However, our perspective is that this matching is not particularly important as this model only intends to capture possible realistic features of the spectral function in order to visualize different scenarios of how the $\epsilon \to 0$ limit may be approached.

In the vacuum-sector we choose to focus on the lightest three two-particle channels: $\pi \pi$, $K \bar K$, $\eta \eta$.
Generally speaking it is difficult to know which functional form and which parameter values to use for $\boldsymbol{\mathcal{K}}(\omega^2)$. We make progress by taking inspiration from existing lattice QCD computations, in particular ref.~\cite{Briceno:2017qmb}, in which the Hadron Spectrum Collaboration computed the two-particle scattering amplitudes in the $I=0$ channel for $\pi \pi$, $K \bar K$, and $\eta \eta$.

\begin{figure}
\centering
\includegraphics[width=\textwidth]{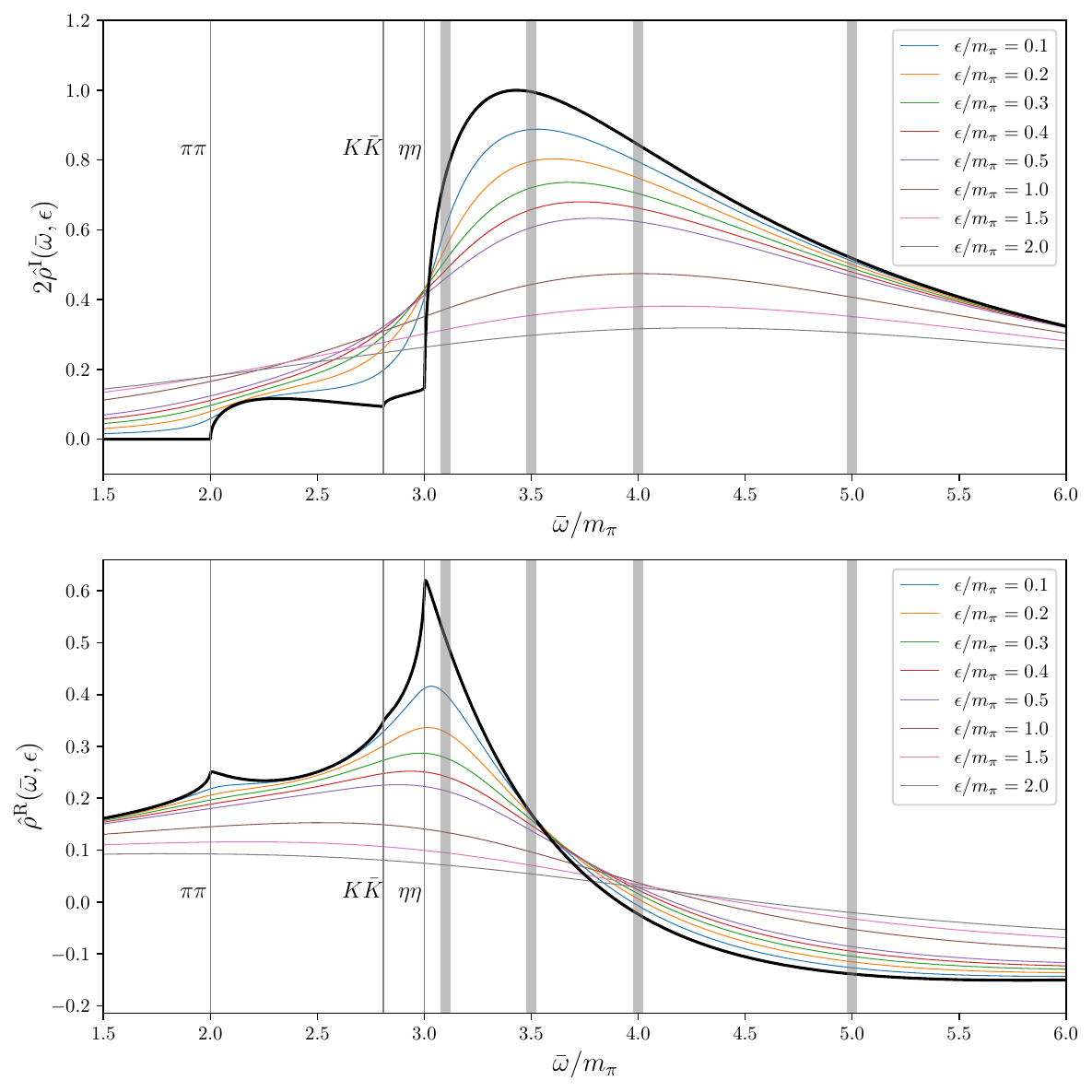}
\caption{Illustration of the model smeared spectral functions $\hat \rho^{\mathrm{I}}(\bar \omega, \epsilon)$ (top panel) and $\hat \rho^{\mathrm{R}}(\bar \omega, \epsilon)$ (bottom panel) vs.~$\bar \omega$, for different values of the smearing parameter $\epsilon$.
As explained in the main text, the form is constructed using a unitarity motivated decomposition including three two-particle channels ($\pi \pi$, $K \bar K$, $\eta \eta$), with inputs partly arbitrary and partly motivated by the known large $\omega$ scaling and by the results of a lattice QCD calculation
from the Hadron Spectrum Collaboration~\cite{Briceno:2017qmb}.
The vertical bars represent four different values of $\bar \omega$ where we envision extracting the unsmeared spectral function.
In practice $\bar \omega = m_D$ is the only reconstruction point of interest, but we include multiple values to reflect our imperfect knowledge of the spectral function. That is, each bar represents a scenario for the form that $\rho(\omega)$ might have in the vicinity of $m_D$, also potentially for calculations with unphysical pion masses.
}
\label{fig:smeared-spectral-function}
\end{figure}

\begin{figure}
\centering
\includegraphics[width=\textwidth]{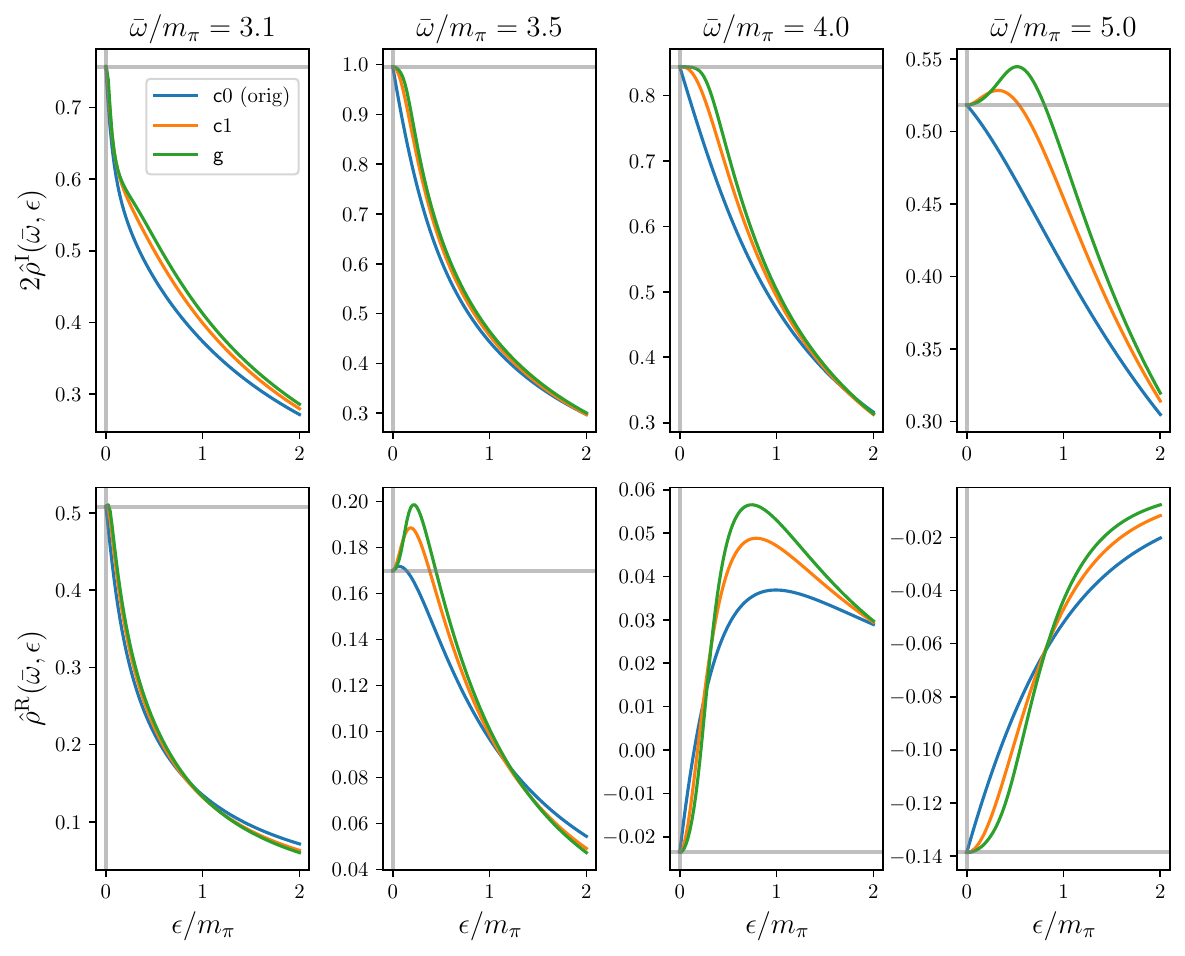}
\caption{Illustration of the $\epsilon$ depedence of $\hat \rho^{\mathrm{I},{\sf x}}(\bar \omega, \epsilon)$ (top panel) and $\hat \rho^{\mathrm{R},{\sf x}}(\bar \omega, \epsilon)$ (bottom panel) for three choices of smearing kernel (as indicated in the legend) and for four different values of $\bar \omega$: $3.1 m_\pi$, $3.5 m_\pi$, $4.0 m_\pi$, and $5.0 m_\pi$, from left to right. In all cases, in order to match the meaning of $\epsilon$, the smearing width is rescaled as $\epsilon_{\sf x} = \epsilon/\alpha_{\sf x}$,  so that $\int_{-\alpha_{\sf x}}^{\alpha_{\sf x}} \dd \omega \, \delta_{\epsilon=1}^{\sf x}(\omega) = \frac{1}{2}$.\label{fig:epsilon-dependence}}
\end{figure}

The work is based on an extraction of the finite-volume spectrum as well as an application of the L\"uscher formula (and extensions) to relate this to the coupled channel amplitudes. The authors find that a good description of the data is achieved with the simple ansatz
\begin{equation}
\boldsymbol{\mathcal K}^{-1}(s) =
\begin{pmatrix}
a + bs & c + ds & e \\
c + ds & f & g \\
e & g & h
\end{pmatrix}
\end{equation}
where $s = \omega^2$ is the Mandelstam variable. The parameters $a, b, c, d, e, f, g, h$ are determined by a fit to the finite-volume spectrum and are given in the supplemental material of ref.~\cite{Briceno:2017qmb}.
Rounding to two significant figures, we take the following values:
\begin{align}
a &= 1.4, & b &= -1.3, & c &= 1.1, & d &= -0.2, \\
e &= 1.0, & f &= 3.1,  & g &= 8.5, & h &= 22 \,.
\end{align}
We also take the masses from this calculation which are heavier than physical ($m_\pi \approx 400~\mathrm{MeV}$), corresponding to $m_K/m_\pi = 1.4$, $m_\eta/m_\pi = 1.5$.
Our remaining input is
\begin{align}
    \boldsymbol{\mathcal T}(s) &=
    C \frac{1}{s} \begin{pmatrix}
    \alpha \\
    \beta \\
    \gamma
    \end{pmatrix} \,,
\end{align}
with $\alpha = \beta = 1$ and $\gamma = 5$ (and $C$ chosen to set the height of the unsmeared spectral function to $1$ at its maximum).
This is essentially arbitrary but the $1/s$ dependence is chosen so that $\rho(\omega) = \mathcal{O}(1/\omega^4)$ as $\omega \to \infty$~\cite{Jiang:2017zwr,Li:2020xrz}.

The results of this model are shown in figures~\ref{fig:smeared-spectral-function} and \ref{fig:epsilon-dependence}.

In figure \ref{fig:smeared-spectral-function} we show the real and imaginary parts of the smeared spectral function $\hat \rho(\bar \omega, \epsilon)$ for different values of the smearing parameter $\epsilon$ using the kernels $\mathcal{K}_\epsilon^{\mathrm{I}}$ and $\mathcal{K}_\epsilon^{\mathrm{R}}$ defined in \cref{eq:kernel_condition_I,eq:kernel_condition_R}.
The vertical bars in the figure represent four different values of $\bar \omega$ where we consider extracting the unsmeared spectral function. In practice only $\bar \omega = m_D$ is relevant, but we include multiple values to reflect different scenarios for the form $\rho(\omega)$ might actually take near $m_D$.

In figure \ref{fig:epsilon-dependence} we show the dependence of the spectral function on the smearing parameter $\epsilon$ for these four values of $\bar \omega$, each for three different smearing kernels. We include the original kernels of~\cref{eq:kernels}, labeled ${\sf c}0$, and two additional kernels, labeled ${\sf c}1$ and ${\sf g}$, given in~\cref{eq:other_kernels_I,eq:other_kernels_R}. Three facets in particular deserve comment. First, a number of the cases show strong dependence on $\epsilon$, as expected due to the close proximity to peaks and thresholds. Second, some cases also show non-monotonic behaviour with $\epsilon$. This is presumably due to the presence of competing peaks and troughs that the smearing kernels overlap differently as $\epsilon$ is varied. Finally, the curve obtained using the ${\sf c}1$ kernel generally lies between those from the ${\sf c}0$ and ${\sf g}$ kernels. This behaviour is not coincidental but reflects the fact that, if $\epsilon$ is defined such that the integral of the $\delta_{\epsilon}^{\sf x}(x)$ between $-\epsilon$ and $\epsilon$ is fixed, then the ${\sf c}s$ kernels asymptote to the ${\sf g}$ kernel as $s \to \infty$. To ensure a consistent interpretation of the smearing width across different kernels, in~\cref{fig:epsilon-dependence} we have rescaled the width as $\epsilon_{\sf x} = \epsilon/\alpha_{\sf x}$, where $\alpha_{\sf x}$ is defined by the condition $\int_{-\alpha_{\sf x}}^{\alpha_{\sf x}} \dd \omega \, \delta_{\epsilon=1}^{\sf x}(\omega) = \frac{1}{2}$\,.

The upshot of these results is that, as already anticipated in the previous section, the scaling regime is only reached when $\epsilon$ is small enough that the overlap to adjacent features is no longer qualitatively changing when decreasing the parameter. Ultimately a more realistic modelling of the spectral function may be required to understand this regime more completely but we speculate that values equal to the pion mass, or else down by a factor of $2$ or $3$, may be sufficient. At the physical point this would correspond to $\epsilon \approx 100$ MeV or less. Finally we note that this also dictates the range of volumes required since $\epsilon L$ must be sufficiently large to ensure that the finite-volume effects are suppressed.
A target of $\epsilon L$ between 3 and 4, corresponding to $L \approx 6$-$8$ fm, seems reasonable. Again, future work is required to understand this more reliably.

\subsection{Considerations for a first lattice QCD calculation}

We conclude this discussion with a summary of how a lattice QCD calculation of the neutral $D$-meson mixing amplitudes might be realistically and effectively performed.

First, as discussed in \cref{sec:renormalization}, we strongly advocate the use a lattice fermion discretization with improved chiral symmetry, such as Domain-Wall Fermions.
This choice is motivated by two key considerations:
\begin{itemize}
    \item[(i)] The relevant operators contributing to the weak Hamiltonian -- namely $Q_{\pm}^{\bar{d}s}$, $Q_{\pm}^{\bar{s}d}$, and the difference $(Q_{\pm}^{\bar{s}s} - Q_{\pm}^{\bar{d}d})$ -- transform as irreducible representations of $SU(4)_L \times SU(4)_R$ with well-defined flavour symmetry properties. Consequently, they renormalize multiplicatively and do not mix with penguin operators.
    \item[(ii)] Preserving chiral symmetry prevents mixing with lower-dimensional operators, which can occur in non-perturbative renormalization schemes, thereby greatly simplifying the renormalization procedure.
\end{itemize}

\begin{figure}[t]
    \centering
    \includegraphics[width=0.7\linewidth]{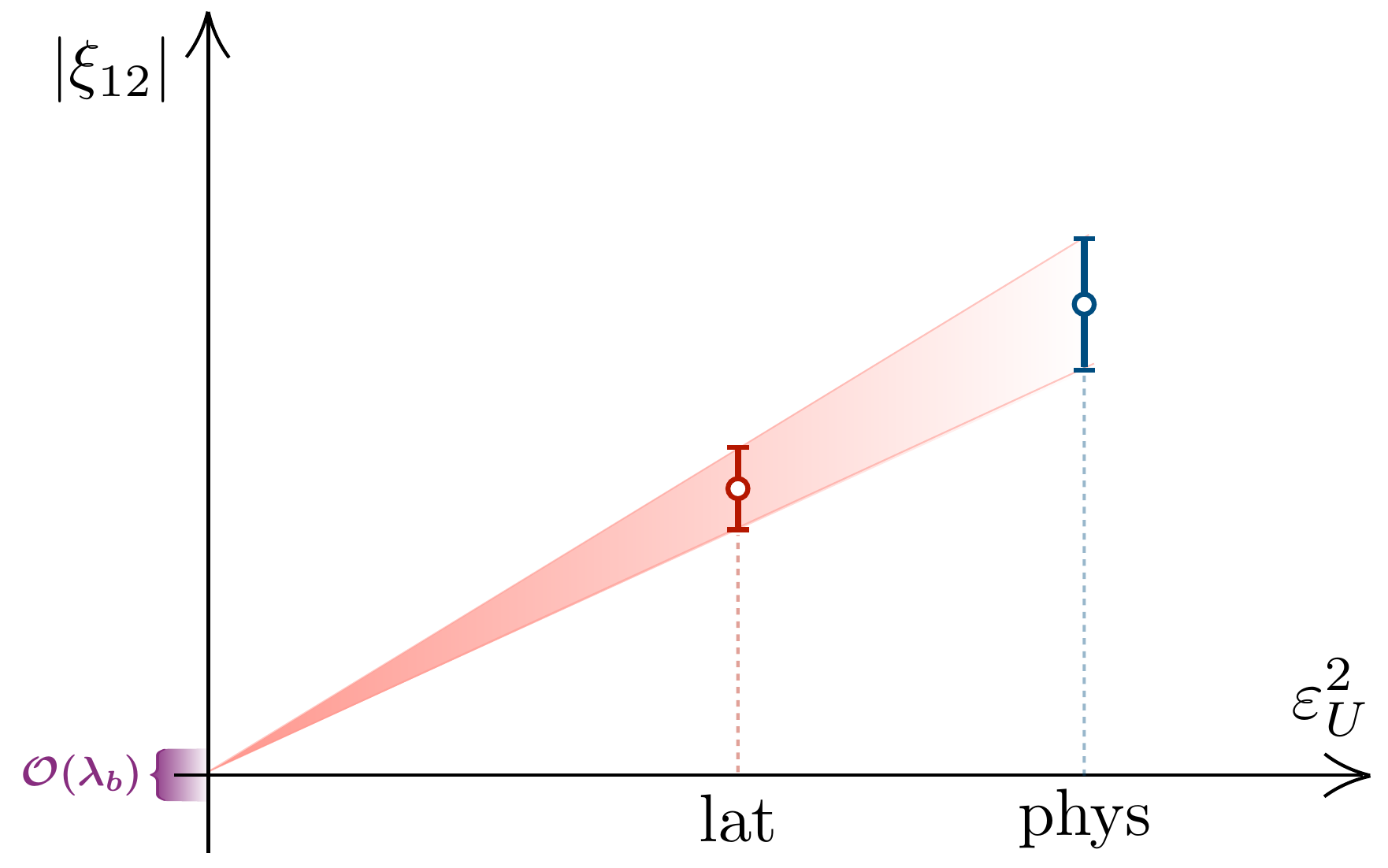}
    \caption{Sketch illustrating a strategy to extract the amplitude $|\xi_{12}|$ (with $\xi \in \{\M,\Gamma\}$) at physical quark masses (blue data point) from lattice QCD calculations performed at heavier-than-physical quark masses (red data point). The simulations are carried out along a trajectory where the combination $(2m_K^2 + m_\pi^2)$ is fixed to its physical value and isospin-breaking corrections are neglected. At the $U$-spin symmetric point, the neutral $D$-meson mixing amplitudes $|\xi_{12}|$ vanish up to $\mathcal{O}(\lambda_b)$ corrections. Deviations from this point are governed by $U$-spin breaking parameter $\varepsilon_U$, which can be estimated using the proxy $\varepsilon_U = m_K^2 - m_\pi^2$.}
    \label{fig:fan-plot}
\end{figure}

Second, we emphasize that the dominant contributions to the neutral $D$-meson mixing amplitudes $\xi_{12}$ (with $\xi \in \{\M, \Gamma\}$) arise at second order in the $U$-spin breaking parameter $\varepsilon_U$, which can be conveniently proxied by the squared meson mass difference $m_K^2 - m_\pi^2 \sim m_s - m_d$. In the $U$-spin symmetric limit, the amplitudes $\xi_{12}$ vanish up to corrections of $\mathcal{O}(\lambda_b)$, which can be safely neglected.
While our formalism is directly applicable at the physical point, a more cost-effective strategy would be to compute the amplitudes $\xi_{12}$ on ensembles with unphysical light quark masses. Neglecting isospin-breaking effects and tuning the ensembles such that the sum of bare quark masses $m_u + m_d + m_s$ remains constant -- for instance, by fixing the combination $2m_K^2 + m_\pi^2$ to its physical value -- the dependence of $\xi_{12}$ on the light-quark masses can be expressed in terms of the $U$-spin breaking parameter $\varepsilon_U$. This allows one to perform a controlled extrapolation to the physical point, which is further constrained by the condition $\xi_{12}(\varepsilon_U = 0) = 0$, as illustrated in~\cref{fig:fan-plot}.

\section{Conclusions}
\label{sec:conclusions}
We have explored the feasibility of computing the long-distance contributions to neutral $D$-meson mixing using lattice QCD and spectral reconstruction techniques. We have provided a comprehensive lattice QCD formalism showing how the mixing amplitudes $\Gamma_{12}$ and $\M_{12}$ can be extracted from first principles using Euclidean lattice correlation functions.

We discussed the renormalization of relevant four-quark operators for both non-chiral and chiral discretizations of lattice fermion actions. For formulations that preserve chiral symmetry, we confirmed that a notably simple operator mixing pattern emerges, analogous to previous findings in lattice studies of kaon mixing. 
All considerations on renormalization done in this work are also directly relevant for the lattice computation of the hadronic decay $D \to K\pi$, as it involves the same set of four-quark operators. 

Given the specific $U$-spin structure of internal $d$ and $s$ quark propagators connecting the two weak Hamiltonian insertions, we identified that certain relevant diagrams can be efficiently computed via the split-even computational strategy. We described how to achieve a significant variance suppression with the use of noise sources and the split-even approach for two out of the four required diagrams, including the quark-disconnected diagram, which would otherwise exhibit significant statistical noise with standard methods. For the remaining two diagrams, we suggested employing point-source techniques to efficiently evaluate the associated correlation function.

We then outlined several numerical approaches for the spectral reconstruction of the mixing amplitudes from lattice correlation functions. Furthermore, we identified the relevant systematic uncertainties inherent to the spectral reconstruction process, including  biases resulting from smearing kernel mismatches, and the extrapolation to zero smearing width. We analysed the influence of the underlying resonance spectrum on the spectral reconstruction, providing guidance on suitable choices for the smearing parameter $\epsilon$. Given the relatively benign nature of the S-wave resonance spectrum near the $D$-meson mass, the mass scale of interest to extract the mixing amplitudes, we anticipate favorable conditions for the reliable application of the spectral reconstruction formalism.

This formalism enables, for the first time, a rigorous first-principles calculation of long-distance contributions to neutral $D$-meson mixing that can be directly compared with experimental data. Certain advantageous features, particular to neutral $D$-meson mixing, such as employing the split-even strategy for disconnected diagrams, significantly enhance computational feasibility compared to previous expectations. This development represents an important advance towards deepening our understanding of the charm-quark sector and potentially identifying signals of new physics through precision studies of neutral $D$-meson mixing.

\section*{Acknowledgments}
We warmly thank Tim Harris for enlightening conversations on the split-even approach. M.D.C. thanks Laurent Lellouch, Alessandro Lupo and Luca Silvestrini for insightful discussions and exchanges. F.E. thanks Maria Laura Piscopo for helpful discussions on the phenomenology of $D$-meson mixing. M.D.C. has received funding from the European Union’s Horizon Europe research and innovation programme under the Marie Sk\l{}odowska-Curie grant agreement No.~101108006. F.E. has received funding from the European Union's Horizon Europe research and innovation programme under the Marie Sk\l{}odowska-Curie grant agreement No.~101106913. M.T.H is supported in part by UK STFC grants ST/X000494/1 and ST/T000600/1. M.T.H. is further supported by UKRI Future Leaders Fellowship MR/T019956/1.

\bibliography{refs}

\end{document}